\documentclass[letterpaper,titlepage,11pt]{article}

\usepackage{tikz} 
\usetikzlibrary{matrix,arrows}

\usepackage{amssymb,amsmath,amscd,cite,array}
\usepackage{times,nicefrac,hhline}
\usepackage[mathscr]{eucal}

\numberwithin{equation}{section}

\newcommand{\oao}[2]{{#1\atopwithdelims[]#2}}

\def\zi{\mathbb{Z}}
\def\er{\mathbb{R}}
\def\ci{\mathbb{C}}
\def\en{\mathbb{N}}

\def\caln{\mathcal{N}}

\def\d{{\partial}}

\def\di{\text{d}}

\def\slr{SL(2,\mathbb{R})}
\def\slc{SL(2,\mathbb{R})/U(1)}

\def\p{\partial}

\def\Id{{\rm 1\kern-.28em I}}

\def\ds{\displaystyle}
\long\def\symbolfootnote[#1]#2{\begingroup%
\def\thefootnote{\fnsymbol{footnote}}\footnote[#1]{#2}\endgroup}
\def\ee{\mathrm{e}}

\begin{document}


\begin{titlepage}

\rightline{\vbox{\small\hbox{\tt CPHT-RR-042-0510}}}
\vskip 1cm

\centerline{\LARGE  Gauge Threshold Corrections for $\mathcal{N}=2$ Heterotic}
\vskip 0.4cm
\centerline{\LARGE Local Models with Flux, and Mock Modular Forms}

\vskip 1.2cm
\centerline{\bf Luca Carlevaro$^{\lozenge,\,\clubsuit}$ and Dan Isra\"el$^{\spadesuit,\,\heartsuit}$\symbolfootnote[2]{Email:
luca.carlevaro@cpht.polytechnique.fr, israel@lpthe.jussieu.fr}}
\vskip 0.5cm
\centerline{\sl $^\lozenge$Centre de Physique Th\'eorique, Ecole Polytechnique,
 91128 Palaiseau, France\footnote{Unit\'e mixte de Recherche
7644, CNRS -- Ecole Polytechnique}}

\vskip 0.3cm
\centerline{\sl $^\spadesuit$Institut d'Astrophysique de Paris,
98bis Bd Arago, 75014 Paris, France\footnote{Unit\'e mixte de Recherche
7095, CNRS -- Universit\'e Pierre et Marie Curie}}

\vskip 0.3cm
\centerline{\sl $\heartsuit$
Laboratoire de Physique Th\'eorique et Hautes Energies, Universit\'e Pierre et Marie Curie,} 
\centerline{4 place Jussieu, 75252 Paris CEDEX 05, 
France\footnote{Unit\'e Mixte de Recherche 7589, CNRS -- Universit\'e Pierre et Marie Curie}}

\vskip 0.3cm
\centerline{\sl $^\clubsuit$LAREMA, D\'epartement de Math\'ematiques, Universit\'e d'Angers,}
\centerline{2 Boulevard Lavoisier, 49045 Angers, France\footnote{Unit\'e mixte de Recherche
6093, CNRS -- LAREMA}
}

\vskip 1.5cm

\centerline{\bf Abstract} \vskip 0.5cm 
We determine threshold corrections to the gauge couplings in local models of $\mathcal{N}=2$ smooth heterotic compactifications 
with torsion, given by the direct product of a warped Eguchi--Hanson space and a two-torus, together with a line bundle. Using 
the worldsheet \textsc{cft} description previously found and by suitably regularising the infinite target space volume 
divergence, we show that threshold corrections to the various gauge factors are governed by the non-holomorphic completion of the 
Appell--Lerch sum. While its holomorphic Mock-modular component captures the contribution of states that 
localise on the blown-up two-cycle, the non-holomorphic correction originates from non-localised bulk states. We infer from this
analysis universality properties for $\mathcal{N}=2$ heterotic local models with flux, based on target space modular invariance and 
the presence of such non-localised states. We finally determine the explicit dependence of these one-loop gauge threshold corrections 
on the moduli of the two-torus, and by  S-duality we extract the corresponding string-loop and E1-instanton corrections to the K\"ahler
potential and gauge kinetic functions of the dual type \textsc{i} model. In both cases, the presence of non-localised bulk states
brings about novel perturbative and non-perturbative corrections, some features of which can be interpreted in the light
of analogous corrections to the effective theory in compact models.

\noindent

\vfill

\end{titlepage}

\tableofcontents

\section{Introduction}

Supersymmetric compactifications of the heterotic string~\cite{Gross:1985fr} were 
soon recognised as a very successful approach to string phenomenology. A crucial role is played by the modified 
Bianchi identity for the  field strength  of the Kalb--Ramond two-form. It should 
include a contribution from the Lorentz Chern--Simons three-form coming from the anomaly-cancellation 
mechanism~\cite{Green:1984sg}, that cannot be neglected in a consistent 
low-energy truncation of the heterotic string:
\begin{equation}
\di \mathcal{H} = \alpha ' \left( {\rm tr}\, \mathcal{R}(\Omega_-) \wedge \mathcal{R}(\Omega_-)
-\mathrm{Tr}_\textsc{v} \, \mathcal{F} \wedge  \mathcal{F}\right)\, .
\label{bianchid}
\end{equation}
Consistent torsionless compactifications can be achieved with an embedding of
the spin connexion in the gauge connexion. For more general bundles, 
the Bianchi identity~(\ref{bianchid}) is in general not satisfied locally, leading to non-trivial 
three-form fluxes, i.e. manifolds with non-zero torsion. These compactifications with torsion were explored in the early days 
of the heterotic string~\cite{Strominger:1986uh,Hull:1986kz}. Their analysis is quite involved, as generically the compactification manifold is not even conformally K\"ahler. In view of this complexity, it is usefull  to describe more quantitatively such flux compactifications with non-compact geometries that can be viewed as {\it local models} thereof. In type \textsc{iib} flux compactifications~\cite{Giddings:2001yu}, an important r\^ole is devoted to throat-like regions of the compactification manifold, whose flagship is the Klebanov---Strassler  background~\cite{Klebanov:2000hb}.

Heterotic torsional geometries, having only \textsc{nsns} three-form and gauge fluxes, are expected to allow for a tractable worldsheet  description. Recently, it was shown in a series of 
works~\cite{Adams:2006kb,Adams:2009zg,McOrist:2010ae,Quigley:2011pv,Blaszczyk:2011ib,Blaszczyk:2011hs,Quigley:2012gq,Adams:2012sh}
that worldsheet theories for such flux geometries can be defined as the infrared limit of some classes of $(0,2)$ gauged linear 
sigma models. This very interesting approach does not however allow for the moment to perform computations of physical quantities in these 
torsional backgrounds, as only quantities invariant under RG-flow can be handled. 

The most studied examples of supersymmetric heterotic flux compactification are elliptic fibrations  $T^2 \hookrightarrow \mathcal{M} \to K3$, where the
 $K3$ base is warped. Those backgrounds, that correspond to the most generic 
$\mathcal{N}=2$ torsional compactifications~\cite{Melnikov:2010pq},  can be equipped with a gauge bundle that is the tensor product of a Hermitian-Yang-Mills bundle over the $K3$ base with a holomorphic line bundle on $\mathcal{M}$. For these geometries, that were found 
in~\cite{Dasgupta:1999ss} using string dualities, a proof of the existence of a family of smooth solutions to the Bianchi identity with flux has only appeared recently~\cite{Fu:2006vj,Becker:2006et,Fu:2008ga}.

Considering as a base space a Kummer surface ({\it i.e.} the blow-up of a  $T^4/\mathbb{Z}_2$ orbifold), an interesting strongly warped regime occurs when the blow-up parameter $a$ of one of the two-cycles is significantly smaller (in string units) than the five-brane charge measured around this cycle, provided small instantons appear in the singular limit. As is shown in~\cite{carlevaro}, one can define a sort of 'near-bolt' geometry, that describes the neighbourhood of one of the 16 resolved $A_1$ singularities, which is decoupled from the bulk. To this end, a {\it double scaling limit} is defined  by sending the asymptotic string coupling $g_s$ to zero, while keeping the ratio $g_s/ a$ fixed in string units, which plays the r\^ole of an effective coupling constant. It  consistently defines a local model for this whole class of $\mathcal{N}=2$ compactifications. More generically, this model can be defined for any value of the five-brane charge.

Remarkably, as we have shown in~\cite{carlevaro}, the corresponding worldsheet non-linear sigma model admits a solvable worldsheet 
\textsc{cft} description, as an asymmetrically gauged \textsc{wzw} model. The existence of a  worldsheet \textsc{cft} first implies that these backgrounds are exact heterotic string vacua to all orders in $\alpha'$, once included the worldsheet quantum corrections to the defining gauged \textsc{wzw} models. Secondly, one can take advantage of the exact \textsc{cft} description in order, for instance, to determine the full heterotic spectrum as was done in~\cite{carlevaro}. It involves BPS and non-BPS representations of the $\mathcal{N}=2$ superconformal algebra, 
that correspond respectively to states localised in the vicinity of the resolved singularity and to a continuum of delta-function normalisable states that propagate in the bulk. 

Having a good knowledge of the worldsheet conformal field theories corresponding to these torsional backgrounds allows to go beyond the large volume limit and tree-level approximation upon which most works on type II flux compactifications are based. In this respect, interesting quantities are gauge threshold corrections, as they both correspond to a one-string-loop effect, which 
only receives fivebrane instanton corrections, and are sensitive to all order terms in the $\alpha'$ expansion, since the compactification manifold is not necessarily taken in the large-volume limit (which does not  exist generically in the heterotic case). In addition, heterotic -- type \textsc{i} duality translates one-loop gauge threshold corrections on the heterotic side to perturbative and multi-instanton corrections to the K\"ahler potential and the gauge kinetic functions on the type \textsc{i}  side. In this respect, provided a microscopic theory is available for a given heterotic model, the method of Dixon--Kaplunovsky--Louis (DKL) is instrumental in retrieving (higher) string-loop and Euclidean brane instanton  corrections to these type \textsc{i} quantities, from a one-loop calculation on the heterotic side, even when the type \textsc{i} S-dual model is unknown. 

This perspective looks particularly enticing from the type~\textsc{i} vantage point, since although remarkable advances have been 
accomplished to understand the perturbative  tree-level physics  of flux compactifications~\cite{Grana}, 
non-perturbative effects and string-loop corrections 
continue to often prove fundamental to lift remnant flat directions in the effective potential 
or ensure a chiral spectrum. Thus, although progresses are still at an early stage, 
the r\^ole of Euclidean brane instanton corrections in  central issues such as moduli 
stabilisation~\cite{Kachru:2003aw,Denef:2005mm,Camara:2007dy}
and supersymmetry breaking~\cite{Argurio:2007qk,Aharony:2007db,Aganagic:2007py,Blumenhagen:2009gk} have been  
intensively studied. In addition, non-perturbative effects  can also induce new interesting couplings in the 
superpotential~\cite{Billo:2002hm,Blumenhagen:2006xt,Ibanez:2006da,Florea:2006si,Akerblom:2006hx,Franco:2007ii,Blumenhagen:2007zk,Billo:2007sw,Bianchi:2007wy,Marchesano:2009rz,Billo':2010bd,Bianchi:2011qh},
while both instanton~\cite{Aparicio:2011jx} and  string-loop corrections~\cite{Becker:2002nn} to the K\"ahler potential of the effective theory prove to be useful to address the problem of the hierarchy of mass scales in large volume  
scenarii~\cite{Balasubramanian:2005zx,Conlon:2005ki}.

For all the above reasons, it appears as particularly appealing to be able to explicitly compute one-loop heterotic gauge threshold corrections and determine their moduli dependence for a  smooth heterotic background, incorporating back-reacted \textsc{nsns} flux. 
To this end, we consider in the present paper a family of {\it non-compact} models giving a {\it local} description of the simplest 
non-K\"ahler elliptic fibration $T^2 \hookrightarrow \mathcal{M} \to K3$, where the fibration reduces to a direct product. 
Locally, the geometry is given by $T^2 \times \widetilde{\textsc{eh}}$, where $\widetilde{\textsc{eh}}$ is the warped Eguchi--Hanson space. These $\mathcal{N} =2$ heterotic backgrounds also accommodate line bundles over the resolved $\mathbb{P}^1$ of the Eguchi--Hanson space, corresponding to Abelian gauge fields which, from the Bianchi identity~(\ref{bianchid}) perspective, induce a non-standard embedding of the gauge connection into the Lorentz connection. For the $Spin(32)/\zi_2$ heterotic theory, the exact \textsc{cft} description for the warped Eguchi--Hanson base with an Abelian gauge fibration has been constructed in~\cite{carlevaro} as a gauged \textsc{wzw} model for 
an asymmetric super-coset of the group $SU(2)_k \times \slr_k$, for which an explicit partition function can be written. 

The presence of a  line bundle in these non-compact backgrounds breaks the $SO(32)$ gauge group to 
$SO(2m)\times \prod_r SU(n_r) \times U(1)^{r-1}$ with $m +\sum_r n_r = 16$, while the $r^\text{th}$ $U(1)$ factor is generically 
lifted by the Green--Schwarz mechanism. One-loop gauge threshold corrections to individual gauge factors can be determined by 
computing  the elliptic index constructed in~\cite{Harvey:1995fq}, which we call {\it modified elliptic genus} as it corresponds to the elliptic genus of the underlying \textsc{cft}, with the insertion of the regularised Casimir
invariant of the gauge factor under consideration. Since the microscopic theory for such heterotic 
$T^2\times \widetilde{\textsc{eh}}$ backgrounds contains as a building block the $\mathcal{N}=2$ super-Liouville theory, a 
careful regularisation of the target space volume divergence has to be considered. This concern is also in order for the 
partition function, for which a holomorphic but non-modular invariant regularisation is usually preferred,
as it results in a natural expression in terms of $\slr_k/U(1)$ characters. For the elliptic genus in contrast, the seminal work~\cite{jan} has shown that the correct regularisation scheme based on a path integral formulation is non-holmorphic but preserves modularity.   In particular, it has the virtue of taking properly into account not only the contribution to the gauge threshold corrections of states that localise on the resolved $\mathbb{P}^1$ of the warped Eguchi--Hanson space (constructed from discrete $\slr_k/U(1)$ representations), but  especially  the contribution of non-localised bulk states, which compensates for an otherwise present holomorphic anomaly.

Taken separately, the $\slr_k/U(1)$ factor in the localised part of the threshold corrections thus transforms as a {\it Mock modular} form,
{\it i.e.}  a holomorphic form which transforms anomalously under S-transformation, but can be completed into a non-holomorphic
modular form, also known as a {\it Maa\ss\, form}, by adding the transform of a what is commonly called a {\it shadow} function.
The concept of Mock modular form~\footnote{or {\it Mock theta functions} as he calls them in a letter to Hardy} goes back to Ramanujan,
but a complete classification of such functions and a definite characterisation of their near-modular properties has only been
achieved recently by Zwegers~\cite{zwegers}, despite many insightful papers written since the twenties on Ramanujan's examples 
(see references in~\cite{Zagier2}).
Recently, Mock modular forms have found their way in string theory. They have in particular been used to address
issues central to wall-crossing phenomena for BPS invariants for systems of D-branes~\cite{Grimm:2010gk}, 
and  to deriving a reliable index for microstate (quarter-BPS state) counting for single- and multi-centered black holes in $\mathcal{N}=4$ string theory~\cite{murthy} (see also in the same line more mathematical works~\cite{Manschot:2011dj,Manschot:2011ym}). 
They also appeared in the computation of D-instanton corrections to the hypermultiplet moduli space of type~\textsc{ii} 
string theory compactified on a Calabi--Yau threefold~\cite{Alexandrov:2012au}, and in the investigation of the 
mysterious decomposition of the elliptic genus
of $K3$ in terms of dimensions of irreducible representations of the Matthieu group $M_{24}$ 
symmetry~\cite{Eguchi:2010ej,Gaberdiel:2010ca,Eguchi:2010fg,Cheng:2011ay,Cheng:2012tq,Eguchi:2012ye}. The theory of Mock modular 
forms is finally at the core of infinite target space volume regularisation issues in non-compact \textsc{cft}s~\cite{jan,Eguchi:2010cb,Ashok:2011cy,Sugawara:2011vg,Ashok:2012qy}, which directly concerns the calculation of gauge 
threshold corrections tackled in this paper.

In the present analysis, we will in particular focus on a family of heterotic torsional local models supporting a line bundle
$\mathcal{O}(1) \oplus \mathcal{O}(\ell)$ with gauge group $SO(28)\times U(1)$ (which
is enhanced to $SO(28)\times SU(2)$ when $\ell=1$). The regularised threshold corrections to the these gauge couplings are shown
to be given in terms of weak harmonic Maa\ss \, forms based on the  non-holomorphic completion of Appell--Lerch sums, a major 
class of Mock modular forms treated by Zwegers.
A deeper physical insight into the shadow function featured in the bulk state contribution is achieved
by investigating the $\ell=1$ model, whose interacting part enjoys an enhanced $(4,4)$ worldsheet superconformal symmetry.
We observe in this particular case that localised effects splits on the one hand into $\nicefrac{4}{\chi(K3)}$  of the gauge 
threshold corrections for a $T^2\times K3$ model, for which there is a rich 
literature both in heterotic and type~\textsc{i} theories~\cite{Bachas:1996zt,Antoniadis:1996vw,Kiritsis:1997hf,bachas1,Bachas:1997xn,Lerche:1998gz,Lerche:1998nx,Antoniadis:1999ge,Kiritsis:2000zi},  
and on the other hand into a Mock modular form $F(\tau)$ encoding the presence of warping due to \textsc{nsns} flux threading the geometry. The
non-holomorphic regularisation mentioned above dictates a completion in terms of non-localised bulk states
which leads to the harmonic Maa\ss\, form $\widehat{F}(\tau)=F(\tau)+g^*(\tau)$, where $g(\tau)$ is the {\it shadow} function 
determined from a holomorphic anomaly equation for $\widehat{F}$. Now, some local models such as the $T^2 \times \widetilde{\textsc{eh}}$
background considered here have a non-trivial boundary at infinity, allowing for non-vanishing five-brane charge,
which would globally cancel  when patching these models together to obtain a warped $K3$ compactification on $T^2 \times \widetilde{K3}$.
The appearance of the Maa\ss\, form $\widehat{F}$ thus results from the combination of the non-compactness of the space 
(with boundary) and the presence of flux with non-vanishing five-brane charge, both things being somehow correlated. 
This analysis can then be generalised to the $\ell >1$ models. However, because of reduced worldsheet supersymmetry  the interpretation 
in terms of $K3$ modified elliptic genera is lost for these theories.

We then carry out a careful analysis of the polar structure of the modified elliptic genus determining these gauge threshold corrections,
which shows that they share the same features with respect to unphysical tachyons and anomaly cancellation
as well-known $\mathcal{N}=2$ heterotic compactifications. This allows us to identify some universality properties
for $\mathcal{N}=2$ heterotic local models with non-localised bulk states. It also sets the stage to compute explicitly
the dependence of these gauge threshold on the $T^2$ moduli, for the 
$\mathcal{O}(1) \oplus \mathcal{O}(1)$ model taken as an example.  The modular integrals can by performed by the 
celebrated {\it orbit method}, which consists in unfolding the fundamental domain of the modular group against the $T^2$ lattice sum.
From these threshold calculations we recover in particular the $\beta$-functions of the effective four-dimensional theory, in
perfect agreement with field theory results based on hypermultiplet counting, previously performed by constructing
the corresponding massless chiral and anti-chiral primaries in the \textsc{cft}~\cite{carlevaro}.

We then consider the type \textsc{i} S-dual theory. Contrary to usual orbifold compactifications
half D5-branes at the orbifold singularities are absent from these local models as the $A_1$ singularity is resolved
and anomaly cancellation is ensured by $U(1)$ instantons on the blown-up $\mathbb{P}^1$.  We  proceed to extract 
the perturbativeand non-perturbative corrections to the K\"ahler potential and the gauge kinetic functions, by the 
DKL method. The contribution from states that localise on the resolved two-cycle yields corrections similar to those
expected for compact models, which separate into string-loop corrections and multi-instanton corrections
due to E1 instantons wrapping the $T^2$. In addition, as for the original heterotic gauge threshold corrections,
non-localised bulk modes bring about novel types of corrective terms, both perturbative and non-perturbative, 
to the  K\"ahler potential and the gauge kinetic functions. Though recently gauge threshold corrections
for local orientifolds in type~\textsc{iib} models have been successfully computed~\cite{Conlon:2009kt,Conlon:2009qa},
this is to our knowledge the first such calculation carried out for local heterotic  models incorporating
back-reacted \textsc{nsns} flux, determining all-inclusively all perturbative and non-perturbative
corrections originating from both localised and bulk states. 

In order to be able to make sensible phenomenological predictions, one should of course properly engineer 
the gluing of sixteen of these heterotic local models into a $T^2 \times \widetilde{K3}$ compactification,
 which would give us a proper effective field theory understanding of bulk state contributions.
This could be of particular interest, on the dual type~\textsc{i} side, to clarify the r\^ole of these
novel bulk state contributions we find in E1-instanton corrections, which include
an infinite sum over descendants of the modified elliptic genus, as functions of the induced  $T^2$ moduli.
These could then be put into perspective with supergravity~\cite{Berg:2005ja,Berg:2005yu} or field theory~\cite{Billo:2007sw}
calculations of Euclidean brane instanton corrections for compact models.

This work is organized as follows. In section~\ref{sec:review} we define the heterotic supersymmetric solutions of interest, and 
recall their worldsheet description. In section~\ref{sec:general} we set the stage for the threshold corrections and provide 
general aspects of the latter. In section~\ref{sec:modgen} we compute the modified elliptic genus that enters into the modular integral. 
Finally in section~\ref{sec:moddepgen} we compute the integral over the fundamental domain in order to recover the moduli dependence, and 
discuss in section~\ref{sec:typeone} the type \textsc{i} dual interpretation in terms of perturbative and non-perturbative corrections. 
Some material about superconformal characters, modular form, and some lengthy computations are given in the various appendices.


\section{Heterotic flux backgrounds on Eguchi--Hanson space}
\label{sec:review}
In this section we briefly descripe the heterotic solution of interest, for which the threshold corrections computations 
will be done, both from the point of view of supergravity and worldsheet conformal field theory. 

\subsection{The geometry}\label{sec:res}
We consider a family of heterotic backgrounds whose transverse geometry is described
by the six-dimensional space $\mathcal{M}_6=T^2 \times \widetilde{\textsc{eh}}$, where the four-dimensional non-compact 
factor $\widetilde{\textsc{eh}}$ is the warped Eguchi-Hanson space, the Eguchi--Hanson space (\textsc{eh})
being the resolution by blowup of a $\ci^2/\zi_2$, 
or $A_1$, singularity.  It provides a workable example of a smooth background with intrinsic torsion induced by the presence 
of \textsc{nsns} three-form flux. In the following, we will be concerned with the heterotic $Spin(32)/\zi_2$ theory, 
but our results can be straightforwardly extended to the $E_8\times E_8$ gauge group.

The two-torus is characterised by two complex moduli, the K{\"a}hler class and the complex
structure, which we denote respectively by $T$ and $U$, related to the 
string frame metric and B-field as:
\begin{equation} \label{T2mod}
T= T_1 + i  T_2 = \frac{B_{12}+i \sqrt{\det G}}{\alpha'}\,,\qquad 
U=U_1+i U_2 = \frac{G_{12}+i \sqrt{\det G}}{G_{11}}\,.
\end{equation}
Accordingly, the full six-dimensional torsional geometry takes the form:
\begin{equation}\label{metric2}
\di s^2_6  = \eta_{\mu\nu}\,\di x^\mu \di x^\nu +  
    \frac{\alpha' T_2}{U_2}\left|\di x^1 + U \di x^2 \right|^2 + 
H(r) \,\di s^2_\textsc{eh}\,. 
\end{equation}
where the torus coordinates have periodicity $(x^1,\,, x^2)\sim(x^1+2\pi,\,, x^2+2\pi)$ and the 
$A_1$ space is locally described by the Eguchi--Hanson (\textsc{eh}) metric:
\begin{equation}\label{EHmetric}
\di s^2_\textsc{eh} =  \frac{ \di r^2}{1-\tfrac{a^4}{r^4}} +
\frac{r^2}{4} \left(
(\sigma^\mathsf{L}_1)^2 + (\sigma^\mathsf{L}_2)^2 +
\Big(1-\frac{a^4}{r^4}\Big) (\sigma^\mathsf{L}_3 )^2 \right)\,,
\end{equation}
here given in terms of the $SU(2)$ left-invariant one-forms:
\begin{equation}
\begin{array}{lll}
{\ds \sigma^\mathsf{L}_1 = \sin\psi \, \di\theta - \cos\psi \sin\theta\, \di\phi }\,, &\quad
 {\ds \sigma^\mathsf{L}_2 = -\big( \cos\psi \, \di\theta + \sin\psi \sin\theta\, \di\phi \big) } \,, & \quad
 {\ds \sigma^\mathsf{L}_3 =  \di\psi + \cos\theta \, \di\phi \,, } 
\end{array}
\end{equation}
with $\theta\in [0,\pi]$ and $\phi,\,\psi\in [0,2\pi]$. Note in particular that the $\psi$ coordinate runs over half of its original span, since for the \textsc{eh} space to be smooth, an extra $\zi_2$ orbifold is necessary to eliminate the bolt singularity at $r=a$.

The \textsc{eh} manifold is homotopic to the blown-up $\mathbb{P}^1$ resolving the original $\ci^2/\zi_2$ singularity. This two-cycle
is given geometrically by the non-vanishing two-sphere $\di s^2_{\mathbb{P}^1} = \frac{a^2}{4}\big(\di\theta^2+ \sin^2\theta\,\di\phi^2\big)$ and is Poincar\'e dual to a closed two-form which has the following {\it local}  description:
\begin{equation}\label{oeh}
\omega = - \frac{a^2}{4\pi}\,\di \left(\frac{\sigma^\textsc{l}_3}{r^2} \right) \,,\qquad  \text{with } \quad
 \int_{\mathbb{P}^1} \omega = 1\,, \quad \text{and } \quad \int_{\textsc{eh}} \omega \wedge \omega = -\tfrac12\,.
\end{equation}
In particular, the last integral yields  minus the inverse Cartan matrix of $A_1$, as expected for a resolved ADE singularity. The second
cohomology thus reduces to $H^{1,1}(\mathcal{M}_\textsc{eh})$, as it is spanned by a single generator $[\omega]$, given by the harmonic and anti-selfdual two-form~(\ref{oeh}). Globally \textsc{eh} can hence be shown to have the topology of the total space of the line bundle $\mathcal{O}_{\mathbb{P}^1}(-2)$.

\subsection{The heterotic solutions} 
The six dimensional space~(\ref{metric2}) can be embedded in 
heterotic supergravity, with a background including an \textsc{nsns} three-form~\footnote{The volume of the three-sphere 
is given in terms of the Euler angles as follows:  $ \text{Vol}(S^3)=\tfrac18\,  
\sigma^\textsc{l}_1\wedge  \sigma^\textsc{l}_2\wedge  \sigma^\textsc{l}_3= \tfrac18 \,\di(\cos\theta)\wedge\di\phi\wedge\di\psi$.} $\mathcal{H}$ and a varying dilaton:
\begin{subequations}\label{solH}
\begin{align}
\mathrm{e}^{2 \Phi(r)} & = g_s^{\,2} \, H(r) =  g_s^{\,2}  \left(  1+ \frac{2\alpha' \mathcal{Q}_5 }{r^2} \right)    \label{solHc}\,, \\
\mathcal{H} & = -H \ast_{\textsc{eh}} \di H = 4\alpha' \mathcal{Q}_5   \left(  1- \frac{ a^4}{r^4} \right)
 \text{Vol}(S^3)\,,   \label{solHb}
\end{align}
\end{subequations}
where $\mathcal{Q}_5$ is the charge of the stack of back-reacted \textsc{ns} five-branes wrapped around the $T^2$ which are recovered in the blowdown limit, opening a throat at $r=0$. When the $A_1$ singularity is resolved the \textsc{ns} five-branes are no longer present and we obtaine a smooth non-K\"ahler geometry threaded by three-form flux, with non-vanishing five-brane charge 
$4\pi^2\alpha' \mathcal{Q}_5=-\int_{\widetilde{\textsc{eh}}} \mathcal{H}$ due to the boundary  $\partial\mathcal{M}_{\textsc{eh}} = \er P^3$.

This background preserves $\mathcal{N}_{\textsc{st}} =(0,2_4)$, resulting from the existence of
a pair of  $Spin(6)$ spinors $\epsilon^{i}$, $i=1,2$ constant with 
respect to only one of the two generalised spin connections $\Omega_{\pm\phantom{a}b}^{\phantom{\pm}a}= 
\omega_{\widetilde{\textsc{eh}}\phantom{a}b}^{\phantom{\textsc{eh}}a} \pm \tfrac12 \mathcal{H}_{\phantom{a}b}^{a}$:
\begin{equation}
\big( \d_{\mu} + \tfrac14\,\Omega_{+\phantom{ab}\mu}^{\phantom{+}ab}\,\Gamma_{ab} \big) \epsilon^i = 0\,, \qquad
i=1,2\,,
\end{equation}
where $\mu$ and $a,b$ are six-dimensional space and frame indices respectively.

\paragraph{Bianchi identity and line bundle}
In addition to satisfying the supersymmetry equations, anomaly cancellation requires a heterotic background 
to solve the Bianchi identity:
\begin{equation}\label{bi}
\di\mathcal{H} = -\alpha' \Big( \text{Tr}_V \mathcal{F}\wedge \mathcal{F} - \text{tr}\,\mathcal{R}(\Omega_-)\wedge
\mathcal{R}(\Omega_-)\Big)   \,.
\end{equation}
For non-zero fivebrane charge $\mathcal{Q}_5$ the \textsc{nsns} three-form~(\ref{solH}b) is not closed.
A non-standard embedding of the Lorentz connection into the gauge 
connection has therefore to be used to satisfy the Bianchi identity. This can be achieved by considering a multi-line bundle
\begin{equation}  
\mathcal{L}=\bigoplus_{a=1}^{16} \mathcal{O}_{\mathbb{P}^1}(\ell_a)
\end{equation}
where the individual line bundles, 
labelled by $a$, are embedded in an Abelian principal bundle 
valued in the Cartan subalgebra of $SO(32)$. The resulting heterotic gauge field, characterised by a 
vector of magnetic charges (or 'shift vector') $\vec{\ell}$, reads:
\begin{equation} \label{gaugebund}
 \mathcal{F} = -2\pi \,\omega \sum_{a=1}^{16}\ell_a \,\textsf{H}^a \,, \qquad   \textsf{H}^a\in \mathfrak{h}(SO(32)) \quad  \text{ with } \quad  \text{Tr}\,\textsf{H}^a\textsf{H}^b=-2\delta^{ab}\,.
\end{equation}
Since the above gauge field is along the anti-selfdual and harmonic two-form of \textsc{eh},
it satisfies the  Hermitian Yang--Mills (or Uhlenbeck--Donaldson--Yau) equations: $J \lrcorner \mathcal{F}=0$ and
$\mathcal{F}^{(0,2)}= \mathcal{F}^{(2,0)}=0$. Hence it does not further break the existing spacetime supersymmetry of 
the background.

Furthermore, it solves the Bianchi identity~(\ref{bi}) in the regime where the gravitational contribution is negligible, {\it i.e.} 
in the large five-brane charge limit :
\begin{equation}
\mathcal{Q}_5= -\frac{1}{4\pi^2 \alpha'} \int_{\er P_3 , \, \infty}\mathcal{H}=\vec{\ell}^{\;2}  \gg 1 \,.
\end{equation}
As we will see later on, in a specific double-scaling limit of the metric~(\ref{metric2}) the background~(\ref{solH}) 
admits an exact worldsheet \textsc{cft} description, even beyond this large-charge limit.

Beyond the large-charge approximation, one can consider corrections resulting
from the integrated Bianchi identity, which are captured by the tadpole equation: 
\begin{equation}\label{tad1}
\frac{1}{4\pi^2\alpha'}\int_{\textsc{eh}} \Big[( \di\mathcal{H}+ 
\alpha'\big(\text{Tr}_{\text{V}}\mathcal{F}\wedge\mathcal{F} - 
\text{tr} \,\mathcal{R}(\Omega_-)\wedge \mathcal{R}(\Omega_-)\big)\Big] = 0 
\quad  \Longrightarrow \quad
 \mathcal{Q}_5 = \vec{\ell}^{\;2}  - 6\,.
\end{equation}
This is particular determines the allowed shift vectors for a given five-brane charge, and the resulting
breaking of the gauge group.

In addition to the tadpole equation, dictated by anomaly cancellation, two more
constraints restrict the value of the shift vector $\vec{\ell}$, namely: 
\begin{itemize}
\item[$i)$] a Dirac quantisation condition  for the adjoint representation of $SO(32)$, requiring the integrated first Chern class of 
the line bundle $\mathcal{L}$ to have only integer or half-integer entries corresponding to bundles with or without vector structure 
respectively:
\begin{equation}\label{bs}
\left\{ \begin{array}{ll}
\vec{\ell}\in \zi^{16} \,,  &  \Rightarrow \text{ bundle with vector structure} \\[6pt]
 \vec{\ell}\in \big(\zi + \tfrac12\big)^{16} &  \Rightarrow \text{ bundle without vector structure} 
\end{array} \right.
\end{equation}
\item[$ii)$]
a so-called 'K-theory' condition  which must be further imposed on the first Chern class of $\mathcal{L}$ 
to ensure that the gauge bundle admits spinors:
\begin{equation}\label{bsb}
  c_1(\mathcal{L})\in H^2(\textsc{eh},2\zi) \quad \Rightarrow  \quad \sum_{a} \ell_a \equiv 0 \text{ mod } 2 \,.
\end{equation}
\end{itemize}

\subsection{The double-scaling limit}

We will now introduce a consistent   {\it double-scaling limit} of the torsional background~(\ref{metric2})--(\ref{solH}), 
which decouples the bulk physics from the physics in the vicinity of the resolved $A_1$ singularity:
\begin{equation}\label{DSL}
g_s \to 0 \quad , \qquad  \lambda = \frac{g_s \sqrt{\alpha'}}{a} \quad \text{fixed}\,.
\end{equation}
This specific regime isolates  the dynamics near the blownup two-cycle, but still keeps the singularity resolved. 
In particular if we wrap five-branes around the two-cycle, their tension will be proportional to 
$\text{Vol}(\mathbb{P}^1)/g_s^2$ and thus held fixed, so that no extra massless degrees of freedom appear in 
the double scaling limit. This procedure results in an interacting theory whose effective coupling constant is set by the double-scaling parameter. Interestingly enough, it has been shown in~\cite{carlevaro} that in this limit the heterotic fluxed background
admits a solvable \textsc{cft}, which we will introduce shortly.

The resulting near-horizon  geometry arising in this regime can best expressed in the new radial coordinate $\cosh \rho = (r/a)^2$:
\begin{equation}\label{DSgeom}
\di s^2_6  = \eta_{\mu\nu}\,\di x^\mu \di x^\nu +  
    \frac{\alpha' T_2}{U_2}\left|\di x^1 + U\di x^2 \right|^2 +  
\frac{\alpha' \mathcal{Q}_5}{2}\Big[ {\rm d}\rho^2 + (\sigma^\mathsf{L}_1)^2+
(\sigma^\mathsf{L}_2)^2 + \tanh^2 \rho \, (\sigma^\mathsf{L}_3 )^2\Big]\,.
\end{equation}
Furthermore, while the dilaton is affected by the near-horizon limit, the gauge field and the three-form, which are localised respectively 
on the blown-up two-cycle and on the $\er P^3$ boundary of \textsc{eh}, remain untouched. Their formulation in the new coordinate are:
\begin{subequations}\label{eqDSL}
\begin{align}
\mathcal{H}& = -4\alpha' \mathcal{Q}_5\,\tanh^2 \rho\, \text{Vol}(S^3) \,, \qquad 
\mathrm{e}^{2 \Phi(\rho)}  =  \frac{2 \lambda^2 \mathcal{Q}_5}{\cosh \rho} \label{dilateq}\\
\mathcal{F}& =    -\frac{1}{2\cosh\rho}\big(\tanh\rho\,\di\rho\wedge\sigma^\mathsf{L}_3 - \sigma^\mathsf{L}_1\wedge\sigma^\mathsf{L}_2\big)\,
\sum_{a=1}^{16}\ell_a \,\textsf{H}^a \,. \label{Feq}
\end{align}
\label{sugrasol}
\end{subequations}
Finally, the tadpole equation correcting the five-brane charge is also modified:
\begin{equation}\label{tad2}
\mathcal{Q}_5|_ {\text{n.h.}}= \vec{\ell}^{\;2}  - 4\,.
\end{equation}
The change with respect to expr.~(\ref{tad1}), namely the jump of $-2$ units in the integrated first
Pontryagin class of the six-dimensional manifold, is due to the decoupling of the boundary of the space,
 because of the now asymptotically vanishing conformal factor $H(\rho)$.

\subsection{The worldsheet \textsc{cft} description}\label{sec:cft}

The exact \textsc{cft} description for the double-scaling limit of the heterotic background~(\ref{DSgeom})--(\ref{eqDSL}) for a line bundle 
$\bigoplus_{a=1}^{16} \mathcal{O}_{\mathbb{P}^1}(\ell_a)$ satisfying the tadpole equation~(\ref{tad2})  has been derived in~\cite{carlevaro}. The 
interacting part is given by an asymmetrically gauged $SU(2)_k\times \slr_{k'}$ super-\textsc{wzw} model 
with $\mathcal{N}_{\textsc{ws}}=(0,1)$ worldsheet supersymmetry:
\begin{equation}
\label{cftback}
\frac{\big(SU(2)_k/\zi_2 \big)\times SL(2,\mathbb{R})_{k'}}{U(1)_\mathsf{L} \times U(1)_\mathsf{R}} \,.
\end{equation}
The gauging in this theory is asymmetric and results from acting on the group elements $(g_1,\, g_2)\in
SU(2) \times SL(2,\mathbb{R})$ as follows:
\begin{equation}\label{gtrans}
(g_1,\, g_2)  \rightarrow \big(g_1\ee^{i\sigma_3\alpha} ,\, \ee^{i\sigma_3\beta} g_2\ee^{i\sigma_3\alpha} \big)\,
\end{equation}
The $SU(2)_k$ factor is also modded out by the $\zi_2$ action $\mathcal{I}:\, g_1 \mapsto -g_1$, which leaves the current algebra 
invariant. This orbifold is at the \textsc{cft} level the algebraic equivalent of the geometric $\zi_2$ orbifold reducing the periodicity of the angular coordinate $\psi$ to $[0,2\pi]$ (see section~\ref{sec:res}). The 16 left-moving Weyl fermions are also 
minimally coupled to the worldsheet gauge fields with charge $\{ \ell_i, i=1,\ldots,16 \}$.  

In order to obtain a gauge-invariant worldsheet action  the following conditions on the levels of the affine superconformal algebras 
are obtained: 
\begin{equation}\label{k-level}
k'=k \,, \qquad     \qquad k=2(\vec{\ell}^{\;2}-1)\,.
\end{equation}
In particular, we recognise in the second constraint the \textsc{cft} equivalent of the tadpole equation~(\ref{tad2}). 
To simplify 
the notations and the computations we will restrict to $U(1)^2$ bundles with shift vector $\vec{\ell} =(1,\ell,0^{14})$. In this subclass 
of models the left superconformal symmetry of the $SL(2,\mathbb{R})/U(1)$ factor is enhanced to 
$\mathcal{N}_{\textsc{ws}}=2$. For this specific choice of shift vector, the condition~(\ref{k-level}) fixes $k=2\ell^2$. 
The K-theory condition~(\ref{bs}), in this case, restricts $\ell$ to be an odd-integer (as we shall see below, this condition 
is also needed in the \textsc{cft}).

Integrating out the worldsheet gauge fields classically, one finds a non-linear sigma model~\cite{carlevaro} whose background 
metric, B-field and dilaton exactly reproduce the double-scaling limit of the torsional background of interest, given in eq.~(\ref{eqDSL}).

\subsubsection*{One-loop partition function}\label{sec:part}

To write down the partition function for $Spin(32)/\mathbb{Z}_2$ heterotic strings in the torsional background~(\ref{eqDSL})
we combine the partition function for the four-dimensional coset \textsc{cft} with the flat space-time part (in the light-cone gauge), the 
remaining 28 free left-moving Majorana-Weyl fermions and 
a toroidal lattice, written in  the Lagrangian formulation:
\begin{equation}\label{latt}
\Gamma_{2,2}(T,U)= \frac{T_2}{\tau_2}\sum_{n_1,n_2,m_1,m_2}\text{exp}
\left[2\pi i T\,\text{det}\,A-\frac{\pi\,T_2}{\tau_2\,U_2}\left|(1,\;U)\;A\begin{pmatrix}\tau\\ -1\end{pmatrix} \right|^2 \right]\,,
\end{equation}
where the matrix $A$ encodes the topologically non-trivial mapping of the string worldsheet onto the target-space torus:
\begin{equation}
A= \begin{pmatrix}n_1 & m_1\\n_2 & m_2\end{pmatrix}\,, \qquad n_i,\,m_i\in \zi,\quad i=1,2\,.
\end{equation}

The representations that appear in the spectrum of the coset theory~(\ref{cftback}) are labelled in particular 
by the spin of $SL(2,\mathbb{R})$ irreducible representations, that fit into two classes:
\begin{itemize}
\item a discrete spectrum of normalisable states localised on the blown-up two-cycle at the resolved $A_1$ singularity. These are 
labelled by a real spin $J$, which runs over the range: $\tfrac12 < J < \tfrac{k+1}{2} $. The corresponding coset representations 
are BPS and have massless ground states. We will denote their contribution to the partition function by $T_{\text{d}}$. 
\item  a continuous spectrum of $\delta$-function normalisable states, which live 
in the weakly coupled asymptotic region  $\varrho\rightarrow\infty$. They are labelled by a continuous $\slr$ 
spin $J=\tfrac12+iP$, with $P\in \er^+$ and correspond to non-BPS  massive representations in the coset. 
We denote their contribution to the partition function by $T_{\text{c}}$. 
\end{itemize}
Combining all together, we obtain the total partition function for all models with line-bundle $\mathcal{O}_{\mathbb{P}^1}(1)\oplus \mathcal{O}_{\mathbb{P}^1}(\ell)$: 
\begin{equation} \label{partfunc}
\begin{array}{rcl}
T & =&   T_{\text{d}}\, + \,T_{\text{c}} \\[4pt]
& = & {\ds \int_{\mathscr{F}}
\frac{\di^2 \tau}{\tau_2^3}\frac{\Gamma_{2,2}(T,U)}{|\eta|^8} 
\, \frac12 \sum_{a,b=0}^1 (-)^{a+b}
\frac{\bar{\vartheta} \oao{a}{b}^2}{\bar{\eta}^2}\frac12 \sum_{\gamma,\delta=0}^1 \sum_{2j=0}^{k-2}
(-)^{\delta\left(2j+\left(\frac{k}{2}-1\right)\gamma\right)} 
\sum_{m \in \mathbb{Z}_{2k}} 
\overline{C}^j_m \oao{a}{b} }
\\[4pt]
&& \qquad {\ds \times\,
\chi_{k-2}^{j+\gamma\left(\frac{k}{2}-2j-1\right)} \frac12 \sum_{u,v=0}^1 
\left( \Gamma_m^{\text{d}}\oao{a}{b}\oao{u}{v}+ \Gamma_m^{\text{c}}\oao{a}{b}\oao{u}{v} \right)
\frac{\vartheta\oao{u}{v}^{14}}{\eta^{14}}\,.}
\end{array}
\end{equation}
The contribution to the partition function~(\ref{partfunc}) of the compact part of the coset \textsc{cft}  
decomposes, on the left-moving side, into the  affine characters $\chi_{k-2}^j$ of the bosonic $SU(2)_{k-2}$~(\ref{su2-char}) affine 
algebra, and on the right-moving side, into the super-parafermion characters $C^j_m \oao{a}{b}$
of the supersymmetric $SU(2)_k / U(1)$~(\ref{relation-char-su2}).  The contributions from $SL(2,\mathbb{R})_k /U(1)$
characters with $\caln_{\textsc{ws}}=(2,2)$ superconformal symmetry are repackaged in expression
$\Gamma_m^{\text{d},\text{c}}\oao{a}{b}\oao{u}{v}$.  Localised states, in particular, are captured by the 
following partition function for discrete $\slr_k/U(1)$ representations:
\begin{multline} \label{Gamd}
\Gamma_m^{\text{d}}\oao{a}{b}\oao{u}{v} =
\sum_{2J=1}^{k}\!\!
\overline{\rm Ch}_{\text{d}} (J,\tfrac{m}{2}-J-\tfrac{a}{2}) \oao{a}{b} 
\sum_{n \in \mathbb{Z}_{2\ell}}\!\! \mathrm{e}^{-i\pi v (n+ \frac{u}{2}) }
\,{\rm Ch}_{\text{d}} \big( J,\ell(n+\tfrac{u}{2})-J -\tfrac{u}{2}\big) \oao{u}{v}\\
\times\delta_{2J-m+a,0}^{[2]}\, \delta_{2J-(\ell-1)u,0}^{[2]}\,,
\end{multline}
with $\delta^{[2]}$ the mod-two Kronecker symbol.\footnote{Note that we have included in the above partition function contributions from the 
'boundary' representation $J=\nicefrac12$. It will be in practice projected out in the partition function with all other half-integer spin states of 
$\slr_k/U(1)$ but we nevertheless include it to make the connection with the elliptic genus of the orbifolded super-Liouville theory 
more palpable.} We refer the reader to Appendix~\ref{appchar} for the definition of extended characters for discrete~(\ref{extdischar}) and 
continuous~(\ref{extcontchar}) $\slr_k/U(1)$ representations. 
It should also be pointed out that when the $SU(2)/\zi_2$ orbifold is combined with the projection by the mod-two Kronecker 
symbol $\delta^{[2]}_{2J-(\ell-1)u,0}$ and the K-theory condition~(\ref{bs}), representations with half-integer spin are 
projected out.

The contribution of $\delta$-function normalisable bulk states is encoded in the partition function for continuous $\slr_k/U(1)$ representations:
\begin{equation}\label{Gamc}
\Gamma_m^\text{c}\oao{a}{b}\oao{u}{v} = \int_0^\infty \!\! \text{d} p\,\, 
\overline{\rm Ch}_\text{c} (\tfrac{1}{2}+i p,\tfrac{m}{2}) \oao{a}{b} 
\sum_{n \in \mathbb{Z}_{2\ell}} \mathrm{e}^{-i \pi v (n+ \frac{u}{2}) }
\,{\rm Ch}_{\text{c}} \big( \tfrac{1}{2}+i p,\ell(n+\tfrac{u}{2}) \big) \oao{u}{v}\,.
\end{equation}

\paragraph{Regularisation of the infinite volume divergence:}
The decomposition of the partition function~(\ref{partfunc}) in terms of characters of discrete and 
continuous representations of the chiral $\mathcal{N}_{\textsc{ws}}=2$ superconformal algebra results from adopting a 
particular regularisation scheme of the infinite volume divergence in target-space.\footnote{To be more precise, 
this regularisation leads in principle to a non-trivial regularised density of continuous representations, see~\cite{Maldacena:2000kv}. However this is not necessary for our present purpose, which is to summarise the full string spectrum in a compact way.} 
This regularisation preserves holomorphicity of the characters; however, as  the infinite volume divergence cannot be factored out as the 
volume of a symmetry group, it breaks modular invariance. Although characters for discrete and continuous representations separately close 
under a T-transformation,  they mix under under an S-transformation. Schematically we have:
\begin{equation}\label{Sdual0}
\begin{array}{l}
\text{(discrete rep.)} \stackrel{S}{\longrightarrow} \text{(discrete rep.) + (continuous rep.)} \\[4pt]
\text{(continuous rep.)} \stackrel{S}{\longrightarrow}  \text{(continuous rep.)} \\[4pt]
\text{(identity rep.)} \stackrel{S}{\longrightarrow} \text{(discrete rep.) + (continuous rep.)} \,.
\end{array}
\end{equation}
Therefore, the full partition function~(\ref{partfunc}) is not modular invariant, but the continuons representation 
term $T_c$ is on its own.

From now on, the one-loop gauge threshold corrections~(\ref{gloop}) that we will tackle shortly  can be formulated in terms of a modified supersymmetric index, similar in spirit to the elliptic genus of the microscopic theory, for which a different kind of regularisation should be prescribed, which is modular invariant but not holomorphic~\cite{jan}.


\subsubsection*{Blowdown limit}
From the perspective of a correspondence between geometrical (supergravity)  and algebraic (\textsc{cft}) data,
we observe that the contribution $T_{\text{d}}$ from discrete representations localises at the 
bolt of the manifold and is thus related, on the geometric side, to the resolution of the $A_1$ singularity.
Consequently, the blowdown limit of the space~(\ref{DSgeom}) will be described at the microscopic level only by 
continuous representations in $T_{\text{c}}$. This is actually in keep with the fact that $T_{\text{c}}$ is by 
itself modular invariant, while extended characters for {it discrete} representations do not close under the action 
of the modular group, and in particular transform into {\it discrete + continuous} extended characters under $S$-transformation.

Correspondingly, in the $a\rightarrow 0$ limit of the supergravity solution~(\ref{solH}), we see genuine coincident heterotic 
fivebranes transverse to the $A_1$ singularity emerging, for which the $T_{\text{c}}$ partition function  
gives a microscopic description of  the near-horizon geometry. The corresponding worldsheet theory is actually a  $\mathbb{Z}_2$ orbifold of 
the Callan--Harvey--Strominger (\textsc{chs}) 
solution~\cite{Callan:1991dj}, together with a linear dilaton of charge $Q=\sqrt{2/\alpha' k}$ :
\begin{equation}\label{chs}
\mathbb{R}^{3,1}\times T^2 \times \mathbb{R}_Q \times SU(2)_{k}/\mathbb{Z}_2\,.
\end{equation}

\subsection{The massless spectrum}

The partition function~(\ref{partfunc}) gives the full spectrum of 
heterotic string on warped Eguchi-Hanson space endowed with the line bundle consisting of two $U(1)$ instantons 
with magnetic charges one and $\ell$. The unbroken gauge group $\mathrm{G}$ is the commutant of 
$U(1)_1 \times U(1)_\ell$ in $SO(32)$. It contains two Abelian factors, but only one of them, 
corresponding to the left $U(1)_{R}$ of the $SL(2,\mathbb{R})/U(1)$ super-coset, remains massless. 
The orthogonal combination, whose embedding in $SO(32)$ is given by $\vec{\ell} \cdot \vec{\mathsf{H}}$, 
is lifted by the Green--Schwarz mechanism. Thus for $\ell \neq 1$ the actual massless gauge group is  $\mathrm{G}=SO(28)\times U(1)_{R}$. 
When $\ell=1$, it is enhanced to $\mathrm{G}=SO(28)\times SU(2)$.

\begin{table}[!!ht]
\hspace{-0.8cm}
\begin{tabular}{|c||c|c||c|}
\hline
$\vec{\ell}$ & Untwisted sector & Twisted sector & Gauge bosons\\
\hline 
$(1,1,0^{14})$&$(\mathbf{28},\mathbf{2})$ &   &   $SO(28)\times SU(2)$  \\ 
  & $+$ & &    (non-normalisable)\,.  \\
&$2(\mathbf{1},\mathbf{1})$ &  &    $U(1)$ with mass $\mathfrak{m}=\tfrac{2}{\sqrt{\alpha'}}\,.$ \\[6pt]
\hhline{=::==::=}
$(1,\ell,0^{14})$ & $ \mathbf{28}_{-1} + (2\ell -1)\mathbf{28}_{\frac1\ell} $ & $ (2\ell^2-1) \mathbf{28}_{1} +(2\ell^2-2\ell+1)  \mathbf{28}_{-\frac1\ell} $ 
& $SO(28)\times U(1)_{R}$\\
$\ell \in 2\en^*+1$ &  $+$ &  $+$ &   (non-normalisable) \,.  \\
   & $  \mathbf{1}_0 + (2\ell -1) \mathbf{1}_{\frac1\ell-1} $ & $(2\ell^2-1) \mathbf{1}_0 +(2\ell^2-2\ell+1)  \mathbf{1}_{1-\frac1\ell} $ 
   & $U(1)$ with mass $\mathfrak{m}=\tfrac{2}{\sqrt{\alpha'}|\ell|}\,.$  \\[6pt]
\hline
\end{tabular}
\caption{\it Spectra of hypermultiplets and gauge bosons for models with integer shift vectors $\vec{\ell}=(1,\ell,0^{16})$.}
\label{tabops}
\end{table}

In table~\ref{tabops} the complete list of massless hypermultiplets charged under $\mathrm{G}$ is given for the $\vec{\ell}=(1,\ell,0^{14})$ theories. Generically, these states are constructed by tensoring the combination $[(\bar c,\bar c)+(\bar a,\bar a)]$ of 
right-moving chiral and anti-chiral primaries of the supercosets $\frac{SU(2)_k}{U(1)} \otimes \frac{\slr_k}{U(1)} $ 
with either a chiral $c_{u/t}$ or anti-chiral $a_{u/t}$ left-moving primary of $\frac{\slr_k}{U(1)} \otimes SU(2)_{k-2}$, 
in the untwisted ($u$ label) or twisted ($t$ label) sector of 
the $\zi_2$ orbifold~(\ref{cftback}) acting on the compact $SU(2)_\mathsf{L}$. The detailed \textsc{cft} construction  of these states can be found in~\cite{carlevaro}.

These hypermultiplets of $d=6$, $\mathcal{N}=1$ supersymmetry 
obtained by 'compactification' of heterotic strings on the warped Eguchi--Hanson space are supplemented by the extra multiplets coming from the compactification on $T^2$ to $d=4$. The latter, being  neutral, do not contribute to the threshold corrections discussed below. 

In the particular case of 'minimal' magnetic charge $\ell=1$, the left superconformal symmetry is enhanced to $\mathcal{N}=4$, hence 
the $U(1)_R$ worldsheet R-symmetry is enhanced to $SU(2)_2$. Since in this case, the action of the $\mathbb{Z}_2$ orbifold is trivialised,
hypermultiplets coming from its twisted sectors are altogether absent, while the 'untwisted' hypermultiplets organise into a doublet
and two singlets of $SU(2)_2$. In the other cases, $i.e.$ for $\ell\in 2\en^* +1$, the emergence of twisted sectors of 
the $\zi_2$ orbifold enhances the spectrum of hypermultiplets.

\paragraph{Hypermultiplet multiplicities and accidental $SU(2)$ symmetry:}  the hypermultiplet multiplicity factors in table~\ref{tabops}
take into account  the $(2j+1)$ state degeneracy  characterising operators  with internal left-moving $SU(2)_{k-2}$ spin $j$. This 
$SU(2)_\mathsf{L}$ symmetry should indeed be regarded as an accidental global symmetry of the local model for which   
one computes the gauge threshold corrections, that can be understood in supergravity as counting KK modes originating 
from the $\mathbb{P}^1$ reduction; in a genuine $T^2\times \widetilde{K3}$ compactification, this symmetry is absent. Another
way of phrasing things is to say that modifications to the worldsheet theory necessary to glue the local model
onto a full-fledged compactification will inevitably break this $SU(2)_\mathsf{L}$ symmetry.


\subsubsection*{Worldsheet non-perturbative effects}
The 'K-theory' constraint~(\ref{bsb}) is actually a necessary condition for the \textsc{cft}~(\ref{cftback})  to make sense, as was shown 
in~\cite{carlevaro}. The super-coset $\slr_k/U(1)$ worldsheet action receives non-perturbative corrections in the form of a dynamically 
generated $\mathcal{N}_{\textsc{ws}}=(2,0)$ Liouville potential. In the present case, the corresponding 
vertex operator is given by the hypermultiplet  which is an uncharged singlet of $SO(32)$ and belongs to the 
twisted sector of the $\mathbb{Z}_2$ orbifold ({\it cf.} table~\ref{tabops}), making the latter mandatory~\footnote{Note that for the $\vec{\ell}=(1^2,0^{14})$ model, the Liouville potential is still present, despite the trivialisation of the $\zi_2$ orbifold. In this case, the corresponding operator sits in the same $(\mathbf{1},\mathbf{1})$ hypermultiplet as the dynamical current--current deformation triggering the blowup.}. Requiring this particular operator 
to be both orbifold and GSO-invariant further imposes respectively that $k\equiv 2\text{ mod }4$ and 
$\sum_{i} \ell_i = 1+\ell \equiv 0 \mod 2$, hence the $\ell\in 2\en^*+1$ condition in table~\ref{tabops}, the latter being nothing else than the K-theory constraint~(\ref{bsb}).

\section{Threshold corrections and the elliptic genus: general aspects}
\label{sec:general}

We consider a generic compactification of  the heterotic string theory to four dimensions, with $\mathcal{N}_{\textsc{st}}=2$ 
space-time supersymmetry and an unbroken gauge group $\mathrm{G}=\prod_{a\leqslant 16} \mathrm{G}_a\subset SO(32)$. 

The one-loop correction to the gauge coupling constants takes the generic form:
\begin{equation}\label{gloop}
\left. \frac{4\pi^2}{g^2_a(\mu^2)}\right|_{\text{1-loop}}= \frac{k_a}{L}+\frac{b_a}{4}\log\left(\frac{M_s^2}{\mu^2}\right)+\frac{\Delta_a(M,\overline{M})}{4}\,,
\end{equation}
where $L$ is the linear multiplet associated to the dilaton, $M_s$ is the string scale, $\mu$ an infrared cutoff that will be discussed below later, $M$ the compactification moduli and $k_a$ the Kac--Moody levels 
determining the normalisation of the gauge group generators. One can alternatively express~(\ref{gloop}) in terms 
the complexified axio-dilaton $S$ multiplet by using the relation:
\begin{equation}\label{univ}
L^{-1}=\text{Im}\,S-\tfrac{1}{4}\Delta_{\text{univ}}(M,\overline{M})\,,
\end{equation}
with $\Delta_{\text{univ}}$ a {\it universal} (group independent) function of the moduli.

The $\beta$-function coefficients $b_a$ are given by a fixed linear combination of the quadratic Casimir
invariants of the gauge group. For $\mathcal{N}_{\textsc{st}} =2$ theories, when $\mathrm{G}_a$ is non-Abelian, these coefficients are determined by
\begin{equation}\label{beta1}
b_a = 2\sum_{\mathbf{R}} n_{\mathbf{R}}\, T_a({\mathbf{R}})-2\,T_a(\text{Adj}_a)\, ,
\end{equation}
where $n_{\mathbf{R}}$ counts the number of matter multiplets in the representation $\mathbf{R}$ of  $\mathrm{G}_a$.

When one of the gauge factors $\mathrm{G}_a$ is Abelian, its $\beta$-function is given by
\begin{equation}\label{beta2}
b_{U(1)} = 2\sum_{\mathbf{R}} n_{\mathbf{R}} \, \eta_{\mathbf{R}}\, \text{dim}(\mathbf{R})\, Q^2_{\mathbf{R}}\,,
\end{equation}
in terms of the $U(1)$ charges $Q_{\mathbf{R}}$ of the representations of the non-abelian factors $\mathrm{G}_a$ which appear in the hypermultiplet spectrum and the respective normalisation $\eta_{\mathbf{R}}$ of their generators. Typically, hypermultiplets which are singlets of $\mathrm{G}_a$ will not contribute to~(\ref{beta1}) but will appear in~(\ref{beta2}) .

\subsection{The modified elliptic genus} 
Heterotic $\caln_{\text{st}}=2$ gauge threshold corrections are determined 
at one-loop by a properly regularised three-point function in the 
worldsheet \textsc{cft} on the torus, integrated over the fundamental domain of the modular group 
$PSL(2,\mathbb{Z})$:
\begin{equation}
\mathscr{F} =\left\{\tau \in \mathscr{H}\;| \, -\frac12 \leqslant \tau_1 < \frac12,\, |\tau| \geqslant 1\right\}
\, ,
\end{equation}
 where $\mathscr{H}$ is the upper half complex plane.

The non-universal part of the threshold~(\ref{gloop}) is the given by the integral over $\mathscr{F} $ of a modification of the supersymmetric index introduced 
in~\cite{Cecotti:1992qh,Cecotti:1992vy,Harvey:1995fq}~\footnote{Another procedure for computing the non-holomorphic modular form  $\widehat{\mathcal{B}}_a$ 
is the so-called {background field} method~\cite{Abouelsaood:1986gd,Bachas:1992bh,Bachas:1996zt}, where a magnetic B-field is turned on along two of the spatial directions in four-dimensional Minkowski space. Expanding 
in the weak field limit the one-loop vacuum energy in powers of $B$, we recover the gauge threshold corrections~(\ref{threshcorr}) as the quadratic term in the expansion, the 
zero order term vanishing because of supersymmetry. We then obtain in the $\overline{\text{DR}}$ renormalisation scheme:
\begin{equation}\label{bfm}
\hat{\mathcal{B}}_a(\tau) = -\frac{i}{\pi}\frac{1}{|\eta|^4}\sum_{(b,c)\neq (1,1)}\partial_{\bar\tau}\left(\frac{\bar{\vartheta}\oao{b}{c}}{
\bar \eta}\right) 
\left[Q^2_a-\frac{k_a}{4\pi\tau_2}\right] Z\oao{b}{c}(\tau)\,,
\end{equation}
where $Z\oao{a}{b}$ is the partition function of the internal six-dimensional theory. This procedure is however not very handy in our case, 
where $\nu$-derivatives of  $\slr_k/U(1)$ characters lack most of the useful identities enjoyed by characters of the \textsc{cft}s associated to heterotic 
toroidal orbifold compactifications.} 
\begin{equation}\label{threshcorr}
\Lambda_a\equiv \frac{b_a}{4}\,\log\frac{M_s^2}{\mu^2}+\frac{\Delta_{a}}{4}=
\frac14\int_{\mathscr{F}}\frac{\di^2\tau}{\tau_2} \,   \hat{\mathcal{B}}_a(\tau)  \, ,
\end{equation}
given by a descendant of an elliptic index modified by the insertion of the (regularised) Casimir operator  of the corresponding gauge group factor:
\begin{equation}\label{I}
\hat{\mathcal{B}}_a(\tau)  = \left. \frac{i}{\eta^2}\, \text{Tr}_{\mathcal{H}^{(22,9)}_{\bar{\text{R}}}}\Big( 
 \Big[Q^2_a-\frac{k_a}{4\pi\tau_2}\Big] \ee^{i\pi \bar{J}^R_0} \bar{J}^R_0\,
q^{L_0-\frac{c}{24}} 
\bar{q}^{\bar{L}_0-\frac{\bar{c}}{24}} \ee^{2 \pi i \vec{\nu}\cdot \vec{J}_0} 
\Big) \right|_{\vec{\nu}=\vec{0}}\, , 
\end{equation}
where $\{ J^i,\, i=1,\ldots,16\}$ denote the Cartan currents of $SO(32)_1$. The trace in $\widehat{\mathcal{B}}_a$ projects onto the ground states of the right-moving twisted 
Ramond sector of the internal six-dimensional $(c,\,\bar c)=(22,9)$ \textsc{cft}. Also, the insertion of the total  right-moving 
$U(1)_{\bar R}$ current zero-mode $\bar{J}^R_0$ is there to remove the extra zero-modes coming from the two-torus \textsc{cft} which would otherwise make the index~(\ref{I}) vanish altogether. 

The quantity $\widehat{\mathcal{B}}_a$ only  depends on the topology of the manifold and of the gauge bundle.
In particular, if we remove the regularised Casimir operator in expression~(\ref{I}),  $-i\eta^2\widehat{\mathcal{B}}_a$ reduces to an elliptic generalisation of the Dirac--Witten index~\cite{Cecotti:1992qh,Cecotti:1992vy}, counting the difference between vector- and hypermultiplets (and including non-physical states violating the level matching condition, which are required by modular invariance of the index). This elliptic genus is thus stable under an arbitrary chiral marginal deformation and is as such invariant under 
deformations of the hypermultiplet moduli.

We remind that the $\frac{k_a}{\tau_2}$ term in~(\ref{I}), which results from a modular invariant regularisation of worldsheet short distance 
singularities appearing when two vertex operators collide, has no analogue in \textsc{qft}~\footnote{Such a term originates from a loop of charged or uncharged string states coupling universal to two external gauge bosons via the dilaton, and corresponds to one-particle reducible diagram~\cite{kiritsisbook}.}. In string theory this term, which is in fact universal, contains in particular the gravitational corrections to the gauge couplings. 

In the class of models coming from a toroidal reduction of a six-dimensional compactification, one can further simplify the expression~(\ref{I}) by using the decomposition of the right $U(1)_R$ current as 
\begin{equation}\label{rsplit}
\bar{J}^{R}=\bar{\jmath}^t + 2\bar{J}^3 \, , 
\end{equation} 
the former being the R-current of the free $T^2$ $\mathcal{N}_\textsc{ws}=(0,2)$ \textsc{cft}\footnote{
The $\jmath_0^t$ insertion in the trace absorbs the zero-modes of the free Weyl fermion in the two-torus \textsc{cft}, ensuring that the index does not vanish.} and the latter being 
the Cartan of the $SU(2)$ R-symmetry of the remaining interacting \textsc{cft} with $(c,\,\bar c)=(22,6)$ which has an $\mathcal{N}_\textsc{ws}=(0,4)$ extended superconformal symmetry.

It can be shown that for any representation $(h,I)$ of the right-moving $\mathcal{N}_{\textsc{ws}}=4$ and $\bar c =6$ superconformal algebra (see appendix~\ref{kappa1}),  the following trace vanishes:
\begin{equation}\label{ind0} \text{Tr}_{ \mathcal {H}_{(h,I)}^{(22,6)}} 
\ee^{2\pi i  \bar J_0^3} \bar J_0^3\, q^{L_0-\frac{c}{24}} \bar q^{\bar L_0-\frac{\bar{c}}{24}}  =0\,.
\end{equation} 
for both the continuous and discrete spectrum of states, 
since discrete representations come in opposite pairs of eigenvalues under $\bar J_0^3$ and since for continuous representations~(\ref{twistcont}) expression~(\ref{ind0}) contains factors  $\theta_i(\tau|0)\theta'_i(\tau|0)$  with $i=1,..,4$, which vanish.

Hence, using the decomposition of the left R-current~(\ref{rsplit}) in the index~(\ref{I}) one obtains that the one-loop gauge threshold corrections factorise as follows:
\begin{equation}\label{decLam}
\Lambda_a =\frac18\int_{\mathscr{F}}\frac{\di^2\tau}{\tau_2} \, \Gamma_{2,2}(T,U)\, \hat{\mathcal{A}}_a (\tau)\,.
\end{equation}
The contribution of the four-dimensional warped $\textsc{eh}$ space and of the the gauge bundle is now encoded in a non-holomorphic Jacobi form obtained by projecting the trace over the Hilbert space of the theory onto the right-moving twisted Ramond ground state of the $(c,\,\bar c)=(20,6)$ \textsc{cft}:~\footnote{Note that the normalisation used here for~(\ref{Ell}) differs from some conventions in the literature by a sign, for instance from that of ref.~\cite{Camara:2008zk}, with conversion $\hat{\mathcal{A}}^{\text{ours}}=- \hat{\mathcal{A}}^{\text{theirs}}$.}
\begin{equation}\label{Ell}
\hat{\mathcal{A}}_a(\tau)= \frac{1}{\eta^4} \text{Tr}_{\mathcal{H}^{(20,6)}_{\bar{\text{R}}}} \left( 
 \left[Q^2_a-\frac{k_a}{4\pi\tau_2}\right] \ee^{2i\pi \bar{J}_0^3} \,
q^{\bar{L}_0-\frac{1}{4}} 
\bar{q}^{L_0-\frac{5}{6}}
\right) \,.
\end{equation}

\subsubsection*{Universality of $\mathcal{N}=2$ threshold corrections}
It has been often emphasised how universal features of $\mathcal{N}_\textsc{st}=2$ heterotic gauge threshold
corrections can be completely determined on the one hand by requiring  the absence of tachyons and cancellation of tadpoles, and
on the other hand from the global  symmetries dictated by the background geometry~\cite{Obers:1999um,Obers:1999es}. 

Thus, by considering its $T^2 \times (T^4/\mathbb{G})$ orbifold limit,
with $\mathbb{G}$ inducing  a breaking of the $SO(32)$ gauge group to $\mathrm{G}=\prod_{a\leqslant 16} \mathrm{G}_a$, one
can show that the one-loop threshold corrections to the gauge couplings $g^{-2}_a$ for the corresponding resolved  heterotic $T^2 \times K3$ 
compactification are fixed  uniquely by the following linear combination~\cite{kiritsisbook}:
\begin{equation}\label{Lam-univ}
 \Lambda_a = \frac18 \int_{\mathscr{F}}\frac{\di^2\tau}{\tau_2} \,  \Gamma_{2,2}(T,U)\,  \big( k_a \hat{\mathcal{C}}   + 2 b_a  \big) \,
\end{equation}
in terms of the following quasi-holomorphic genus:
\begin{equation}\label{doub}
\hat{\mathcal{C}} = 
\frac{1}{12} \! \left(-\frac{{\widehat{E}}_2 E_{10}}{\eta^{24}} + j -1008 \right)
\equiv
 \frac{D_{10}E_{10}  - 528 \eta^{24} }{ 20\eta^{24} }  
\end{equation}
with the Klein invariant $j = E_4^3 / \eta^{24}$ and $D_{10}E_{10}$ the modular covariant derivative~(\ref{DE2}). 

In particular, in the first expression of $\hat{\mathcal{C}}$, the combination $-\widehat{E}_2E_{10}  + j\eta^{24}$ is fixed by 
requiring no $q^{-1}$ pole to be present in~(\ref{doub}), which would signal the presence of a tachyon. 
Such a would-be tachyon being uncharged under the gauge group, the potential single pole coming from $\eta^{-24}$ should
not appear in the gauge threshold correction.  
Nevertheless, gauge threshold corrections for $\mathcal{N}_\textsc{st} =2$ heterotic  compactifications allow for a  $(\tau_2 q)^{-1}$ behaviour 
of $\hat{\mathcal{A}}_a$~(\ref{decLam}), as $q\rightarrow 0$, stemming from the \textsc{ir} regulator 
in $\widehat{E}_2$. This pole, associated to an unphysical tachyon, will be referred to as 'dressed' pole in the following, in contrast to the 'bare' $ q^{-1}$ pole, which should be absent from a gauge threshold correction. In consequence, $\hat{\mathcal{C}}$ is fixed by the linear combination of 
two modular forms of weight 12: the quasi-holomorphic modular form $D_{10}E_{10}$ and the cusp form $\eta^{24}$, a feature which we
will also observe for non-compact models.

In addition, gauge and gravitational anomaly cancellation in six-dimensional vacua fixes the constant term in $\hat{\mathcal{A}}_a$ and fixes the coefficients of the linear combination~(\ref{Lam-univ})to be  the $\beta$-functions $b_a$ and the levels $k_a$ of the corresponding Kac--Moody algebras.  Another way to look at the decomposition~(\ref{Lam-univ}) is to observe that the $\hat{\mathcal{C}}$
dependent piece is \textsc{ir}-finite when integrated over $\mathscr{F}$ thanks to the regulator $\tau_2^{-1}$, while the constant $b_a$ contribution exhibits an \textsc{ir} divergence, signaling the presence of massless states. These are precisely the massless hypermultiplets and the vector multiplet in the four-dimensional effective field theory  which contribute to the  $\beta$-functions~(\ref{beta1}). 

As a consequence of these universality properties,  the two-by-two difference of threshold corrections for different gauge factors
satisfy, for such heterotic $\mathcal{N}_\textsc{st}=2$ vacua, the relation:
\begin{equation}\label{relN2}
\frac{\Lambda_{a_1}}{k_{a_1}} - \frac{\Lambda_{a_2}}{k_{a_2}} = \frac{1}{4} \left(\frac{b_{a_1}}{k_{a_1}} - \frac{b_{a_2}}{k_{a_2}} \right) \int_{\mathscr{F}}\frac{\di^2\tau}{\tau_2} \,  \Gamma_{2,2}(T,U) \,.
\end{equation}

A reformulation of the the threshold correction $\Lambda_a$ associated to a $T^2\times T^4/\mathbb{G}$ heterotic vacuum is particularly useful to understand the topology of the gauge bundle 
supported by the string compactification. Merging the combination~(\ref{Lam-univ}) into a single contribution yields~\cite{Stieberger:1998yi}:
\begin{equation}\label{refthresh}
\Lambda_a =
 \frac{k_a}{8} \int_{\mathscr{F}}\frac{\di^2\tau}{\tau_2} \, \Gamma_{2,2}(T,U)\,
 \frac{1}{12 \eta^{24}}\left(- {\widehat{E}}_2  E_{10}+\frac{n_a}{24} E_6^2+\frac{m_a}{24} E_4^3\right) \,,
\end{equation}
with the identification:
\begin{equation}
n_a = 14 -\frac{b_a}{3 k_a} \,, \qquad  m_a =  10 + \frac{b_a}{3 k_a}  \,.
\end{equation}
Then, the tadpole equation is reproduced by the constraint:
\begin{equation}\label{tadK3}
n_a + m_a = \chi(K3) = 24\,.
\end{equation} 
One  can achieve some insight into the topology of the gauge bundle after resolution in the smooth $K3$ limit of the  $T^4/\mathbb{G}$ orbifold   by rewriting $n_a=12 + t_a$ and $m_a =12 - t_a$.  In particular, the various $\beta$-functions depend on $t_a$ as follows:
\begin{equation}\label{t-ins}
b_a = 3k_a(2 -  t_a) \, ,
\end{equation}
where $t_a$  is the number of $SU(2)$ instantons now present in the resolved $T^2 \times K3$ geometry\footnote{In these models tadpole cancellation usually requires 
the presence of a certain number of small instantons hidden at orbifold singularities. Performing a slight resolution of the singularities brings out these instantons in the open, in the guise of $SO(2)$ instantons embedded in $SO(32)$. But when realizing a full blow-up to a a smooth K3 geometry, these $U(1)$ instantons cannot be defined anymore on the blown-up $\mathbb{P}^1$'s and are replaced by  $SU(2)$ instantons with instanton number $t_a$.}.
Depending on the value of $t_a$, partial or total Higgsing of the gauge group $\mathrm{G}$ is possible.


\subsection{Threshold corrections for local models}

Before embarking, in the next section, on discussing the intricacies of  how to evaluate
gauge threshold corrections for $T^2 \times \widetilde{\textsc{eh}}$ models,
it is worthwhile to put them in a wider perspective. The non-compact nature of these backgrounds will have drastic consequences, both at the physical and 
mathematical levels, as we will discuss below.

\subsubsection*{Vector and hyper multiplets}

In order to built vertex operators corresponding to gauge bosons in space-time, one needs, as far as right-movers are concerned, 
to tensor a standard vector operator of the free $\mathbb{R}^{3,1}$ theory with an operator of dimension zero in the internal CFT. The 
latter is necessarily built on the {\it identity representation}~(\ref{Idrep0})  of the $\slr /U(1)$ coset (with spin $J=0$), since the conformal weights for
the $SU(2)/U(1)$ coset theory are non-negative.

As the identity representation of $\slr /U(1)$ is non-normalisable, we readily see that vector multiplets do not appear in the spectrum obtained from the 
partition function~(\ref{partfunc}).  In means that, assuming that these local models can be glued to a full-fledged compactification with flux, the 
wave-functions corresponding to the gauge bosons are not localised in the throat regions that are decoupled from the bulk by the double-scaling limit~(\ref{DSL}). Hence, they 
cannot be considered as fluctuating fields in the path-integral.

We can then interpret the result of the computation that we perform here as a one-loop correction to the gauge couplings in the effective four-dimensional theory from hypermultiplets whose higher-dimensional wave-function is localized in a particular region of the compactification manifold with strong warping, near a resolved $A_1$ singularity -- provided the  gauge group $\mathrm{G}$  is not further broken by global effects in the full theory.

Since vector multiplets, being 'frozen', are expected  not to contribute to the $\beta$-functions, the factor~(\ref{beta1}) in the one-loop correction~(\ref{gloop}) will thus be modified as:
\begin{equation}\label{beta1'}
b^{\text{loc}}_a = 2\sum_{\mathbf{R}} n_{\mathbf{R}}\, T_a({\mathbf{R}})\,.
\end{equation}
In the class of models studied here, with shift vectors of the form $\vec{\ell}=(1,\ell,0^{14})$, whose 
spectrum is given in table~\ref{tabops}, the $\beta$-functions for the gauge group factors are given accordingly 
by:
\begin{equation}\label{hyp}
\begin{array}{lll}
\ell = 1\,: & \qquad  b_{SO(28)}^{\text{loc}} = 4 \,, &\qquad    b_{SU(2)}^{\text{loc}} = 56 \,, \\[4pt]
\ell = 2\en^* +1\,: &\qquad  b_{SO(28)}^{\text{loc}} = 8\ell^2 \,,& \qquad{\ds b_{U(1)_{R}} = 4(29 \ell^2-2\ell +29) } \,.
\end{array}
\end{equation}
The useful Casimir invariants are $T(\Box) =1$ for $SO(2N)_1$, and  $T(\Box) =\frac{k_{SU(2)}}{2} =1$ for the $SU(2)_2$
factor. The level of the latter is fixed by its embedding into the $SO(32)_1$ gauge algebra and is determined by identifying
its Cartan with the $U(1)_{R}$ charge generator, which generically has level $k_{U(1)_R} = \frac{k+2}{k}$.


\subsubsection*{Perspectives on non-holomorphicity}
\label{sec:persp}

For the heterotic local models considered here, one observes some deviations from the standard computation of threshold 
corrections for $T^2 \times K3$ compactifications. These are not peculiar to one-loop gauge threshold corrections, but 
can already be found at the level of the elliptic genus. They are due both to the non-compactnes of target space and to 
the presence of non-zero five-brane charge at infinity. 

The modified elliptic genus~(\ref{Ell})  for the four-dimensional  warped Eguchi--Hanson theory will schematically take the form:
\begin{equation} \label{Agross}
\hat{\mathcal{A}}_a(\tau)  = \hat{\mathcal{A}}_{a}^{\text{d}}(\tau) + k_a \mathcal{R}^{\text{c}}(\tau) =
 \sum_{g=0}^{g_{\text{max}}} \frac{1}{\tau_2^g} \left( \sum_{n=-1}^\infty c^{\text{d}}_{g n}\, q^{n}
 + \sum_{m \in \zi} c^{\text{c}}_{g m}(\tau_2)\, q^{m}\right)\,.
\end{equation}
We can already give an overview of some prominent features of~(\ref{Agross}) which will be made more precise in the following:
\begin{itemize}
\item the $\hat{\mathcal{A}}_{a}^{\text{d}}$ contribution in~(\ref{Agross}) arises from states which are obtained from 
discrete $\slr_k/U(1)$ representations, {\it i.e.} from states which localise on the blown-up $\mathbb{P}^1$. As such, it retains some characteristics of its compact $K3$ analogues~(\ref{refthresh}): it is quasi-holomorphic and,  as we require no charged tachyon to appear in the spectrum, a 'bare' $q^{-1}$ pole at infinity is absent from its Fourier expansion ($c^{\text{d}}_{0, -1}=0$). However as for K3 compactifications, $ \hat{\mathcal{A}}_{a}^{\text{d}}$ generically has poles dressed by \textsc{ir} regulators, namely $(\tau_2^{g} q)^{-1}$, which are the only source of non-holomorphicity.
The maximal power $g_{\text{max}}$ for
such non-holomorphic factors is fixed by supersymmetry, as it relates to the regularisation of worldsheet divergences caused by 
$g$ pairs of vertex operators colliding at the corners of the moduli space and giving rise to a massless state.
Mathematically this translates as the presence of $\widehat{E}_2^g$ factors in $\hat{\mathcal{A}}_{a}^{\text{d}}$. For a background preserving $\caln_{\textsc{st}} =2$ in four dimensions, the effective action starts with two legs, entailing $g_{\text{max}}=1$.~\footnote{In contrast, for an $\caln_{\textsc{st}} =1$ background, threshold corrections would derive from an effective action with four legs, inducing $g_{\text{max}}=2$ and $\widehat{E}_2^2$ factors in~(\ref{Agross}).}
The term $\hat{\mathcal{A}}_{a}^{\text{d}}$ however differs from its $K3$ counterpart in that it transforms anomalously under S-transformation. It actually  transforms as a {\it Mock modular form}, which will be discussed below~\footnote{It is not {\it strictly speaking} a Mock modular form since it contains a finite number of non-holomorphic terms, but  can be recast as a sum of Mock modular forms multiplied by almost-holomorphic Jacobi forms, as we will shortly see.}. 

\item This anomalous behaviour of $\hat{\mathcal{A}}_{a}^{\text{d}}$ comes from considering only 
the contribution of BPS representations to the index, as we are instructed to do in the compact case. The usual argument fails here, as 
the fermionic zero-modes are compensated by the infinite-volume divergence. Indeed, by resorting to a modular invariant regularisation of 
this divergence --~that adds extra non-holomorphic contributions to the index~--  one obtains the additional term 
$\mathcal{R}^{\text{c}}$, which  decomposes on a continuous spectrum of states and will be shown to be independent of the gauge group, and universal for a fixed value of the five-brane charge $\mathcal{Q}_5$.\footnote{Discussions on non-holomorphicity of the elliptic genus 
are also central to the question of deriving a reliable index for micro-states counting for systems of multi-centered  
black-holes~\cite{murthy}.}
This non-holomorphic completion seems at first sight to exhibit an infinite number of poles in $q$, with arbitrarily large order. However, in the theory of non-holomorphic Jacobi forms reviewed below, $\mathcal{R}^{\text{c}}$ contains the transform of the {\it shadow} of a Mock modular form. This dictates a specific form for the functions $c_{g m}(\tau_2)$ (see~\cite{zwegers}). In particular, as a sum $\mathcal{R}^{\text{c}}$ can be shown to be absolutely and uniformly convergent for $\tau \in \mathscr{H}$ (upper half complex plane) in such a way  as only to possess a single 'dressed' pole at $\tau_2 \rightarrow\infty$. In this case however the non-holomorphic regulator comprises additional exponential terms which are a distinguishable feature of non-localised states, the real part of this polar term
being generically  bounded by $\frac{1}{\tau_2}\sum_{n \in I} c_n \ee^{-\frac{\pi n}{2k(k-2)} \tau_2}  \ee^{2\pi \tau_2}$, where $I \subset \en$ is a finite set.
Thus $\mathcal{R}^{\text{c}}$ has a polar structure even less divergent at $\tau_2 \rightarrow \infty$  than the $(\tau_2q)^{-1}$ cusp of $\hat{\mathcal{A}}_{a}^{\text{d}}$. Hence, to determine explicitely the moduli dependence of the threshold corrections~(\ref{decLam}), 
one can proceed as for toroidal orbifolds.
\end{itemize}


\subsection{A brief review on Mock modular forms}
\label{sec:brief}
In the previous section,  we mentioned that the gauge threshold correction~(\ref{thresh2}) incorporates contributions from non-localised states, which enter into the function  $\mathcal{R}^{\text{c}}$ and recombine into the transform of the shadow function of some Mock modular form. We find it useful to recall here some facts about Mock modular forms and their isomorphism to weak harmonic Maa\ss\, forms, and in particular to clarify the notion of shadow. In this perspective, we synthesise among other things  the illuminating presentation of~\cite{Zagier2}.

Disregarding possible dependence on elliptic variables, a Mock modular form  $h$ of weight $r$ is  a function of  
the upper half-plane $\mathscr{H}=\{ \tau \in \ci |  \tau_2 \geqslant 0 \}$, which {\it almost} transforms as a modular form of corresponding weight. The space of all such forms, which we call $\mathbb{M}_r$, contains as subspace the space $M_r^{!}$ of weak 
holomorphic modular forms of weight $r$, which are allowed to have exponential growth, that is $q^{-N}$ singularities, at cusps.   
Then, associated to a Mock modular form $h\in \mathbb{M}_r$ there exists a  {\it shadow} $g= \mathcal{S}[h]$, which is an ordinary 
holomorphic modular form of  weight $2-r$. As such it has expansion
\begin{equation}
g(\tau)=\sum_{\nu\geq 0}b_\nu q^\nu\,,
\end{equation}
 where $\nu$ runs over some arithmetic progression in $\mathbb{Q}$.

The shadow map $\mathcal{S}$ is $\er$-linear in $h$ and can be given by defining an associated function $g^*$, which is the following transform of $g$:
\begin{equation}\label{gstar}
\begin{array}{rcl}
g^*(\tau) & = & {\ds  \left(\tfrac{i}{2}\right)^{r-1} \int_{-\bar\tau}^{i\infty}  \frac{\overline{g(-\bar\tau)}}{(z+\tau)^{r}}\, \di z } \\[12pt]
& = &  {\ds \frac{\bar b_0}{(r-1)(4\tau_2)^{r-1}}   +
\pi^{r-1} \sum_{\nu>0}\nu^{r-1}\bar{b}_\nu \,
\Gamma\big(1-r,4\pi\nu\tau_2\big) \,q^{-\nu} \, ,}
\end{array}
\end{equation}
where $\nu$ belongs to an arithmetic progression in $\mathbb{Q}$, and  $\Gamma(x,s)$ is the upper incomplete gamma function:
\begin{equation}\label{gam}
\Gamma(s,x) = \int_x^\infty t^{s-1} e^{-t}\di t \, , \qquad x>0 \,.
\end{equation}
The function $g^*$ is such that the combination
\begin{equation}\label{hhat}
 \hat h(\tau)= h(\tau) + g^*(\tau)
\end{equation}
transforms, for all $\gamma \in {\scriptsize \begin{pmatrix} \bullet & \bullet \\ c & d \end{pmatrix} }\in \Gamma$,
a suitable subgroup of $SL(2,\zi)$, as a modular form of weight $r$:
$$\hat h(\gamma\tau) = \rho(\gamma) (c\tau+d)^r \hat h(\tau)\, , $$
where $\rho$ is a character of $\Gamma$. As $\mathcal{S}$ is surjective and in addition
vanishes when $h$ is (a weakly holomorphic) modular form, we have the  following exact sequence over $\er$:
\begin{equation}\label{es1}
\begin{tikzpicture}[description/.style={fill=white,inner sep=2pt}]
 \matrix (m) [matrix of math nodes, row sep=3em, column sep=2.5em, text height=1.5ex, text depth=0.25ex]
{ 0 &  M_r^{!}  &  \mathbb{M}_r  & M_{2-r} & 0 \\};
\path[->,font=\scriptsize] 
(m-1-1) edge  (m-1-2)
(m-1-2) edge (m-1-3)
(m-1-3) edge node[auto] {$\mathcal{S} $} (m-1-4)
(m-1-4) edge  (m-1-5);
\end{tikzpicture} 
\end{equation}
and $\mathbb{M}_r$ can be regarded as an extension of a space of classical modular forms.

As the non-holomorphicity of $\hat h$ is integrally encoded in the shadow function $g^*$,  we can reverse the perspective and obtain $h$ by acting with Cauchy--Riemann operator $\partial/ \partial\bar\tau$ on $\hat h$, which by combining~(\ref{hhat}) and~(\ref{gstar}) gives:
\begin{equation}\label{cauch}
\frac{\d\hat h}{\partial\bar\tau}  = -\frac{2i}{(4\tau_2)^r}\, \overline{g(\tau)}\,,
\end{equation}
by which we recover $h = \hat h - g^*$. Through this procedure we can establish a canonical isomorphism 
$\mathbb{M}_r \cong \widehat{\mathbb{M}}_r$ between the space $\widehat{\mathbb{M}}_r$ of non-holomorphic weak modular forms of weight $r$, to which $\hat h$ belongs, and the space of Mock modular forms of corresponding weight.
 
We can now push further and show that the space $\widehat{\mathbb{M}}_r$ is actually the space of weak harmonic 
Maa\ss forms. To this end, we define $\mathfrak{M}_{r,l}$ the space of modular forms of weight $(r,l)$, {\it i.e.}
which transform as $F(\gamma\tau) = \rho(\gamma) (c\tau+d)^r (c\bar\tau+d)^l  F(\tau)$ for $\gamma \in \Gamma \subset SL(2,\er)$, such that $\mathfrak{M}_{r} =\mathfrak{M}_{r,0}$ reduces to the space of real-analytic modular forms in $\tau\in \mathscr{H}$ of weight $r$. 
In addition we introduce an operator $\tau_2^s \partial_{\bar\tau}$ which sends
$\mathfrak{M}_{r,l} \stackrel{\cong}{\longrightarrow} \mathfrak{M}_{r,l+2}  \longrightarrow \mathfrak{M}_{r-s,l-s+2}$, where the first map is an isomorphism for $s\in\zi$.  

Applying this operator to $\mathfrak{M}_{r} =\mathfrak{M}_{r,0} $ and  further acting with the holomorphic derivative we obtain the commutative diagram:
\begin{center}
\begin{tikzpicture}[descr/.style={fill=white,inner sep=2.5pt}] 
\matrix (m) [matrix of math nodes, row sep=3em, column sep=3em] 
{\mathfrak{M}_{r} =\mathfrak{M}_{r,0}   &\mathfrak{M}_{r,2} & \mathfrak{M}_{0,2-r} & \mathfrak{M}_{2,2-r} \\
  & & \overline{M_{2-r}}   &  \\ }; 
\path[->,font=\scriptsize] 
(m-1-1) edge node[auto] {$ \d/\d_{\bar\tau}  $} (m-1-2)
(m-1-2) edge node[auto] {$ \tau_2^r $} (m-1-3)
 (m-1-2)  to node [below] {$\cong$}  (m-1-3)
(m-1-3) edge node[auto] {$ \d/\d_{\tau} $} (m-1-4)
(m-2-3) edge node[auto] {$ \cup $} (m-1-3)
(m-2-3) edge node[below] {$ 0 $} (m-1-4)
; 
\end{tikzpicture} 
\end{center}
It follows from this diagram that $\widehat{\mathbb{M}}_r$ is defined as the space of real-analytic modular forms $F\in \mathfrak{M}_{r}$ such that $\tau_2^r \partial_{\bar\tau}F$ belongs to $\overline{M_{2-r}}$, in other words for which it is antiholomorphic:
\begin{equation}\label{whm}
\widehat{\mathbb{M}}_r = \left\{ F \in \mathfrak{M}_r \;\left|\;  \frac{\d}{\d\tau}\Big(\tau_2^r\frac{\d}{\d\bar\tau} F(\tau)\Big) =0  \right. \right\}\,.
\end{equation}
Now since $\partial_{\tau}\big(\tau_2^r \partial_{\bar\tau}(\bullet)\big)$ is up to an additive constant proportional to the weight $r$ Laplace operator $\Delta_r$, namely:
\begin{equation}\label{lapl}
\Delta_r F =  \frac14 \frac{\d}{\d\tau}\Big(\tau_2^r\frac{\d}{\d\bar\tau} F\Big) + \frac{r}{2} \Big( 1- \frac{r}{2} \Big) F
\end{equation}
$\widehat{\mathbb{M}}_r$ is thus the space of real-analytic modular forms which are allowed exponential growth at cusps and that are harmonic with $\frac{r}{2} \big( 1- \frac{r}{2} \big)$ eigenvalue under the weight $K$ Laplacian. This is precisely
the definition of {\it weak harmonic Maa\ss\, forms} according to Bruiner and Funke, which completes the identification.

\subsubsection*{Appell-Lerch sums} 

The simplest and most familiar example of a Mock modular form is the almost modular
Eisenstein series $E_2$, whose shadow $g(\tau)=-\frac{12}{\pi}$ is a constant. Using formula~(\ref{hhat}) for a weight $2$ Mock modular form, we get the well known non-holomorphic completion $\widehat{E}_2= E_2-\frac{3}{\pi\tau}$.

In this work, we will be particuliarly interested in a more involved class of Mock modular forms, the {\it Appell-Lerch sums}. 
The Appell--Lerch sums of level $K$ are functions of the upper half plane $\tau \in \mathscr{H}$ and depend on
two elliptic variables $u\in \ci$ and $v \in \ci / (\zi + \zi \tau)$:
\begin{equation}\label{applerch}
A_{K}(u,v|\tau)= a^{\frac{K}{2}} \sum_{n\in \zi} (-)^{Kn}\, \frac{q^{\frac{K}{2}n(n+1)} b^n}{1-aq^n} \,, \quad
\text{with } \,a= \ee^{2\pi i u}\,, \; b= \ee^{2\pi i v}\,.
\end{equation}
The investigation of the near modular behaviour of these functions can be deduced 
from the transformation properties of the level one $A_1$ sum, since for an arbitrary level $K$ we can reexpress:
\begin{equation}\label{Ak}
\begin{array}{rcl}
A_{K}(u,v|\tau)&=&{\ds  \sum_{m=0}^{K-1} a^m A_1\!\left(Ku,v +m\tau+\tfrac{K-1}{2}|K\tau\right) 
 }\\[15pt]
&\equiv& { \ds \tfrac{a^{\frac{K-1}{2}}}{K}\sum_{m \in \zi_K} a^m A_1\!\left(u, \tfrac{v+m}{K} + \tfrac{(K-1)\tau}{2K}\left|\tfrac{\tau}{K}\right.\right)} \,.
\end{array}
\end{equation}
We can thus concentrate on the level one case. In particular, the almost modularity of $A_1$ consists in its failure to transform as modular form under S-transformation:
\begin{equation}
A_1\!\left(\tfrac{u}{\tau},\tfrac{v}{\tau}|-\tfrac{1}{\tau}\right) = \tau \ee^{\pi i \frac{(2v-u)u}{\tau}} 
\left[ A_1(u,v|\tau) - \tfrac{1}{2} M(u-v|\tau)\, i \vartheta_1(v|\tau)\right]
\end{equation}
where the second term on the rhs contains the function $M$ of $\tau\in \mathscr{H}$ and $\nu\in\ci$ first  studied
by Mordell, which is defined in terms of the integral:
\begin{equation}\label{Mfun}
M(\nu|\tau) = \int_\er \di x\, \frac{q^{\frac{x^2}{2}}\, \ee^{-2\pi x\nu}}{\cosh(\pi x)}\,.
\end{equation}
There is a clear reminiscence of this behaviour in the S-transformation of discrete $\slr_k/U(1)$ 
characters~(\ref{Sdual0}), which will be made explicit in a moment.

To construct the non-holomorphic completion of the Appell-Lerch sums~(\ref{applerch}), it then suffices to consider the level $K=1$ example, in which case it actually proves more convenient to normalise this sum by a $\vartheta$-function:
\begin{equation}\label{appfun0}
\mu(u,v|\tau) = -\frac{i}{\vartheta_1(v|\tau)} A_1(u,v|\tau)\,,
\end{equation}
also called the {\it Appell function}.  Then,
by studying the modular transformation properties of the function $M$~(\ref{Mfun}) and by noticing that the near modularity of the Appell function $\mu$ only depends on the difference $u-v$, Zwegers was able to construct its function $g^*$:
\begin{equation}\label{Zwegers}
R(\nu|\tau)= \sum_{n\in\zi} (-)^n\left( \text{sgn}\left(n+\tfrac12\right)-E\left(\left[n+\tfrac12+\tfrac{\nu_2}{\tau_2}\right]\sqrt{2\tau_2}\right)\right)
z^{-(n+\frac12)}q^{-\frac12(n+\frac12)^2}\, ,
\end{equation}
where $\nu_2 = \text{Im}\,\nu$, and $E(z)$ is the error function, defined as follows:
\begin{equation}\label{error}
E(z)=2\int_0^z e^{-\pi w^2}\,\di w \,, \qquad  z \in \ci\,, 
\end{equation}
which is an odd and entire function of $z$.
Since the argument of $E$ in~(\ref{Zwegers}) is real, we can alternatively express $R(\nu|\tau)$  in terms of the incomplete gamma 
function~(\ref{gam})
\begin{equation}
E(x)  =  \text{sign}(x)\left[ 1-  \tfrac{1}{\sqrt{\pi}}\,\Gamma\big(\tfrac12,\pi x^2\big) \right]\,,   \qquad x \in \er\,,
\end{equation}
by means of the following identity:
\begin{equation}\label{erfcG}
\text{erfc}(\sqrt{\pi}|x|) = 2 \int_{|x|}^\infty \ee^{-\pi u^2} \,\di u = \int_{x^2}^\infty v^{-\frac12} \ee^{-\pi v}\, \di v = \frac{1}{\sqrt{\pi}}\,\Gamma\big(\tfrac12,\pi x^2\big)\,,
\end{equation}
where $\text{erfc}(\sqrt{\pi}x) = 1-E(x)$ is the complementary error function. One sees that $R(\nu|\tau)$ is indeed of the form propounded
in~(\ref{gstar}).

The Appell function can thus be completed into a non-holomorphic Jacobi form of two elliptic variables:
\begin{equation}\label{Ahat}
\widehat{\mu}(u,v|\tau) = \mu(u,v|\tau) - \frac{1}{2} R(u-v|\tau)\,,
\end{equation}
which is furthermore a harmonic Maa\ss  \ form for the Laplace operator $\Delta_{\nicefrac12}$~(\ref{lapl}),
and thus transforms as a Jacobi form of weight $\nicefrac12$.

In particular for $u=v=\nu$, the non-holomorphic completion of the Appell function of one elliptic variable, which we denote by  $\widehat{\mu}(\nu| \tau) \equiv \widehat{\mu}(\nu,\nu| \tau)$ in the following, reads
\begin{equation}
\widehat{\mu}(\nu| \tau) = -\frac{i}{\vartheta_1(\nu|\tau)} \sum_{n=0}^\infty  (-)^n \frac{q^{\frac12 n(n+1)} z^{n+\frac12}}{1-zq^n}
-
\sum_{n=0}^\infty (-)^n \text{erfc} \big( (n+\tfrac12)\sqrt{2\pi \tau_2}\big) q^{-\frac12 (n+\frac12)^2}
\end{equation}
and is characterised by  a shadow function which can be  extracted from the relation~(\ref{cauch}):
\begin{equation}
\frac{\partial \widehat{\mu}(\nu|\tau)}{\d\bar\tau} = \frac{i}{2\sqrt{2}}\frac{\overline{\eta(\tau)}^3}{\sqrt{\tau_2}}\,.
\end{equation}
The shadow of $\mu(\nu|\tau)$ is thus the holomorphic modular form $g(\tau) = -\frac{1}{2\sqrt{2}} \eta(\tau)^3$ with weight $\nicefrac32$,
as expected for a Mock Jacobi form of weight $\nicefrac12$.

The full modular transformation properties of $\widehat{\mu}$ are neatly given by:
\begin{equation}\label{modpropmu}
\widehat{\mu}(u,v|\tau+1) = \ee^{-\frac{\pi i}{4}} \,\widehat{\mu}(u,v|\tau) \,, \qquad
 \widehat{\mu}\left(\tfrac{u}{\tau},\tfrac{v}{\tau}\big|-\tfrac{1}{\tau}\right) = -\ee^{-\frac{\pi i}{4}}\sqrt{\tau}\, \ee^{-\pi i \frac{(u-v)^2}{\tau}}\widehat{\mu}(u,v|\tau)
\end{equation}
from which we deduce its index to be ${\footnotesize \begin{pmatrix} -\nicefrac 12 & \nicefrac12\\ \nicefrac12 & -\nicefrac 12 \end{pmatrix}}$.
Transformations of $\widehat{\mu}$ under shifts in the elliptic variables can also be worked out (note that $\widehat{\mu}$ is symmetric in $u$ and $v$):
\begin{equation}
\widehat{\mu}(u+1,v|\tau) = a^{-1}bq^{-\frac12} \widehat{\mu}(u+\tau,v|\tau) =- \widehat{\mu}(u,v|\tau)\,.
\end{equation}
In addition, $\widehat{\mu}$ satisfies:
\begin{equation}\label{diffmu}
\begin{array}{rcl}
{\ds \widehat{\mu}(u+\lambda,v+\lambda|\tau) - \widehat{\mu}(u,v|\tau)} & =  &
{ \ds \mu(u+\lambda,v+\lambda|\tau) - \mu(u,v|\tau)  } \\[6pt]
&=& {\ds - \frac{\eta(\tau)^3 \vartheta_1(u+v+\lambda|\tau) \vartheta_1(\lambda|\tau) }{\vartheta_1(u|\tau) \vartheta_1(v|\tau)  \vartheta_1(u+\lambda|\tau)\vartheta_1(v+\lambda|\tau) } \,,} \\[6pt]
 && \hspace{4cm}  u,\, v,\, u+\lambda,\, v+ \lambda \notin \zi\tau +\zi\,.
\end{array}
\end{equation}

By using the reformulation of the Appell--Lerch sums at arbitrary level $K$ in terms of $A_1$, as in~(\ref{Ak}), and the non-holomorphic completion~(\ref{Zwegers}), we can generalise the construction of similar corrective terms for all sums $A_K$:
\begin{equation}\label{Appell-comp}
\begin{array}{rcl}
{\ds \hat{A}_K(u,v|\tau)} &=& {\ds A_K(u,v|\tau) -
\tfrac{1}{2}\sum_{m=0}^{K-1} a^m\, R\!\left(Ku-v-m\tau-\tfrac{K-1}{2}\big|K\tau\right)
\,i\vartheta_1\!\left(v+m\tau+\tfrac{K-1}{2}\big|K\tau\right)}\\[12pt]
&=&
{\ds A_K(u,v|\tau) - \tfrac{a^{\frac{K-1}{2}}}{2K}\sum_{m\in\zi_K}
\!R\!\left(u-\tfrac{v+m}{K}-\tfrac{(K-1)\tau}{2K}\big |\tfrac{\tau}{K}\right)
\,i\vartheta_1\!\left(\tfrac{v+m}{K}+\tfrac{(K-1)\tau}{2K}\big|\tfrac{\tau}{K}\right)\,.
}
\end{array}
\end{equation}
One can then show that these non-holomorphic Appell--Lerch sums indeed transform  under the modular group as  a Jacobi form
of two elliptic variables:
\begin{equation}\label{modA}
{\ds \hat{A}_K(u,v|\tau+1) = \hat{A}_K(u,v|\tau)\,,} \qquad
{\ds \hat{A}_K\left(\tfrac{u}{\tau},\tfrac{v}{\tau}\big|-\tfrac{1}{\tau}\right) = \tau e^{\pi i\frac{(2v-Ku)u}{\tau}}
\hat{A}_K(u,v|\tau)\,,}
\end{equation}
and display the following elliptic transformations:
\begin{equation}\label{elltrans}
\begin{array}{ll}
{\ds \hat{A}_K(u+1,v|\tau) = (-)^K\hat{A}_K(u,v|\tau)\,,} &
{\ds \qquad \hat{A}_K(u,v+1|\tau) = \hat{A}_K(u,v|\tau)\,, }\\[4pt]
{\ds \hat{A}_K(u+\tau,v|\tau) = (-)^Ka^K b^{-1}q^{\frac{K}{2}}\hat{A}_K(u,v|\tau)\,,} & 
{\ds \qquad  \hat{A}_K(u,v+\tau|\tau) = a^{-1}\hat{A}_K(u,v|\tau)\,, }
\end{array}
\end{equation}
which makes them into non-holomorphic Jacobi form of weight 1 and index 
${\footnotesize \begin{pmatrix} -\nicefrac K2 & \nicefrac12\\ \nicefrac12 & 0 \end{pmatrix}}$.

\section{Computations of the gauge threshold corrections}
\label{sec:modgen}
After the preliminary discussions of section~\ref{sec:general} we are now ready to get to the heart of the matter, namely the actual 
computation of the threshold corrections.  We need to consider each gauge factor separately, namely the $U(1)$ (enhanced to $SU(2)$ for $\ell=1$) and the $SO(28)$ factor,  since the former comes from the R-symmetry of the interacting CFT and the latter from the remaining free left-moving fermions. We will 
start with the $SO(28)$ case, which is simpler, and consider in more detail the special case $\ell=1$ for which the superconformal 
symmetry is enhanced.

\subsection{The $SO(28)$ gauge threshold corrections: discrete representations}
\label{sec:so28discr} 

In this section, we consider the one-loop corrections to the $SO(28)$ gauge coupling~(\ref{Agross}). For the sake of clarity,
we start by computing the contribution $\hat{\mathcal{A}}_{SO(28)}^{\text{d}}$ from discrete (BPS) representations that localises on the resolved $A_1$ singularity, which can be determined algebraically from the partition function~(\ref{partfunc}).  As stressed before, this contribution is not modular-invariant by itself, and needs a non-holomorphic completion, namely $\mathcal{R}^{\text{c}}$ in~(\ref{Agross}) coming from non-BPS non-localised states to be free of modular anomalies. 

Keeping in mind that the Kac--Moody level of this orthogonal factors is $k_{SO(28)}=1$, the contribution to the modified elliptic genus~(\ref{Ell}) which localises on the resolved singularity is obtained by projecting the right-moving sector of the internal four-dimensional theory~(\ref{cftback}) onto its twisted Ramond ground state, while summing over all states in the right-moving sector. To facilitate the calculation we split the genus into left- and right-moving contributions:
\begin{equation}
\hat{\mathcal{A}}_{SO(28)}^{\text{d}}(\tau) = \sum_{2j=0}^{k-2}\sum_{2J=2}^k A_{\textsc{l}}^{(j,J)} A_{\textsc{r}}^{(j,J)}\,.
\end{equation}
The right-movers part yields a Witten type index identifying the $\slr_k/U(1)$ discrete spin and the $SU(2)_k/U(1)$ one:
\begin{eqnarray}
\label{index}
{\ds A_{\textsc{r}}^{(j,J)}}  & = & {\ds  \sum_{m\in\zi_{2k}}\!\!\overline{C}^{j}_{m} \oao{1}{1}(\tau,0)\,
   \overline{{\rm Ch}}_{\text{d}} (J,\tfrac{m}{2}-J-\tfrac{1}{2}; \tau,0) \oao{1}{1}\,
   \delta_{2J-m,1}^{[2]} } \notag \\
   \allowdisplaybreaks
   & = & {\ds  \sum_{m\in\zi_{2k}} \!\! \big(\delta_{m,2j+1}-\delta_{m,-(2j+1)}\big)\,\delta_{\tfrac{m}{2}-J+\tfrac{1}{2},0}^{[k]}\,\delta_{2J-m,1}^{[2]}}\\ 
  &  = & {\ds  \left(\delta_{j,J-1}^{[k]}-\delta_{j,-J}^{[k]}\right)  \delta_{2j,2J}^{[2]}}\notag  \\
   &  = & {\ds  \delta_{j,J-1} }\,. \notag
\end{eqnarray}
In particular, we observe from the second line of the above expression that this index counts representations built on 
right-moving anti-chiral primaries of $\slr/U(1)$, see appendix~\ref{appchar}. 
As the extended discrete $\slr/U(1)$ character in expression~(\ref{index}) takes into account all winding sectors 
of the model by incorporating all $\zi_{2k}$ orbits of spectral flow, the latter condition selects all states with:
\begin{equation}
m -2J = -1 \,\text{mod}\, 2k\,.
\end{equation}

To determine  the contributions $A_{\textsc{l}}^{(j,J)}$ from the left-moving sector, we observe that the quadratic
Casimir operator acts on $SO(28)$ characters as $Q^2\,\chi_{SO(28)}(\nu_1,..,\nu_{14}|\tau)= -\frac{1}{4\pi^2}\,
\partial^2_{\nu_1} \chi_{SO(28)}(\nu_1,..,\nu_{14}|\tau) $. Hence we obtain
\begin{multline}
A_{\textsc{l}}^{(j,J)} = 
\frac{1}{2\bar\eta^4} 
 \sum_{\gamma,\delta=0}^1 (-)^{\delta(2j+[\frac k2-1]\gamma)}
\,\chi_{k-2}^{j+\gamma(\frac k2-2j-1)}  \sum_{u,v=0}^1  
\sum_{n \in \mathbb{Z}_{2\ell}}\!\! \mathrm{e}^{-i\pi v (n+ \frac{u}{2}) } 
\,{\rm Ch}_{\text{d}} \big( J,\ell(n+\tfrac{u}{2})-J -\tfrac{u}{2}\big) \oao{u}{v} \\[4pt]
\times \left(-\frac{1}{4\pi^2}\right)
\left[\frac{\vartheta ''\oao{u}{v}}{\vartheta\oao{u}{v}}+\frac{\pi}{\tau_2}\right]
\frac{\vartheta\oao{u}{v}^{14}}{\eta^{14}}\, \delta_{2J,(\ell-1)u}^{[2]}\,,
\end{multline}
We note that the $\zi_2$ orbifold~(\ref{cftback}) and the K-theory condition~(\ref{bsb}) combine to project
out half-integer $SU(2)_{k-2}$ and $\slr_k/U(1)$ spins $j$ and $J$, which are identified through~(\ref{index}).

If we tried to use this algebraic method to determined the contribution of continuous $\slr_k/U(1)$ representations to the modified index on the basis of how they enter into the partition function, for which a non-modular invariant  regularisation has been adopted,
we would obtain zero. The reason is that non-localised states behave like the 'untwisted' sector of an orbifold compactification, hence 
do not contribute to the index because of their fermionic zero-modes. As we shall see shortly, continuous $\slr_k/U(1)$ representations nontheless enter into the modified elliptic genus, if we adopt a non-holomorphic regularisation of the path integral.

Collecting both left- and right-moving contributions from localised states, and leaving for the moment the term 
$\mathcal{R}^\text{c}$ unspecified, the one-loop threshold correction 
to the $SO(28)$ gauge coupling for arbitrary five-brane charge $\mathcal{Q}_5 = k/2 =\ell^2$ reads:
\begin{multline}\label{thresh2}
\Lambda_{SO(28)}[\mathcal{Q}_5] = \\[4pt] \frac{1}{96} \int_{\mathscr{F}}\frac{\di^2\tau}{\tau_2} \, \Gamma_{2,2}(T,U)\,
\left[ \sum_{(u,v)\neq (0,0)} \Phi_k \oao{u}{v} \frac{\vartheta \oao{u}{v}^{14}}{\eta^{14}} \frac{ 
\widehat{E}_2 + (-1)^v \vartheta \oao{u+v+1}{u}^4 - (-1)^u \vartheta \oao{v}{u+v+1}^4 
}{\eta^4} 
 +12\, \mathcal{R}^{\text{c}}[\mathcal{Q}_5] \right] \,.
\end{multline}
The contributions from the $(c,\bar c)=(6,6)$ interacting \textsc{cft} with is encoded
in the localised elliptic indices with mixed left / right boundary conditions:
\begin{equation}\label{genera}
\begin{array}{rcl}
\Phi_k \oao{u}{v} (\nu|\tau) &  = & {\ds  
\text{Tr}_{\mathcal{H}^{(6,6)\,\text{discrete}}_{\ast\otimes \overline{\textsc{r}}}} \left( \ee^{\pi i(2 \bar{J}_0^3
+v J_0^R)}
q^{L_0-\frac{c}{24}} 
\bar{q}^{\bar{L}_0-\frac{\tilde{c}}{24}} z^{J_{R}}\right) }\\[10pt]
& = &{\ds   \sum_{J=1}^{k/2} \big(\chi_{k-2}^{J-1} +\chi_{k-2}^{k/2-J} \big)\sum_{n \in \mathbb{Z}_{2\ell}}\! \mathrm{e}^{-i\pi v (n+ \frac{u}{2}) } \, {\rm Ch}_\text{d} \big( J,\ell n-J;\nu |\tau \big) \oao{u}{v} \, ,}
\end{array}
\end{equation}
where $\ast$ stands for NS when $u=0$ and for R when $u=1$.

The elliptic indices~(\ref{genera}) can be obtained by spectral flows of  what is commonly known as the {\it elliptic genus} of the 
$(c,\bar c) =(6,6)$  \textsc{cft} underlying the solution~(\ref{DSgeom}). This topological
invariant is obtained by projecting the trace on the (discrete representation) Hilbert space onto the $\tilde{\textsc{r}}\otimes \tilde{\bar{\textsc{r}}}$ ground state of the \textsc{cft}. It is nothing but 
\begin{equation}\label{ellsupl}
\Phi_k (\nu|\tau) \equiv \Phi_k \oao{1}{1} (\nu|\tau)\,.
\end{equation}
It will be convenient for later us to  package the contribution of $\slr_k/U(1)$ characters with spin $J$ in the single function:
\begin{equation}\label{Yfact}
\mathcal{Y}_k^J(\nu|\tau) = \sum_{n\in\zi_{2\ell}} (-)^{n+1} \,
i{\rm Ch}_\text{d} \big( J,\ell(n+\tfrac{1}{2})-J -\tfrac{1}{2};\nu|\tau\big) \oao{1}{1}\,.
\end{equation}
In terms of $\Phi_k$, the remaining genera~(\ref{genera}) are
easily recovered by spectral flow:
\begin{equation}\label{ex1}
\Phi_k\oao{1-a}{1-b}(\nu|\tau) =  (-)^b\ee^{\frac{\pi i (\ell+1)}{2}(a+b+ab)} q^{\frac{k+2}{16}a^2} z^{-\frac{k+2}{4\ell}a}\sum_{J=1}^{k/2} \big({\chi}^{J-1} + {\chi}^{k/2-J} \big)  \, \mathcal{Y}_k^J\big(\nu-\tfrac{\ell(a\tau + b)}{2}\big|\tau\big)\,,
\end{equation}
with $\ell \in 2\en^* +1$.

As these elliptic genera are restricted to localised states, they can be given the same interpretation as for compact models.  More specifically,  $\Phi_k \oao{0}{v}$ are elliptic generalisations of the Dirac index, which keep track of how antisymmetric tensor representations of the $SO(2)^2 \subset SO(4)$  embedding of the line bundle $\mathcal{O}_{\mathbb{P}^1}(1) \oplus \mathcal{O}_{\mathbb{P}^1}(\ell)$  are counted,
whether with a plus ($v=0$) or minus ($v=1$) sign~\cite{Witten:1986bf,Harvey:1995fq}. In contrast, the index $\Phi_k \oao{1}{0}$ 
captures the coupling of the elliptic generalisation of the Dirac index to the spinor bundles associated to 
$\mathcal{O}_{\mathbb{P}^1}(1) \oplus \mathcal{O}_{\mathbb{P}^1}(\ell)$.

Starting from the discrete representations contribution to the elliptic genus~(\ref{thresh2}), using equations~(\ref{chi-lim}) 
and~(\ref{idchar}),  one reproduces the inverse cusp form $\eta^{-24}$ characteristic of the polar behaviour of $\caln_{\textsc{st}}=2$ heterotic gauge threshold  corrections~(\ref{refthresh}), related to the would-be tachyon
(see section~\ref{sec:persp} for discussion). This confirms that the contribution of localised states to the gauge threshold 
correction~(\ref{thresh2}) is similar in nature to what is expected for a  genuine heterotic compactification.

\subsection{Infinite volume regularisation and non-holomorphic completion of the  Appel--Lerch sum}

We have shown above how to express the contribution to the gauge threshold correction of states localised on the resolved singularity,
see eq.~(\ref{thresh2}), in terms of a combination of holomorphic $\slr / U(1)$ characters, given by eq.~(\ref{Yfact}). As we discussed 
above, the modular properties of these characters, given in appendix~\ref{appchar}, imply that the result is not a modular form as 
it should. 

This problem can be traced back to the partition function~(\ref{partfunc}), from which the elliptic genus has been extracted, which 
displays a holomorphic anomaly, since the infinite-volume divergence has been remove in a rather cavalier way, which preserves the splitting 
of the theory into holomorphic and anti-holomorphic characters of the chiral algebra but spoils modular invariance. 

\subsubsection*{Completing the elliptic genus}
A modular-covariant regularization of the $\slr/U(1)$ elliptic genus has been developped first in~\cite{jan} and subsequent~\cite{Ashok:2011cy,Ashok:2012qy}. The idea behind this work, which is summarised in appendix~\ref{sec:ellgen}, was to reformulate the elliptic genus directly in terms of a path integral. The poles in the zero-mode integral, corresponding 
to the infinite target-space volume divergence, were regularised in a way that preserves modular invariance, thus giving an
unambiguous prescription to evaluate the elliptic genus.  As explained in more details in appendix~\ref{sec:ellgen},
the result of this evaluation splits into a holomorphic contribution coming from the discrete representations, 
which can be resummed into the Appell--Lerch sum $A_{2k}$, and 
a non-holomorphic contribution coming from continuous representations:
\begin{equation}\label{regut}
\begin{array}{rcl}
{\ds \widehat{\mathcal{Z}}_k(\nu|\tau)} & = &{\ds  \sum_{2J=1}^{k} \text{Ch}_d(J,-1;\tau,\nu\big) \oao{1}{1}  
- \frac{i}{\pi}\sum_n \int_{\er-i\varepsilon} \frac{\di p}{2ip+n} \,\text{Ch}_c\left(\tfrac12+ip,\tfrac{n}{2};\nu\big|\tau\right)\oao{1}{1}\,\bar{q}^{\frac{p^2}{k}+\frac{n^2}{4k}} } \\[14pt]
& = & {\ds - \hat{A}_{2k}\left(\tfrac{\nu}{k},2\nu\big|\tau\right)  \frac{i\theta_1(\nu|\tau)}{\eta(\tau)^3} \,.}
\end{array}
\end{equation}
In this non-holomorphic regularisation of the  infinite target-space volume divergence, continuous representations 
supply the precise counter-term needed to cancel the holomorphic anomaly, which is none other than the (transformed) shadow function $R(u|\tau)$~(\ref{Zwegers}), summed as
in expression~(\ref{Appell-comp}).

In the cases considered here a similar procedure can be carried out. However, since we have already computed the discrete 
representations contribution, $i.e.$ the Mock Jacobi form of interest, it will suffice to use the {\it shadow map} $\mathcal{S}$ dictated by theregularisation scheme~(\ref{regut}) in order to get a genuine modular form. To this end we rewrite the contribution of 
discrete $\slr_k/U(1)$ representations~(\ref{Yfact}) as:
\begin{align}\label{sj}
\mathcal{Y}_k^J(\nu|\tau) & =  {\ds \sum_{n\in\zi_{2\ell}} (-)^{n+1} \,
i{\rm Ch}_\text{d} \big( J,\ell(n+\tfrac{1}{2})-J -\tfrac{1}{2};\tau,\nu) \oao{1}{1} }\notag\\
   &= {\ds \sum_{n\in\zi_{2\ell}} (-)^{n+1} \sum_{m\in \zi}
   \frac{q^{\frac12\left(2\ell m+n+\frac12\right)^2-\frac1k\left(J-\frac12\right)^2}
   z^{\frac1\ell(2\ell m+n+\frac12)}}{1-zq^{\ell(2\ell 
   m+n)+\frac{\ell+1}{2}-J}}\,\frac{i\vartheta_1(\tau,\nu)}{\eta(\tau)^3} } \notag \displaybreak[2]\\
   & =  {\ds   q^{\frac18-\frac1k\left(J-\frac12\right)^2}
   \sum_{s\in \zi} (-)^{s+1} \frac{q^{\frac12 s(s+1)}
   z^{\frac1\ell\left(s+\frac12\right)}}
   {1-zq^{\ell s+\frac{\ell+1}{2}-J}}\,\frac{i\vartheta_1(\tau,\nu)}{\eta(\tau)^3}}\notag \\
&= 
{\ds -q^{-\frac{1}{k}\left(J-\frac{\ell+1}{2}\right)^2}\,\tfrac{1}{\ell} \sum_{m=0}^{\ell-1} e^{-\pi i\frac{m}{\ell}} A_1\!\Big(\tfrac{1}{\ell}\big(\nu+\left( \tfrac{\ell+1}{2}-J\right)\tau +m\big),\tfrac{\nu}{\ell}\Big|\tau \Big)\,\frac{i\vartheta_1(\tau,\nu)}{\eta(\tau)^3}\,.}
\end{align}
To express the result in terms of a level 1 Appell sum we used the following identity~\footnote{We are particularly grateful to S
.~Zwegers for suggesting this formula.}:
\begin{equation}
\frac{1}{1-aq^{\ell n}} = \frac{1}{\ell}\sum_{m=0}^{\ell-1} \frac{1}{1-e^{2\pi i\frac{m}{\ell}} a^{\frac{1}{\ell}} q^{n}}\,,\qquad a= e^{2\pi i u}\,.
\end{equation}
Then, the regularisation of the infinite target-space volume divergence
goes through~\footnote{
It is interesting to note that initial $\zi_{2\ell}$ orbifold of the $\slr_k/U(1)$ theory in~(\ref{sj}) 
is rewritten in terms of a $\zi_\ell$ orbifold of the Appell sum
$A_1$. As can be seen by combining~(\ref{ID}) and~(\ref{zorb}), $A_1$ encodes the discrete representation ({\it i.e.}  holomorphic)  contribution to the elliptic genus of  the $(\slr_2/U(1))/\zi_2$ orbifold:
$$
\tfrac12 \sum_{\gamma,\delta \in\zi_2} z^{2\gamma}\, q^{\frac{\gamma^2}{2}} \,\mathcal{Z}^\text{d}_2(\nu+\gamma\tau+\delta|\tau)
=    - A_1(\nu|\tau) \,\frac{i\vartheta_1(\nu|\tau)}{\eta(\tau)^3}\,,
$$ 
$\mathcal{Z}^\text{d}_2$ being the localised part of the elliptic genus~(\ref{ell-Liouv0}). 
By virtue of a relation similar to~(\ref{Ak}), the holomorphic piece in the elliptic genus of  the $(\slr_k/U(1))/\zi_{2\ell}$  
theory can thus be rewritten, for $k=2\ell^2$, in terms of a $\zi_{\ell}$ orbifold of the  $(\slr_2/U(1))/\zi_{2}$ theory, and eventually of 
the Appell sum $A_1$. } like in eq.~(\ref{regut}). Using~(\ref{appfun0}) and its completion into a Ma\ss\, form~(\ref{Ahat}), the full expression for the $\slr_k/U(1)$ factor~(\ref{sj}) can be nicely repackaged in a sum of non-holomorphic Appell functions:
\begin{equation}\label{Rcorr0}
\widehat{\mathcal{Y}}_k^J(\nu|\tau)= q^{-\frac{1}{k}\left(J-\frac{\ell+1}{2}\right)^2}\,\tfrac{1}{\ell} \sum_{m=0}^{\ell-1} e^{-\pi i\frac{m}{\ell}} 
\widehat{\mu}\Big(\tfrac{1}{\ell}\big(\nu+\left( \tfrac{\ell+1}{2}-J\right)\tau +m\big),\tfrac{\nu}{\ell}\Big|\tau \Big)\,\frac{\vartheta_1(\frac\nu\ell|\tau)\,\vartheta_1(\nu|\tau)}{\eta(\tau)^3}\,.
\end{equation}
From the above expression, we obtain the regularised expression of the elliptic genus~(\ref{ellsupl}), now also including continuous representation resummed in the (transform) shadow function $R(u|\tau)$~(\ref{Zwegers}), as the following weight 0 Ma\ss\, form:
\begin{equation}
\label{Rcorr}
\widehat{\Phi}_k (\nu|\tau)=  \sum_{J=1}^{k/2} \big( \chi_{k-2}^{J-1}+ \chi_{k-2}^{k/2-J}\big) \,
\widehat{\mathcal{Y}}_k^J(\nu|\tau) \,.
\end{equation}

\subsubsection*{Spectral flow and gauge threshold corrections}
Using this result one can recover the full set of regularised genera~(\ref{ex1}) for the $(c,\bar c)=(6,6)$ theory by spectral flowing 
the elliptic genus~(\ref{Rcorr}):
\begin{multline}
\label{Gencorr}
\widehat{\Phi}_k\oao{u}{v}(\nu|\tau)=   \sum_{J=1}^{k/2} \big( \chi_{k-2}^{J-1}+ \chi_{k-2}^{k/2-J}\big) \,
q^{-\frac{1}{k}\left(J-\frac{\ell+1}{2}\right)^2}  \,
 \tfrac{1}{\ell}\sum_{m=0}^{\ell-1} e^{-\pi i\frac{m}{\ell}}\times
\\
\times  \widehat{\mu}
\Big(\tfrac{1}{\ell}\big(\nu+( \tfrac{\ell u+1}{2}-J) \tau + \tfrac{\ell (v-1)}{2}+m\big),
\tfrac{\nu}{\ell}+\tfrac{(u-1)\tau +v-1}{2}\Big|\tau \Big)\,
\frac{\vartheta\oao{u}{v}(\frac\nu\ell|\tau)\,\vartheta\oao{u}{v}(\nu|\tau)}{\eta(\tau)^3}\,.
\end{multline}
The  threshold corrections to the $SO(28)$ gauge coupling for arbitrary $\mathcal{Q}_5= k/2$ units of
five-brane flux then reads:
\begin{multline}\label{fullthr}
\allowdisplaybreaks
\Lambda_{SO(28)}[\mathcal{Q}_5] = \frac{1}{96} \int_{\mathscr{F}}\frac{\di^2\tau}{\tau_2} \, \Gamma_{2,2}(T,U)
\,\sum_{J=1}^{k/2} \big(\chi_{k-2}^{J-1}+ \chi_{k-2}^{k/2-J} \big) \,
q^{-\frac{1}{k}\left(J-\frac{\ell+1}{2}\right)^2} \times \\[4pt]
\times 
\sum_{(u,v)\neq (1,1)}      \,
 \frac{1}{\ell}\sum_{m=0}^{\ell-1} e^{-\pi i\frac{m}{\ell}}\, \frac{
{\widehat{\mu}
\Big(\big(\tfrac{\ell u+1}{2}-J \big) \tfrac{\tau}{\ell} +\tfrac{v-1}{2}+\tfrac{m}{\ell}\big),
\tfrac{(u-1)\tau +(v-1)}{2}\Big|\tau \Big)}}{\eta} \times
 \\[-4pt]
\times
\frac{\Big({\widehat{E}}_2+(-)^v \vartheta\oao{u+v+1}{u}^4 - (-)^u \vartheta\oao{v}{u+v+1}^4\Big)\vartheta\oao{u}{v}^{16}}{\eta^{20}} \,.
\end{multline}
We  can in particular extract the contribution of non-localised bulk states from expression~(\ref{fullthr}):
\begin{equation}\label{nlocSO28}
\begin{array}{rcl}
{\ds \mathcal{R}^{\text{c}}[\mathcal{Q}_5]} & = & 
{\ds - \frac{1}{12} \sum_{J=1}^{k/2} \big(\chi_{k-2}^{J-1}+ \chi_{k-2}^{\frac k2-J} \big)
\, \frac{1}{\ell}\sum_{m=0}^{\ell-1}e^{-\pi i\frac{m}{\ell}}  q^{-\frac{1}{k}\left(J-\frac{\ell+1}{2}\right)^2}  
R\!\left(
\big( \tfrac{\ell+1}{2}-J\big)\tfrac{\tau}{\ell} +\tfrac{m}{\ell} \big| \tau 
\right)\,
\frac{ \widehat{E}_2E_8-E_{10}}{\eta^{21}}} \\[12pt]
 & = & 
 {\ds \frac{1}{16}\sum_{J=1}^{k/2} \big(\chi_{k-2}^{J-1}+ \chi_{k-2}^{k/2-J} \big) \sum_{n\in \zi} (-)^n \, 
\text{sgn}\big(n+\tfrac12 \big)\,
 \text{erfc}\, \left(\left| (n+1)k -2J +1 \right| \sqrt{\tfrac{\pi\tau_2}{k}}\right) \times } \\[10pt]
&& \hspace{8.5cm} {\ds  \times \, q^{-\frac{1}{4k}((n+1)k -2J +1)^2}\,\frac{D_8 E_8}{\eta^{21}} } \\[10pt]
  & = & 
  {\ds  \frac{1}{8}\sum_{J=1}^{k/2} \big(\chi_{k-2}^{J-1}+ \chi_{k-2}^{k/2-J} \big) \sum_{n=0}^\infty (-)^n
\text{erfc}\,  \left( |n k  + 2J - 1|  \sqrt{\tfrac{\pi \tau_2}{k}}\right)  
  q^{-\frac{1}{4k}( n k  + 2J - 1)^2}\,\frac{D_8 E_8}{\eta^{21}} } \,.
\end{array}
\end{equation}
The intermediate steps that bring us from the first to the second line are explicitly given in Appendix~\ref{NLT1}.
To go from the second to the third line, we have exploited the $\zi_2$ symmetry of the $SU(2)_{k-2}$ factor.

As anticipated in expression~(\ref{Agross}), the bulk state contribution~(\ref{nlocSO28}) is (up to a factor $k_a$)
universal, {\it i.e.} gauge group independent for both $SO(28)$ and $U(1)$ factors. This will become clearer
in section~\ref{sec:u1}.

\subsubsection*{Polar structure}  
It is worth spending some time discussing the polar behaviour of the modified elliptic genus appearing in the 
gauge threshold~(\ref{fullthr}), thereby clarifying its physical signification. 
Firstly, the polar structure of the contribution of localised states 
$\hat{\mathcal{A}}_{SO(28)}^\text{d}$ has already been addressed in 
section~\ref{sec:so28discr}. It has been shown that it reproduces the inverse cusp form $\eta^{-24}$ in the denominator of 
$\hat{\mathcal{A}}_{SO(28)}^\text{d}$. Further analysing  the Fourier expansion of the localised part of expression~(\ref{thresh2}), we can
show that this expression has no more than a dressed single pole $(\tau_2 q)^{-1}$, also characteristic of heterotic $K3$ compactifications.

Turning to the contribution from bulk states to the threshold corrections, given by eq.~(\ref{nlocSO28}), 
we observe, following~\cite{zwegers} that, for $\tau_2 >0$, $n\geqslant 0$ and $J\in[1,.., \frac k2]$ :
\begin{equation}
\label{pole2}
\begin{array}{rcl}
{\ds \left|(-)^n  \text{erfc}\!\left( \tfrac{1}{2\ell} (n k  + 2J - 1) \sqrt{2\pi \tau_2} \right)  
  q^{-\frac{1}{4k}( n k  + 2J - 1)^2}\right|  }  & \leqslant & {\ds  \ee^{- \frac{\pi (nk+2J-1)^2}{k} \tau_2} 
\left| q^{-\frac{1}{4k}( n k  + 2J - 1)^2}\right|} \\[6pt]
 & =&{\ds  \ee^{-\frac{\pi (nk+2J-1)^2}{2k} \tau_2} \,.}
  \end{array}
\end{equation}
Hence all terms in the sum over $n$ in~(\ref{nlocSO28}) are exponentially suppressed as $\tau_2\rightarrow \infty$. 
Thus, including the $\eta^{-3}$ factor coming from $\chi_{k-2}^J(\tau)$, the only pole at $\tau_2 \rightarrow \infty$ in 
expression~(\ref{nlocSO28}) comes from:
\begin{equation}
\frac{D_8 E_8}{\eta^{24}} = \frac{4}{\pi  q \tau _2}-960+\frac{2016}{\pi  \tau _2} + \mathcal{O}(q) \,.
\end{equation}
Taking into account the Fourier expansion of the characters $\chi_{k-2}^J(\tau)$, {\it cf.} eq.~(\ref{chi-lim}), we see that 
$\mathcal{R}^\text{c}$ only has a $q^{-1}$ cusp at $\tau_2\rightarrow \infty$, whose real part is bounded by:
\begin{equation}\label{bound}
\frac{1}{\tau_2} \,\sum_{n_1\in [1,..,2\sqrt{2}\ell[} \sum_{|n_2|\in [1,..,2\sqrt{k-2}[ } c_{n_1n_2}\,\ee^{-\frac{\pi}{2} \left(\frac{n_1^2}{k} + \frac{n_2^2}{(k-2)}\right) 
\tau_2}  \,\frac{1}{|q|}\, ,
\end{equation}
with $n_1$ and $n_2$ following some progression in $\zi$. More specifically we have $n_1 =nk+2J-1 \geqslant 0$ and $n_2 = 2(k-2) m +2J-1$ for $m\in \zi$. The contribution from bulk states thus has a similar polar behaviour as the localised part $\mathcal{A}_{SO(28)}^\text{d}$, with a simple pole 'dressed' by a regulator; the difference being that the regulator is now exponentially suppressed for $\tau_2\rightarrow \infty$, which we interpret as the signature of an unphysical tachyon appearing in the spectrum of non-localised states.

Thus we conclude that by considering a regime where $T_2 >1$ we can compute the integral~(\ref{fullthr}) by unfolding the 
fundamental domain $\mathscr{F}$ against the lattice sum $\Gamma_{2,2}(T,U)$, similarly in every respect to calculations of heterotic gauge thresholds for 
 toroidal orbifold compactifications~(\ref{refthresh}).

\subsection{Threshold corrections for $\mathcal{Q}_5=1$ and $\mathcal{N}=4$ characters}
\label{workex}

After having discussed the $SO(28)$ threshold for a generic value of the fivebrane charge, we would like to discuss here 
in detail the particular case $\mathcal{Q}_5=1$, which is somehow degenerate, but displays interesting features. In this case, 
the worldsheet supersymmetry of  the $(6,6)$ CFT is further enhanced to $\mathcal{N}_{\textsc{ws}}=(4,4)$, 
so that the result can be nicely repackaged, as we shall see, into  $\mathcal{N}_{\textsc{ws}}=4$ superconformal 
characters at level $\kappa=1$. This will help making contact with the known threshold corrections for $T^2\times K3$.

For $k=2$, the contributions to~(\ref{thresh2}) from discrete representations greatly simplifies. In particular,
as the $\slr_k/U(1)$ spin can now only take the value $J=1$, the $\zi_2$ orbifold which selects integer spins 
in~(\ref{thresh2}) becomes trivial.  In addition, the $SU(2)_{k-2}$ theory reduces now  to the identity. Then:
\begin{multline}\label{threshk2}
\Lambda_{SO(28)}[1] = \frac{1}{96} \int_{\mathscr{F}}\frac{\di^2\tau}{\tau_2} \, \Gamma_{2,2}(T,U) 
\Big[ \sum_{(u,v)\neq (1,1)}\sum_{n \in \mathbb{Z}_{2}}\! 
\ee^{-i \pi v(n+\frac{u}{2})} {\rm Ch}_\text{d} \big( 1, n-1 \big) \oao{u}{v}\, \times\\ \times \, 
\frac{ 
\left(\widehat{E}_2 + (-1)^v \vartheta \oao{u+v+1}{u}^4 - (-1)^u \vartheta \oao{v}{u+v+1}^4 \right) \vartheta\oao{u}{v}^{14}
}{\eta^{18}} 
+ 12\,{ \mathcal{R}^{\text{c}}[1]}  \Big]\, .
\end{multline}
We will discuss now how to rephrase this result in terms of $\mathcal{N}=4$ characters. We refer the reader to the Appendix~\ref{appchar2}, in particular to subsection~\ref{kappa1}, for details on the subject.

Representations of the $\caln_{\textsc{ws}}=4$ superconformal algebra at level $\kappa$ are distinguished by two quantum numbers $(h,I)$, namely their conformal weight and their spin. Unitary representations are:
\begin{itemize}
\item BPS representations labeled by discrete quantum numbers $(h,I)$ and with massless ground states,
which are obtained by saturating the unitary bounds, {\it i.e.} by setting: 
$$
\begin{array}{lll}
 h=I \;\text{ in the \textsc{ns} sector}\,, &\quad
 h=\frac{\kappa}{4} \;\text{ in the \textsc{r} sector}\,, &\quad
 \text{ for spin range } 0 \leqslant   I \leqslant \frac\kappa2
\end{array}
$$
\item non-BPS massive representations with discrete spin values $I$ but continuous conformal weight $h$ bounded from below:
$$
\begin{array}{lll}
 h> I \; \text{ with } \;0  \leqslant  I  \leqslant   \tfrac{\kappa-1}{2} \;\text{ in the \textsc{ns} sector}\,, &\quad
 h> \frac{\kappa}{4} \;  \text{ with }  \; \tfrac12  \leqslant  I  \leqslant   \tfrac{\kappa}{2}\;  \text{ in the \textsc{r} sector}
\end{array}
$$
\end{itemize}

Focusing on discrete representations, we can exploit the branching relations of the $\caln_{\textsc{ws}}=4$ 
characters at level $\kappa=1$ into $\caln_{\textsc{ws}}=2$ representations with $c=6$ in order to  rewrite the localised 
elliptic genus $\Phi_2$ in terms of the $\caln_{\textsc{ws}}=4$ character for the only normalisable
BPS (discrete) representation in the twisted Ramond sector, defined in eq.~(\ref{massless1}):
\begin{equation}\label{twisted-N4}
\Phi_2 (\nu|\tau) =  \sum_{n\in\zi_2} e^{-i\pi(n+\frac12)}\, {\rm Ch}_\text{d} \big( 1,n-1;\nu|\tau\big) \oao{1}{1} =
  {\rm ch}_{1,\frac14,0}^{\widetilde{\textsc{r}}}\big(\nu|\tau\big)
\end{equation}
The other elliptic indices with mixed boundary conditions~(\ref{genera}) can then be obtained by spectral flow, as previously 
explained, namely
\begin{equation}\label{branch2}
\Phi_2\oao{1-a}{1-b}(\nu|\tau) = (-)^{a(1+b)} q^{\frac{a^2}{4}}z^{-a}
\,{\rm ch}_{1,\frac14,0}^{\widetilde{\textsc{r}}}\big(\nu-\tfrac{a\tau+b}{2}\big|\tau\big) \,, \qquad a,b \in[0,1]\,,
\end{equation}
with different values of the 'spin structure' reproducing all the characters for normalisable BPS representations listed in~(\ref{massless1}), for instance $\Phi_2 \oao{0}{0} (\nu|\tau)=  {\rm ch}^\textsc{ns}_{1,\frac12,\frac12}(\nu|\tau \big)$.  This corresponds pictorially to circumnavigating the orbit under spectral flow of the $(h,I)=(\nicefrac12,\nicefrac12)$ representation in the \textsc{ns} sector, as illustrated by the diagram in Figure~\ref{fig:spec}.\footnote{We should emphasise that the orbit for the other existing discrete representation with $(h,I)=(0,0)$ in the \textsc{ns} sector does not contribute to the localised part of the threshold correction, since these representations are non-normalisable, as is the corresponding identity representation of $\slr_2 /U(1)$.}
Identities~(\ref{branch2}) belong to the more general case of branching relations of $\caln_{\textsc{ws}}=4$
super-conformal representations into $\caln_{\textsc{ws}}=2$ ones~\cite{eguchi1}. We shall see shortly how these relations can be exploited 
to rephrase the gauge threshold corrections for $\mathcal{Q}_5=1$ in a more suggestive way.
\begin{figure}
\begin{center}
\begin{tikzpicture}[description/.style={fill=white,inner sep=2pt}]
 \matrix (m) [matrix of math nodes, row sep=5em, column sep=5em, text height=2ex, text depth=1ex]
{{\rm ch}_{1,\frac14,0}^{\widetilde{\textsc{r}}} &  {\rm ch}_{1,\frac12,\frac12}^{\widetilde{\textsc{ns}}} \\
{\rm ch}_{1,\frac14,0}^{\textsc{r}}  &  {\rm ch}_{1,\frac12,\frac12}^{\textsc{ns}}  \\};
\path[->,font=\scriptsize] 
(m-1-1) edge node[auto] {$ \nu\mapsto\nu-\tfrac{\tau}{2} $}  (m-1-2)
(m-1-1) edge  node[below,sloped]{$\nu\mapsto\nu-\tfrac{1}{2}$} (m-2-1)
(m-2-1) edge  (m-2-2)
(m-1-2) edge  (m-2-2)
;
\end{tikzpicture}
\end{center}
\caption{Spectral flow of $\caln_{\textsc{ws}}=4$ characters at level $\kappa=1$ : $(h,I)=(\nicefrac12,\nicefrac12)$ orbit}
\label{fig:spec}
\end{figure}

Since  we are dealing with a degenerate case where the $SU(2)_{k-2}$ factor in the $\widetilde{\textsc{eh}}$ \textsc{cft} is the identity, the localised 
part of the elliptic genus is directly given by expression~(\ref{sj}) for $k=2$ and takes the simple form:
\begin{equation}\label{ID}
\Phi_2 (\nu|\tau) 
= - A_1(\nu|\tau)\,\frac{i\vartheta_1(\nu|\tau)}{\eta(\tau)^3} 
\equiv \mu(\nu|\tau)\,\frac{\vartheta_1(\nu|\tau)^2}{\eta(\tau)^3}\,,
\end{equation}


\subsubsection*{$\mathcal{N}=4$ characters and $\mathcal{N}=2$ Liouville theory}
We have seen above that the genera $\Phi_2\oao{u}{v}$  organise into an 
orbit under spectral flow, comprising the $\caln_{\textsc{ws}}=4$ character ${\rm ch}_{1,\frac14,0}^{\widetilde{\textsc{r}}}$. 
Now, from~(\ref{twisted-N4}), we observe that this character is precisely the holomorphic part of the elliptic genus of a $\zi_2$ orbifold of the $\mathcal{N}=2$ Liouville theory at level $k=2$ (see appendix~\ref{sec:ellgen}), as has already been pointed out in~\cite{Eguchi:2004yi}. From eq.~(\ref{ell-Liouv}) one finds indeed:
\begin{equation}\label{partorb}
{\rm ch}_{1,\frac14,0}^{\widetilde{\textsc{r}}}(\nu|\tau) = 
\tfrac12 \sum_{\gamma,\delta\in\zi_2} z^{2\gamma}\, q^{\frac{\gamma^2}{2}} \,\mathcal{Z}_2^{\text{d}}(\nu+\gamma\tau+\delta|\tau) 
\end{equation}
As already explained, by using the regularisation scheme~(\ref{regut}) we can compute the non-holomorphic completion of the
localised part of the elliptic genus, by which we determine the complete elliptic genus of the orbifolded $\mathcal{N}=2$ Liouville theory at level $k=2$:
\begin{equation}\label{zorb}
\widehat{\Phi}_2(\nu|\tau) = \tfrac12 \sum_{\gamma,\delta \in\zi_2} z^{2\gamma}\, q^{\frac{\gamma^2}{2}} \,\widehat{\mathcal{Z}}_2(\nu+\gamma\tau+\delta|\tau)  
 =  \widehat{\mu}(\nu|\tau)\,\frac{\vartheta_1(\nu|\tau)^2}{\eta(\tau)^3} \, .
\end{equation}
in keep with the general formula~(\ref{sj}). By making use of the modular and elliptic properties of the Appell function~(\ref{modpropmu}), 
we find that $\widehat{\Phi}_2$ transforms as Jacobi form weight 0 and index 1, in accordance with the corresponding transformations of the $(\slr_k/U(1))/\zi_2$ theory, see~(\ref{ellM}) and~(\ref{ellSL}).\footnote{The only trace of the $\zi_2$ orbifold is found in the extension of the allowed range for the shift parameters in the elliptic transformations~(\ref{ellSL}), which is now $\mu,\,\lambda\in \zi$ instead of $2\zi$.}

By spectral flow one obtains the remaining regularised genera:
\begin{equation}\label{branch3}
\widehat{ \Phi}_2\oao{u}{v}(\nu|\tau) =
\widehat{\mu}\big(\nu+\tfrac{(u-1)\tau+(v-1)}{2}\big|\tau\big) \,\frac{\vartheta\oao{u}{v}^2(\nu|\tau)}{\eta(\tau)^3}\,,
\end{equation}
leading to the $SO(28)$ threshold corrections for five-brane charge $\mathcal{Q}_5=1$:
\begin{multline}\label{thresh4}
\Lambda_{SO(28)}[1]  =  \frac{1}{96} \int_{\mathscr{F}}\frac{\di^2\tau}{\tau_2} \,\Gamma_{2,2}(T,U)\, \times \\[10pt]
\times \sum_{(u,v)\neq(1,1)} \frac{\widehat{\mu}\!\left(\tfrac{(u-1)\tau+(v-1)}{2}\big|\tau\right)
\left(\widehat{E}_2 + (-1)^v \vartheta \oao{u+v+1}{u}^4 - (-1)^u \vartheta \oao{v}{u+v+1}^4 \right) \vartheta\oao{u}{v}^{16}}{\eta^{21}}
\,.
\end{multline}
As for the general $\mathcal{Q}_5 > 1$ case~(\ref{nlocSO28}), from the decomposition of the elliptic genus~(\ref{zorb}) into discrete and continuous $\slr_2/U(1)$ representations,
we may single out the contribution of non-localised bulk modes:
\begin{equation}\label{conR}
\mathcal{R}^{\text{c}}[1] =  -\frac{R(\tau)\big(\widehat{E}_2 E_8 -E_{10}\big)}{12\,\eta^{21}} = 
 \frac{R(\tau)\, D_{8} E_8}{16\,\eta^{21}}
\end{equation}
which factorises in terms of the (transform) shadow function $R(\tau)\equiv R(0|\tau)$ and
the covariant derivative $D_8E_8$~(\ref{DE2}). This contribution is universal (up to a $k_a$ factor) for both 
$SO(28)$ and $SU(2)$ thresholds.

\subsubsection*{Local thresholds vs. $K3$ thresholds}
We shall now exploit the $\mathcal{N}_{\textsc{ws}}=4$  superconformal algebra at level $\kappa=1$ that appears for 
$\mathcal{Q}_5=1$ in order to make contact with the well-known threshold corrections for $T^2\times K3$ compactifications. In the same vein, we will show in the next section that the $SU(2)$ threshold for $\mathcal{Q}_5=1$ can be cast in the same universal form.

The relation between $\mathcal{N}_{\textsc{ws}}=4$ characters at level $\kappa=1$ and $K3$ characters can be illustrated by considering the S-transformation of say the twisted Ramond character for (normalisable) discrete representations~(\ref{massless1}):
\begin{equation}\label{mordell}
\begin{array}{rcl}
{\ds {\rm ch}_{1,\frac14,0}^{\tilde{\textsc{r}}}\!\left(\tfrac{\nu}{\tau}|-\tfrac{1}{\tau} \right)} & = &
{\ds  \ee^{2\pi i\frac{\nu^2}{\tau}}\left[
  {\rm ch}_{1,\frac14,0}^{\tilde{\textsc{r}}}(\nu|\tau)
  -\frac12 
  \int_\er   \frac{\di x}{\cosh\pi x}\,{\rm ch}_{1,\frac{x^2}{2}+\frac38,\frac12}^{\tilde{\textsc{r}}}(\nu|\tau)
   \right]  } 
 \\[12pt]
  &=&
 {\ds \ee^{2\pi i\frac{\nu^2}{\tau}}\left[
  {\rm ch}_{1,\frac14,0}^{\tilde{\textsc{r}}}(\nu|\tau)
  -\frac12 M(0|\tau) \frac{\vartheta_1(\nu|\tau)^2}{\eta(\tau)^3} \right] } \,.
  \end{array}
\end{equation}
As can alternatively be inferred from combining identities (\ref{twisted-N4}) and~(\ref{ID}), the transformation law~(\ref{mordell}) indicates that ${\rm ch}_{1,\frac14,0}^{\tilde{\textsc{r}}}(\nu|\tau)$ is a Mock Jacobi form. In particular, the extra piece appearing in the RHS of~(\ref{mordell}) and breaking modular covariance can be reexpressed in terms of continuous twisted Ramond $\mathcal{N}_{\textsc{ws}}=4$ characters~(\ref{twistcont}) or alternatively repackaged into the Mordell integral~(\ref{Mfun}). 

From eq.~(\ref{mordell}) we note in particular that the continuous $\mathcal{N}_{\textsc{ws}}=4$ representations  which contribute
to the Mordell integral have conformal weight in the range $h = \frac{x^2}{2}+\frac38 > \frac14$, which falls within the $h> \frac{\kappa}{4}$ bound imposed by unitarity on non-BPS representations (see Appendix~\ref{sec:class}). Now, the
integral $M(0|\tau)$ has been shown by Mordell to be S-invariant, which follows from rewriting it as~\cite{mordell}:
\begin{equation}\label{Mordellrel0}
\tfrac12 M(\tau)=h_3(\tau)+h_3(-1/\tau)\,,
\end{equation}
with the function $h_3$ given by:
\begin{equation}
\label{Mordellrel}
 h_3(\tau) =  \frac{1}{\vartheta_3(0|\tau)\eta(\tau)} \sum_{n\in \zi} 
 \frac{q^{\frac12 (n-\frac12)(n+\frac12)} }{1+q^{n-\frac12} } 
 =  \frac{iq^{\frac18}\,A_1\big(-\tfrac{\tau +1}{2},-\tfrac{\tau +1}{2}\big|\tau \big)}{\vartheta_3(0|\tau)\eta(\tau)} 
  \,.
\end{equation}
By spectral flow of the Appell function, we may define two other such functions:
\begin{equation}
h_2(\tau) = \frac{i A_1\big(-\tfrac{1}{2},-\tfrac1 2\big|\tau \big)}{\vartheta_2(0|\tau)\eta(\tau)}  
\quad , \qquad 
h_4(\tau) =
\frac{q^{\frac18}\,A_1\big(-\tfrac{\tau}{2},-\tfrac{\tau}{2}\big|\tau \big)}{\vartheta_4(0|\tau)\eta(\tau)} 
 \, ,
\end{equation}
for which there is a relation to the Mordell integral analogous to~(\ref{Mordellrel}):
\begin{equation}
\tfrac12 M(\tau)=h_4(\tau)+h_2(-1/\tau)\,.
\end{equation}
Using the functions $h_i$, the localised part of the elliptic genus~(\ref{zorb}) can in particular be rewritten
in three different ways:
\begin{equation}
\Phi_2 (\nu|\tau)
= \left(\frac{\vartheta_i(\nu|\tau)}{\vartheta_i(\tau)}\right)^2 + h_i(\tau)\,\left(\frac{\vartheta_1(\nu|\tau)}{\eta(\tau)}\right)^2\,, \qquad  i=2,3,4
\end{equation}
Then, combining the three above expressions, we can reformulate the regularized elliptic genus $\widehat{\Phi}_2$ as follows:
\begin{equation}\label{ellF}
\widehat{\Phi}_2 (\nu|\tau)
=  \frac{1}{24}\,{\rm ch}_{\text{K}3}(\nu|\tau)\oao{1}{1} + \frac{1}{12}\,\widehat{F}(\tau)\,\frac{\vartheta_1(\nu|\tau)^2}{\eta(\tau)^3}
\end{equation}
in terms of the twisted Ramond character corresponding to the elliptic genus of the $K3$ surface:
\begin{equation}\label{chK3Rt}
{\rm ch}_{\text{K}3}(\nu|\tau) \oao{1}{1}=
 8\left[ \left(\frac{\vartheta_3(\nu|\tau)}{\vartheta_3(\tau)}\right)^2 + \left(\frac{\vartheta_4(\nu|\tau)}{\vartheta_4(\tau)}\right)^2
+\left(\frac{\vartheta_2(\nu|\tau)}{\vartheta_2(\tau)}\right)^2\right]
\end{equation}
This character can for example be determined by \textsc{cft} methods from its $T^4/\zi_2$ orbifold limit~\cite{Eguchi:1988vra}.\footnote{ 
It reproduce in particular $\chi(\text{K3}) = {\rm ch}_{\text{K}3}^{\widetilde{\textsc{r}}}(0|\tau)=24$, as expected.}

The function $\widehat{F}$ that appears in eq.~(\ref{ellF}) is a weak Maa\ss \ form of weight $\nicefrac12$, which decomposes as follows:
\begin{equation} \label{decomF}
\begin{array}{rcl}
{\ds \widehat{F}(\tau)} & = & {\ds F(\tau) - 6 R(0|\tau) } \\
 & = & {\ds 4\,\eta(\tau)\! \sum_{i=2,3,4}\! h_i(\tau) - 
12 \sum_{n=0}^\infty (-)^n\text{erfc}\big((n+\tfrac12)\sqrt{2\pi\tau_2}\big)   q^{-\frac12(n+\frac12)^2} } \,,
\end{array}
\end{equation}
where $\text{erfc}(x)$ is the complementary error function~(\ref{erfcG}).
Its holomorphic part has the following Fourier expansion:
\begin{equation}\label{mockF}
q^{1/8} F(\tau)  = 1-45q-231q^2 -770q^3 -2277q^4 +\mathcal{O}(q^5)\,.
\end{equation}
This Mock modular form $F$ clearly has the same shadow as $12\,\widehat{\mu}(\nu|\tau)$, since its completion $\widehat{F}$
satisfies the holomorphic anomaly differential equation:
\begin{equation}\label{shadFF}
\frac{\p \widehat{F}(\tau)}{\p \bar\tau} = \frac{3\sqrt{2}i}{\sqrt{\tau_2}} \, \overline{\eta(\tau)}^3
\end{equation}

The Fourier coefficients~(\ref{decomF}) actually appear in the Rademacher expansion of the elliptic genus of (non-)~compact $K3$ surfaces~\cite{Eguchi:2008gc}, and were in particularly shown to be relevant to the counting of half BPS states for string theory compactified on such surfaces and hence to a microscopic determination of the black hole entropy for these configurations~\cite{Eguchi:2009cq}. They also found a more recently application  in the derivation of
BPS saturated one-loop amplitudes with external legs stemming from half BPS short multiplets
for type \textsc{ii} string theory compactified on $T^2 \times K3$~\cite{Hohenegger:2011us}. Here we see a novel occurence of the Rademacher expansion 
of $\mathcal{N}_\textsc{ws}=4$ characters, where the Fourier coefficients of $F$ now encode the contribution of three-form flux to 
gauge threshold correction~(\ref{thresh4}).

Then, by spectral flowing expression~(\ref{ellF}), one obtains for the sectors with even spin-structure:
\begin{equation}\label{ref1}
\widehat{\Phi}_2\oao{a}{b}(\nu|\tau)
 = 
\frac{1}{24}\,{\rm ch}_{\text{K}3}(\nu|\tau) \oao{a}{b}+ \frac{1}{12} \widehat{F}(\tau) 
\frac{\vartheta\oao{a}{b}(\nu| \tau)^2}{\eta(\tau)^3}\,,
\end{equation}
with the  $K3$ characters given by expressions:
\begin{equation}\label{k3char}
{\rm ch}_{\text{K}3}(\nu|\tau) \oao{a}{b}  = 
8\left[ (-1)^{a+1} 
\left(\frac{\vartheta\oao{1-a}{1-b} (\nu|\tau)}{\vartheta_3(\tau)}\right)^2 + 
\left(\frac{\vartheta \oao{1-a}{b} (\nu|\tau)}{\vartheta_4(\tau)}\right)^2
+(-1)^{a+1} \left(\frac{\vartheta \oao{a}{1-b} (\nu|\tau)}{\vartheta_2(\tau)}\right)^2\right]
\end{equation}
Using~(\ref{ref1}) the $SO(28)$ threshold correction
for $\mathcal{Q}_5 = 1$~(\ref{thresh4}) can be recast as follows:
\begin{equation}\label{thresh5}
\Lambda_{SO(28)}[1] = {\ds  \frac{1}{48} \int_{\mathscr{F}} \frac{\di^2\tau}{\tau_2} \,\Gamma_{2,2}(T,U) \,\frac{1}{12\eta^{24}}\!
\left(
-\Big[{\widehat{E}}_2 E_{10}-\frac{2}{3}{E}_6^2 -\frac{1}{3} E_4^3\Big]
+ \eta^3 \widehat{F}\big({\widehat{E}}_2  E_8- E_{10} \big)
\right) } 
\end{equation}

We would like to make the following comments on the structure of the threshold correction~(\ref{thresh5}) and its relationship with 
$K3$ characters. 
\begin{itemize}
\item[$i)$]  The first contribution on the RHS of~(\ref{thresh5}), stemming from localised states, reproduces the gauge threshold 
corrections~(\ref{refthresh}) for an $SO(32)$ heterotic compactification on  $T^2 \times (T^4/\zi_2)$, with an orbifold action determined 
by the shift vector $\vec{v}=(1,1,0^{14})$.
Upon blowing up the singularity, one obtains a $T^2 \times K3$ compactification with $SU(2)$ instanton number $t=4$ (see eq.~(\ref{t-ins}) and above),
which we can explicitly read off the first term in~(\ref{thresh5}). The $SU(2)$ background breaks the gauge group symmetry to
$SO(28) \times SU(2)$,~\footnote{In the $T^2\times K3$ model, the gauge group may be further Higgsed and broken down to the terminal group $SO(8)$.} as is also the case for the non-compact model considered here, where however the breaking is due to the presence of $U(1)$ instantons in the background.
The hypermultiplet spectrum for this $T^2\times K3$ compactification is given for instance in~\cite{Aldazabal:1997wi} and reads  $10(\boldsymbol{28},\boldsymbol{2}) + 65(\boldsymbol{1},\boldsymbol{1})$. For the $T^2 \times \widetilde{\textsc{eh}}$ background under scrutiny, 
he hypermultiplet multiplicities are instead
$(\boldsymbol{28},\boldsymbol{2}) + 2(\boldsymbol{1},\boldsymbol{1})$, as given by table~\ref{tabops}. 
This  reduction results, on the on hand, from considering a {\it single} resolved $A_1$ singularity and, on the other hand, from having five-brane flux supported by $U(1)$ gauge instantons threading  the geometry, which is featured in the second term in expression~(\ref{thresh5}). 

\item[$ii)$]   The second contribution on the RHS of~(\ref{thresh5}), originating from both localised and bulk states, is both the sign 
that we are dealing with a non-compact space, and that we are considering a non-K\"ahler geometry with three-form flux, characterised by 
non-zero fivebrane charge $\mathcal{Q}_5$ at infinity; these two aspects are tied together, since the net fivebrane charge on a compact 
manifold has to vanish. The appearance of the Maa\ss \ form $\widehat{F}$, and in particular its decomposition~(\ref{decomF}) into a Mock modular form and its shadow function, can  be understood as follows. The elliptic genus that we computed for the $T^2 \times \widetilde{\textsc{eh}}$ contains a contribution localised on the blown-up $\mathbb{P}^1$, due to the presence of flux;  it is precisely encoded in the Mock modular form $F$~(\ref{mockF}). This expression alone would be anomalous under modular transformations. However the non-compact \textsc{cft} 
at hand displays, alongside localised states,  a continuous spectrum of bulk modes which cancel this holomorphic anomaly, at the price of introducing an extra non-holomorphic contribution in the elliptic genus~(\ref{ellF}). This feature, peculiar to non-compact models, explains  the appearance of the Maa\ss\ form $\widehat{F}= F - 6R$ in the threshold~(\ref{thresh5}) with a contribution of the (transform) shadow function $R$ corresponding to an infinite tower of non-localised massive non-BPS states~(\ref{comp-nl}). In a compact $T^2 \times \widetilde{K3}$ 
model, we instead expect the extra contribution of localised states due to the flux to be cancelled by a contribution from the 
bulk of the globally tadpole-free compactification, without spoiling the holomorphicity of the genus. 

\item[$iii)$] It is also worth rediscussing the polar structure of the  $\widetilde{\textsc{eh}}$ modified elliptic genus~(\ref{thresh5}), since it exhibits some differences with respect to the bulk contribution,
compared to the $k>2$ case discussed previously. But first, we note that in the localised part of the modified elliptic genus the 
$q^{-1}$ pole coming from the $K3$ and the localised flux contibutions exactly compensate, as can be shown from the following Fourier expansion:
\begin{equation}
\begin{array}{rcl}
\hat{\mathcal{A}}_{SO(28)}^{\text{d}}[1] & = & {\ds -\frac{1}{72\eta^{24}} \Big[\widehat{E}_2 E_{10}-\tfrac{2}{3}E_6^2 - \tfrac13 E_4^3
-\eta^3F(\widehat{E}_2 E_8- E_{10})\Big] \,,} \\[12pt]
 & = & {\ds 8 -\frac{29}{\pi  \tau _2}
 + \left(6960-\frac{7955}{\pi  \tau _2}\right) q 
  + \text{O}(q)\,.}
\end{array}
\end{equation} 
Analysing the contribution from bulk states, encoded in the shadow function $R(\tau) D_8 E_8 / \eta^{21}$, is even simpler as for the $k>2$ cases discussed in~(\ref{pole2}). We first consider the sum $q^{1/8}R(\tau)$. The terms of this sum~(\ref{erfcG}) are bounded, for any $n\in \en$ and for $\tau_2  >0$, by:
\begin{equation}\label{boun}
 \left|  (-)^n\text{erfc}\big((n+\tfrac12)\sqrt{2\pi\tau_2}\big)   q^{-\frac12n(n+1)} \right|   \leqslant   \ee^{-2\pi (n+\frac12)^2\tau_2} \Big| \ee^{-i\pi n(n+1)\tau}   \Big| =  \ee^{-\pi \left( n (n+1) +\frac12  \right) \tau_2}     \,.
\end{equation}
All these terms are exponentially supressed for $\tau_2 \rightarrow \infty$. Since $\frac{D_8 E_8}{q^{1/8} \eta^{21}} =  \frac{4}{\pi  q \tau _2}-960+\frac{2004}{\pi  \tau _2}+...$\,, the contribution 
$R(\tau) D_8 E_8 / \eta^{21}$ also has a  'dressed' pole of order one, with an exponentially decaying regulator
characteristic of bulk states, as we already emphasised for the $k>2$ cases. In particular,  for $n=0$, the real part of this pole diverges as 
$\tau_2^{-1} \ee^{\frac32\pi \tau_2}$ for $\tau_2 \rightarrow\infty$, while it is completely suppressed for all terms with $n>0$. Since $\hat{\mathcal{A}}_{SO(28)}^{\text{d}}[1]$ is regular at $\tau_2 \rightarrow \infty$, we observe that $\mathcal{R}^\text{c}[1]$ contains the only 'dressed' pole related to an unphysical tachyon, with a 'dressing' acting as regulator for both the \textsc{ir} divergence stemming from the Casimir $\text{Tr}\,Q_{SO(28)}^2$ and the infinite volume divergence.
\end{itemize}

\subsection{The $U(1)_R$ and $SU(2)$ gauge threshold corrections}
\label{sec:u1}
Having determined the regularised elliptic genera~(\ref{Gencorr}) for warped Eguchi-Hanson \textsc{cft}, we can compute the threshold corrections corresponding to the $U(1)$ or $SU(2)$ gauge coupling depending on whether we consider an arbitrary value $\ell\in 2\en^*+1$ of the Abelian magnetic charge  or the particular value $\ell=1$. This is made easy by the fact that this gauge symmetry corresponds to 
the left $U(1)$ R-symmetry of the $\slr/U(1)$ coset, or has its Cartan generator determined by it in the $\ell=1$ case. Hence the elliptic variable $\nu$ in the genus $\widehat{\Phi}_k\oao{a}{b}(\nu)$ keeps precisely track of its charges; the corresponding (regularised) Casimir operator then acts as a derivative 
with respect to this variable. The Kac--Moody levels, that enter into the regularisation of the Casimir in~(\ref{I}), are in this case:
\begin{equation}
k_{U(1)}=1+\frac2k=1+\frac{1}{\ell^2} \,,\qquad  \qquad  k_{SU(2)} =2 \,.
\end{equation}
The threshold corrections to the $U(1)$ and $SU(2)$ gauge couplings are then given by descendants of the  genera~(\ref{Gencorr}) as:
\begin{equation}\label{genU1}
\Lambda_A[\mathcal{Q}_5]   =  - \frac{1}{32\pi^2} 
\int_{\mathscr{F}}\frac{\di^2\tau}{\tau_2} \,\Gamma_{2,2}(T,U) \!
\sum_{(a,b) \neq (1,1)}  
\frac{{\vartheta\oao{a}{b}(0|\tau)^{14} } }{{\eta(\tau)} }
 \frac{1}{\eta^4}\left[ \d_\nu^2+\frac{\pi(\ell^2+1)}{ \ell^2\tau_2}  \right] 
 {\widehat{\Phi}_k\oao{a}{b} (\nu\big|\tau\big)}
 \Big|_{\nu=0}\,,
\end{equation}
with  $A=U(1)$ for $\ell\in 2\en^* +1$ and $A=SU(2)$ for $\ell=1$.

Working out expression~(\ref{genU1}) explicitly, one obtains for $\ell\in 2\en^* +1$:
\begin{multline}\label{fullthrU1}
\Lambda_{U(1)}[\mathcal{Q}_5] = \frac{1}{4k} \int_{\mathscr{F}}\frac{\di^2\tau}{\tau_2} \, \Gamma_{2,2}(T,U)
\,\sum_{J=1}^{k/2} \big({\chi}_{k-2}^{J-1}+ {\chi}_{k-2}^{k/2-J} \big)  \, q^{-\frac{1}{k}\left(J-\frac{\ell+1}{2}\right)^2}
\times   
 \\[4pt] 
 \times
 \frac{1}{\ell}\sum_{m=0}^{\ell-1}   \ee^{i\pi\frac{m}{\ell}} \hspace{-0.4cm}
  \sum_{(a,b)\neq (1,1)}  (-)^{a+b}  \left[ 
 \frac{ \eta\,{\vartheta\oao{a+1}{0}\left((\frac{\ell+1}{2}-J)\frac{\tau}{\ell}+ \frac{m}{\ell}\big| \tau\right)} \,\,{\vartheta\oao{0}{b+1}\left( (\frac{\ell+1}{2}-J)\frac{\tau}{\ell}+ \frac{m}{\ell}\big| \tau\right)} }{{\vartheta\oao{a}{b}\left( (\frac{\ell+1}{2}-J)\frac{\tau}{\ell}+ \frac{m}{\ell}\big| \tau\right)}^3}
   \,+\right.
 \\[4pt]
 +
\Big(\frac{k+2}{24}\Big)\,
\frac{ {\widehat{\mu}
\Big(\big(\tfrac{\ell a+1}{2}-J \big) \tfrac{\tau}{\ell} +\tfrac{b-1}{2}+\tfrac{m}{\ell},\tfrac{(a-1)\tau +(b-1)}{2}\Big|\tau \Big)}}{\bar \eta}
\times
\\[4pt]
\times \left.
\frac{({\widehat{E}}_2+(-)^b \vartheta\oao{a+b+1}{a}^4 - (-)^a \vartheta\oao{b}{a+b+1}^4)}{ {\eta}^{4} } \right] 
\frac{\vartheta\oao{a}{b}^{16} }{\eta^{16}}\,.
\end{multline}
In particular, the second line of the above expression comes from the second derivative~\footnote{Note that in~(\ref{genU1}) all (mixed) terms containing simple derivatives of theta functions vanish since $\partial_\nu\vartheta_{2,3,4}(0|\tau) =0$.} 
$$\d_{\nu}^2 \, \mu\,
\left(\frac{\nu}{\ell}+\big(\tfrac{\ell a+1}{2}-J \big) \tfrac{\tau}{\ell} +\tfrac{b-1}{2}+\tfrac{m}{\ell},\frac{\nu}{\ell}+\tfrac{(a-1)\tau +(b-1)}{2}\big|\tau \right)\Big|_{\nu=0}\, ,$$ which is computed in appendix~\ref{NLT2}. Moreover, the contribution from bulk states is entirely captured in the last two lines of~(\ref{fullthrU1}) and is given by:
\begin{multline}
k_{U(1)}\, \mathcal{R}^{\text{c}}[\mathcal{Q}_5]=  \Big(\frac{k+2}{8k}\Big) \sum_{J=1}^{k/2} \big(\chi_{k-2}^{J-1}+ \chi_{k-2}^{k/2-J} \big) \sum_{n=0}^\infty (-)^n
 \text{erfc}\!\left( \tfrac{1}{2\ell} (n k  + 2J - 1) \sqrt{2\pi \tau_2} \right)  \times
 \\
\times \,  q^{-\frac{1}{4k}( n k  + 2J - 1)^2}\,\frac{D_8 E_8}{\eta^{21}} 
\end{multline}
where $\mathcal{R}_k^{\text{c}}$ is non-holomorphic completion~(\ref{nlocSO28}) also appearing in the $\Lambda_{SO(28)}[\mathcal{Q}_5]$ threshold. This is in accordance with the general form taken by the modified elliptic genera that we outlined in~(\ref{Agross}).
In particular, it shows that for this class of models the contribution to the gauge threshold corrections coming from bulk states is independent of the gauge group and only depends on the five-brane charge.

\subsubsection*{The $\mathcal{Q}_5=1$ case and $\mathcal{N}=4$ symmetry} 
As previously, we consider in more detail the $\mathcal{Q}_4=1$ case, which has enhanced $\mathcal{N}_\textsc{ws} = 4$ left superconformal symmetry
and second gauge factor $SU(2)$. By using expressions~(\ref{ref1}), equation~(\ref{genU1}) yields in this case
\begin{equation}\label{thresh6}
\begin{array}{rcl}
\Lambda_{SU(2)}[1] & = & {\ds  \frac{1}{48} \int_{\mathscr{F}} \frac{\di^2\tau}{\tau_2} \,\Gamma_{2,2}(T,U) \,\frac{1}{6\eta^{24}}\!
\left(
-\Big[{\widehat{E}}_2 E_{10} +\frac{4}{3} {E}_6^2-\frac{7}{3} E_4^3\Big]
+  \eta^3{\widehat{F}}\big({\widehat{E}}_2  E_8- E_{10} \big)
\right) }  \,.
\end{array}
\end{equation}
Again, the first term in~(\ref{thresh6}) is the gauge threshold correction~(\ref{refthresh}) for a $T^2\times K3$ 
compactification, with this time $t=-44$.

\subsubsection*{Universality properties}

By comparing $\Lambda_{SO(28)}[1]$ and  $\Lambda_{SU(2)} [1]$, given in eqs.~(\ref{thresh5}) and~(\ref{thresh6}),  
we observed that these threshold corrections  satisfy some universality properties when the underlying \textsc{cft}  exhibits enhanced 
$\mathcal{N}_{\textsc{ws}} = 4$ left super-conformal symmetry . 

More generically, we would like to consider $\mathcal{N}_{\textsc{ts}}=2$ six-dimensional local models with non-zero five-brane, based on $T^2$ times a smooth geometry corresponding to the warped resolution of a $\ci^2/\mathbb{G}$ singularity, the action $\mathbb{G}$ leaving the gauge group $\prod_{a} \mathrm{G}_a$ unbroken. If their \textsc{cft} description displays
$\mathcal{N}_{\textsc{ws}} = 4$ left super-conformal symmetry and allows for non-localised bulk states, we propose that such
theories have threshold corrections $\Lambda_a$~(\ref{decLam}) to the couplings of the various gauge factors $\mathrm{G}_a$ determined by
the four-dimension modified elliptic genus:
\begin{equation}\label{univDD}
\hat{\mathcal{A}}_a = 
\frac{k_a}{6}
\left( 6(2-t_a)  + \frac{1}{20\eta^{24}} \left[
{D_{10} E_{10}} -528\eta^{24}
+ c_r g\widehat{F} D_8E_8 \right]
\right)
\end{equation}
where $\widehat{F}$ is a Maa\ss\, form of weight $r$, $g$ its shadow of weight $2-r$, and $c_r$ is a weight dependent constant. 
In the particular warped $\ci^2/\zi_2$ resolution
considered until now, $\widehat{F}$ is the weight $\nicefrac12$ Maa\ss\, form~(\ref{decomF}) with shadow function $g= -\frac{1}{2\sqrt{2}}\eta^3$,
which is a weight $\nicefrac32$ holomorphic Jacobi form, and $c_{\nicefrac12} =5\sqrt{2} $. In this particular case, expression~(\ref{univDD}) yields an alternative formulation of expressions~(\ref{thresh5}) and~(\ref{thresh6}). 
We thus observe that the three first terms in~(\ref{univDD}) reproduce $\nicefrac{4}{\chi(K3)}$ of the $K3$ modified elliptic genus, see~(\ref{Lam-univ})--(\ref{doub}), with $\beta$-functions $b_a= 3k_a(2-t_a)$. Mind that these are not the full $\beta$-functions for the torsional local models under consideration, which receive an additional contribution from the constant part of the  flux induced term $\widehat{F} D_8E_8/\eta^{24}$. Also the normalisation factor
$\nicefrac{4}{\chi(K3)}$ comes from considering a {\it single} resolved $A_1$ singularity, instead of the global $K3$ geometry. It is in keep with
hypermultiplet counting for a double-scaled geometry as currently investigated in~\cite{Grootwp}

For a general local model with five-brane charge which satisfies the above conditions, the four-dimension modified elliptic genus is  then completely determined by a certain linear combination of three modular forms of weight 12: 
the quasi-holomorphic modular form $D_{10}E_{10}$, the cusp form $\eta^{24}$ and the non-holomorphic modular form $g\widehat{F}D_8E_8$, where $\hat{F}$ is the weak Maa\ss \ form capturing the effects due to \textsc{nsns} three-form flux, $g$ is its shadow function and $D_8E_8$ is a universal contribution. 
The coefficients of this linear combination are fixed by the absence of charged tachyons in the spectrum and the 
tadpole equation~(\ref{tad1}) with now non-vanishing charge
$\mathcal{Q}_5$. 
Consequently, the difference of two such gauge thresholds satisfies the relation 
\begin{equation}\label{diffthresh}
 \frac{\Lambda_a[\mathcal{Q}_5]}{k_a} -  \frac{\Lambda_b[\mathcal{Q}_5]}{k_b}  =
  \frac{3(t_b-t_a)}{24} \int_{\mathscr{F}}\frac{\di^2\tau}{\tau_2} \, \Gamma_{2,2}(T,U)\,,
\end{equation}
which, interestingly enough, is $\nicefrac{4}{\chi(K3)}$ times 
what is expected for gauge threshold corrections for $T^2\times K3$ models~(\ref{Lam-univ}), which generically satisfy the relations~(\ref{relN2}). The fact that this universal feature of $\mathcal{N}_\textsc{st}=2$ heterotic compactifications carries over to the local non-compact models under consideration is clearly ascribable in this case to their displaying enhanced $\mathcal{N}_\textsc{ws}=4$ left-moving superconformal symmetry, as for $T^2\times K3$ compactifications with the standard embedding.


In particular, formula~(\ref{diffthresh}) holds for the $T^2\times \widetilde{\textsc{eh}}$ background with $\mathcal{Q}_5=1$. In this case,
the difference between  the $\Lambda_{SU(2)}[1]$ and  $\Lambda_{SO(28)} [1]$ thresholds yields a factor $6$ multiplying the integral on the RHS of eq.~(\ref{diffthresh}), see footnote below.

In the higher $\mathcal{Q}_5 >1$ cases, the underlying \textsc{cft} only has $\mathcal{N}_\textsc{ws}=2$ left-moving superconformal symmetry, hence the difference between the $U(1)$ and $SO(28)$ threshold corrections is more complicated~\footnote{The $\mathcal{Q}_5=1$ case~(\ref{diffthresh}) with $\frac{3(t_{SO(28)}- t_{SU(2)} )}{24} = 6$  can be recovered from expression~(\ref{diffthresh2}) by setting $\ell=1$, replacing the $SU(2)_{k-2}$ contributions by the identity and using the identity for $\vartheta$-functions~(\ref{kounnas}).}:
\begin{multline}\label{diffthresh2}
 \frac{\Lambda_{U(1)}[\mathcal{Q}_5]}{k_{R}} -  \frac{\Lambda_{SO(28)}[\mathcal{Q}_5]}{k_{SO(28)}} 
 =  \frac{1}{4(k+2)} \int_{\mathscr{F}}\frac{\di^2\tau}{\tau_2} \, \Gamma_{2,2}(T,U)
\,\sum_{J=1}^{k/2} \big({\chi}_{k-2}^{J-1}+ {\chi}_{k-2}^{k/2-J} \big)  \, 
q^{-\frac{1}{k}\left(J-\frac{\ell+1}{2}\right)^2}
 \times \\[4pt]
\times
 \frac{1}{\ell}\sum_{m=0}^{\ell-1}   \ee^{i\pi\frac{m}{\ell}} \hspace{-0.4cm}
  \sum_{(a,b)\neq (1,1)}  (-)^{a+b} \,
 \frac{{\vartheta\oao{a+1}{0}\left( (\frac{\ell+1}{2}-J)\frac{\tau}{\ell} + \frac{m}{\ell}\big| \tau\right)} \,\,{\vartheta\oao{0}{b+1}\left( (\frac{\ell+1}{2}-J)\frac{\tau}{\ell}+ \frac{m}{\ell}\big| \tau\right)} }{{\vartheta\oao{a}{b}\left( (\frac{\ell+1}{2}-J)\frac{\tau}{\ell}+ \frac{m}{\ell}\big| \tau\right)}^3}
 \frac{{\vartheta}\oao{a}{b}^{16}}{\eta^{15}} \,,
\end{multline}
and in particular does not abide by the rule~(\ref{relN2}) characterizing toroidal orbifold compactifications,
as $SU(2)_{k-2}$ right-moving characters now intermingle with characters of a 
compact $\mathcal{N}_\textsc{ws}=2$ \textsc{cft} with $c=1+\frac k2$.
Nonetheless,  since the contribution of bulk states is, up to a multiplicative Kac--Moody level, gauge group independent, the difference of thresholds~(\ref{diffthresh2}) shares the common feature with the $\mathcal{Q}_5=1$ case 
of being a purely localised effect, and could thus in principle be compared
with corresponding expressions  for $\mathcal{N}_\textsc{st}=2$ heterotic compactifications.

\section{The moduli dependence}
\label{sec:moddepgen}

In order to determine the explicit dependence of the $SO(28)$ and $SU(2)$ or $U(1)$ threshold corrections on the $T^2$ moduli,
we have to carry out the integrals~(\ref{fullthr}) and~(\ref{fullthrU1}) or~(\ref{thresh6}) over of the fundamental domain $\mathscr{F}$ of 
the modular group. Since both integrands $\tau_2\Gamma_{2,2}(T,U)\hat{\mathcal{A}}_a$ are invariant under the full modular group $\Gamma$,
we are entitled to compute these integrals by unfolding the $T^2$ lattice sum, a method pioneered by Dixon-Kaplunovsky-Louis (DKL)
to evaluate threshold corrections for heterotic $\mathcal{N}_\textsc{st}=2$ compactifications~\cite{Dixon:1990pc}.\footnote{This
method actually goes back to the study of string thermodynamics~\cite{McClain:1986id,Ditsas:1988pm,Kutasov:1990sv}.}

More recently, an alternative method~\cite{Angelantonj:2011br,Angelantonj:2012gw} has
been developed to evaluate these integrals, which keeps manifest the T-duality invariance of the result
under the $O(2,2;\zi)$ group of the Narain lattice. Generalising an idea developped
in~\cite{Cardella:2008nz, Cardella:2010bq, Angelantonj:2010ic} which proposes to  unfold the integral domain against the (modified) elliptic genus rather than against the torus lattice sum, these authors have shown how this procedure could be extended
to any BPS-saturated amplitudes in string theory compactifications of the form $\int_\mathscr{F} \frac{\di\tau^2}{\tau_2^2}\,\tau_2^{d/2} \Gamma_{d+k,d}(G,B,Y)\hat{\mathcal{A}}$, by rephrasing $\hat{\mathcal{A}}$ in terms of a certain class of non-analytic Poincar\'e-type series.

Given that $\hat{\mathcal{A}}$  generically  includes non-holomorphic terms such as $(\widehat{E}_2)^g\Phi_{12-g}$, with $\Phi_{12-g}$ a combination of products of holomorphic Eisenstein series of weight $12-g$, and may exhibit poles in $q$ 
related to unphysical tachyons,  the authors of~\cite{Angelantonj:2012gw} have shown that all modified elliptic genera of
interest can be appropriately rewritten as a linear combination of Niebur--Poincar\'e series~\cite{nieb,hej} 
$\mathcal{F}(s, \alpha, r)$, with $\text{Re}(s)>1$ lying within the radius of absolute
convergence of the series. Considering at first a genus $\mathcal{A}$ which can be any weakly holomophic  modular form,
these authors have shown that by specialising to Niebur-Poincar\'e series with $s=1-\frac r2$ and $r<0$ and taking
a suitable linear combination of those series whose coefficients are determined by the principal part of 
$\mathcal{A}$, one can reproduce $\mathcal{A}$ exactly, even though  
$\mathcal{F}(1-\frac r2, \alpha, r)$ taken individually are generically weak harmonic Maa\ss\ forms. 
This analysis extends to genera $\hat{\mathcal{A}}$ which are weak  {\it almost} holomorphic modular form, 
by consider a combination of Niebur-Poincar\'e series with $s=1-\frac r2 + n$ and $n\in \en$.
This applies in particular to gauge threshold corrections for heterotic  $\mathcal{N}_\textsc{st} =2$ compactifications~(\ref{doub}) discussed previously. Then absolute convergence of the Niebur-Poincar\'e series for these specific values of its weight $r$ allows to properly
unfold  them  against integration domain $\mathscr{F}$  and compute the integral in a way that
keeps manifest the $O(k+d,d;\zi)$ invariance inherited from the Narain lattice. 

Since for  $s=1-\frac{r}{2}$ with $r<0$ in particular  the Niebur--Poincar\'e series belong to the space 
$\widehat{\mathbb{M}}_r$  of weak harmonic Maa\ss\ forms~(\ref{whm}), we can in principle consider  rephrasing
the four-dimensional genera for warped Eguchi-Hanson, eq.~(\ref{fullthr}) and~(\ref{fullthrU1}), by use of this method. However  we will prefer evaluating these integrals by the traditional 'orbit method' of DKL, since the result is more directly interpretable, for large $T_2$, in terms of perturbative and Euclidean brane instanton corrections in the type \textsc{i} S-dual theory. The procedure~\cite{Angelantonj:2011br,Angelantonj:2012gw} albeit yielding a compact and elegant result for string amplitudes, is less suited to study this corner of the moduli space.

Nevertheless, the non-compactness of the heterotic background we are considering, manifested in 
the contribution~(\ref{nlocSO28}) of non-localised bulk states to the gauge threshold corrections, will entail novel results for 
these integrals for each class of orbits of the modular group. For the zero orbit piece, in particular we will even have to resort to
results established in~\cite{Angelantonj:2012gw} by the procedure we elaborated on above, in order to exactly determine the flux contribution to the tree level correction to the heterotic gauge couplings.

In the following, we will restrict ourselves to working out explicitly the gauge  threshold corrections~(\ref{thresh5}) and~(\ref{thresh6}) in the model with $\mathcal{Q}_5=1$ unit of flux. The dependence of the resulting threshold corrections on the $(T,U)$ moduli of the two-torus will be qualitatively the same as in the general $\mathcal{Q}_5=k/2 > 1$ case. At first sight a discrepancy might arise as one considers the seemingly more involved structure of the bulk modes contribution in the general $\mathcal{Q}_5 > 1$ case.  However, by comparing expressions~(\ref{nlocSO28}) and~(\ref{decomF}), it appears that despite a mixing with $SU(2)_{k-2}$ characters the contribution of continuous $\slr_k/U(1)$ representations is very similar to the non-holomorphic completion~(\ref{decomF}) for the $\mathcal{Q}_5 = 1$ case, with a sum shifted by the $SU(2)_{k-2}$ spin.

\subsection{The orbit method}
The orbit method allows to compute integrals over the fundamental domain $\mathscr{F}$ by trading the sum over
the winding modes in the $T^2$ partition function~(\ref{latt}) for an unfolding of $\mathscr{F}$.

Following DKL~\cite{Dixon:1990pc}, we decompose the set of matrices $A$ in the $T^2$ lattice sum~(\ref{latt}), encoding  
the maps from the worldsheet to the target space, into orbits
of the modular group $\Gamma \cong PSL(2,\zi)$, characterised as follows:
\begin{enumerate}
\item[i)] \underline{Invariant \it{or} zero orbit:}
\begin{equation}\label{A0}
A= \begin{pmatrix} 0 & 0 \\ 0 & 0\end{pmatrix} \,.
\end{equation}

\item[ii)] \underline{Degenerate orbits:} $\text{det}\,A=0$ and $A\neq 0$, parametrised by:
\begin{equation}\label{Adeg}
A= \begin{pmatrix} 0 & j \\ 0 & p\end{pmatrix} \,
\end{equation}

with $(j,p)\sim (-j,-p)$ and $AV=AV'$ iff $V=T^nV'$, for some $n\in \en$ and $V,\,V'\in \Gamma$.

\item[iii)] \underline{Non-degenerate orbits:} $\text{det}\,A \neq 0$:
\begin{equation}\label{Anondeg}
A= \begin{pmatrix} k & j \\ 0 & p\end{pmatrix} \,
\end{equation}
with $d>j\geq 0$, $p\neq 0$ and $AV=AV'$, for $V,\,V'\in \Gamma$.
\end{enumerate}

Since distinct elements of degenerate and non-degenerate orbits are in one-to-one correspondence 
with modular transformations mapping  the $PSL(2,\mathbb{Z})$ fundamental domain $\mathscr{F}$ inside, respectively, the strip  
$\mathscr{S} =\left\{\tau \in \mathscr{H}\;| \, -\frac12 \leqslant \tau_1 < \frac12,\, \tau_2 \geqslant 0\right\}$, 
and the double cover of the upper half-plane $\mathscr{H}$, 
the gauge threshold corrections~(\ref{decLam}) can be expressed as follows:
\begin{equation}\label{intgr}
\begin{array}{rcl}
\Lambda_a & = &  \Lambda_a^0 + \Lambda_{a}^{\text{deg}} + \Lambda_a^{\text{non-deg}} \\[4pt]
        & = & {\ds  \frac{T_2}{8} \left[ 
        \int_{\mathscr{F}} \frac{\di^2\tau}{\tau_2^2}\hat{\mathcal{A}}_a +
        \int_\mathscr{S} \frac{\di^2\tau}{\tau_2^2}\sum_{(j,p)\neq (0,0)}
        e^{-\frac{\pi T_2}{\tau_2 U_2}|j+pU|^2}\hat{\mathcal{A}}_a   \right.
         } 
        \\[6pt] & &  \qquad 
        +\, {\ds  2   \left.    \int_{\mathscr{H}} \frac{\di^2\tau}{\tau_2^2}\sum_{k>j\geq 0}\sum_{p\neq 0}
        e^{2\pi k pi T-\frac{\pi T_2}{\tau_2 U_2}|k\tau-j-pU|^2}
        \hat{\mathcal{A}}_a
         \right]  \,.}
\end{array}
\end{equation}
If the modified elliptic genus $\hat{\mathcal{A}}_a$ exhibits a $q^{-1}$ pole, which is typically the case for expressions~(\ref{thresh5}), (\ref{thresh6}),
(\ref{fullthr}) and~(\ref{fullthrU1}) as was pointed out in~(\ref{bound}) and~(\ref{boun}), the unfolding procedure~(\ref{intgr}) is subject to a caveat.
When such a pole is present, convergence of  the original threshold integral dictates a prescription for its evaluation, namely that we integrate first
over $\tau_1$, discarding all Fourier modes of $\hat{\mathcal{A}}_a$ except the zero modes, and only then over $\tau_2$. In general, the modular transformations $\gamma_i$  that bring the matrix $A$ into the forms~(\ref{Adeg}) and~(\ref{Anondeg}) characteristic of
degenerate and non-degenerate orbits translate the latter into a highly a complicated, $\gamma_i$ dependent prescription for the integration domains of the unfolded threshold integral, which usually invalidates the decomposition~(\ref{intgr}). Then, when  unphysical tachyons are present in $\hat{\mathcal{A}}_a$ the identity~(\ref{intgr}) only holds when the integral over $\mathscr{F}$ on the LHS is independent of the integration order, which is
the case whenever the integration of the $(n_1,n_2) \neq (0,0)$ terms in the Lagrangian lattice sum~(\ref{latt}) is absolutely convergent. 
If $\hat{\mathcal{A}}_a$ contains a $q^{-1}$ pole, this is the case when $T_2 > 1$, so that expression~(\ref{intgr}) is only valid in this regime.

\subsection{Moduli dependence of the $SO(28)$ threshold corrections} 
\label{sec:moddep}

We give hereafter the threshold corrections to the $SO(28)$ gauge coupling for the model~(\ref{thresh5}) with $\mathcal{Q}_5=1$. 
The details of the evaluation of the integrals corresponding to the three classes of orbits of $\Gamma$ are given  in Appendix~\ref{Th}. 
We will nontheless discuss later on some salient features of how the moduli dependence of the flux contributions can be established,
as it is an interesting novel result. The following expression is valid in the region $T_2> 1$ as discussed before\footnote{Note however that the continuous state contributions in the last two lines of~(\ref{inthresh}) only take this form in the large volume limit of the $T^2$, while for finite
volume these expressions are more involved, as we will see later on.}:
\begin{equation}\label{inthresh}
\begin{array}{rcl}
\Lambda_{SO(28)}[1] & = & {\ds  \frac{29\pi}{144} \,T_2   -  \big(  \log|\eta(U)|^4 + \log(\mu^2\,T_2 U_2) \,+ \gamma \big)
- \frac{29\pi}{360} \frac{E(U,2)}{T_2}  +} \\[10pt]
  & &  {\ds - \frac{1}{8} \sum_{n=0}^\infty   (-)^n \left[  \frac{\pi}{18}d_1\!\big(\tfrac{n(n+1)}{2}\big) \, \mathcal{E}_{(n+\frac12)\sqrt{2T_2}} (U,1) \,-
 \phantom{ \frac{\mathcal{E}_{(n+\frac12) \sqrt{2T_2}} (U,2) }{T_2} } \right.  }
     \\[12pt]
   &&{\ds \left. \quad- \frac{\sqrt{2} \zeta(3)}{\pi^2} \big(n+\tfrac12 \big)d_2\!\big(\tfrac{n(n+1)}{2}\big)  \, 
   \frac{\mathcal{E}_{(n+\frac12) \sqrt{2T_2}} (U,3/2) }{\sqrt{T_2}} 
     -  \frac{\pi}{90}d_2\!\big(\tfrac{n(n+1)}{2}\big) \, \frac{\mathcal{E}_{(n+\frac12) \sqrt{2T_2}} (U,2) }{T_2}  \right] 
   } 
    \\[12pt]
     && {\ds  + \frac{1}{4} \left( \sum_{k>j\geqslant 0} \sum_{p>0} \frac{1}{kp}\,
     \ee^{2\pi i\mathcal{T}}
      \left[ 
     \hat{\mathcal{A}}_{SO(28)}(\mathcal{U})+\frac{\hat{\mathcal{A}}_K(\mathcal{U})}{\mathcal{T}_2}  \right. \right.}
     \\[12pt] &&
       {\ds \left. \left. \qquad \qquad+ \sum_{r=0}^\infty  \frac{1}{(r+2)!}\frac{1}{(\mathcal{T}_2)^{r+2} } \, 
   (-iD)^{r+2} (\mathcal{U}_2^2 \bar\partial_{\mathcal{U}})^r 
     \hat{\mathcal{A}}_H(\mathcal{U})  
       \right] + \, \text{c.c.} \right) \,.
     }
\end{array}
\end{equation}
In the following, we will discuss the physical implication of the various terms appearing in the above result, after giving proper
definitions of the expressions entering into it.

\subsubsection*{The zero orbit contribution}

We first observe that in accordance with the double-scaling limit~(\ref{DSL}), the expression~(\ref{inthresh}) does not depend on the 
blow-up modulus $a$, which can be rescaled away in the near-horizon geometry~(\ref{DSgeom}). We are however aware of the possibility for worldsheet instantons to wrap the blown-up $\mathbb{P}^1$ and to contribute accordingly to the threshold correction~(\ref{inthresh}). As we will show, these terms can actually be found in the zero orbit contribution to $\Lambda_{SO(28)}[1]$, {\it i.e.} the first expression on the first line of~(\ref{inthresh}) proportional to $T_2$. 

To identify these worldsheet instanton contributions,
it is more handy to reason in terms of the type \textsc{i} S-dual theory on $T^2 \times \widetilde{\textsc{eh}}$, 
which has space-time filling D9 branes supporting the unbroken gauge group. 
When the singularities in the background geometry are resolved, the gauge kinetic functions of the gauge factors receive a 
tree-level (disk) contribution of the type~\cite{Camara:2008zk}:
\begin{equation}
 \sim \sum_{\text{two-cycles}} \sqrt{\text{det}(P[g+ \mathcal{F}])}\, T\,,
\end{equation}
with $P[...]$ the pull-back to the blown-up two-cycles. For smooth $K3$ models, such contributions to the gauge kinetic functions
typically arise from $SU(2)$ gauge instantons attached to the blown-up $\mathbb{P}^1$'s, see discussion following eq.~(\ref{t-ins}).
Here, in contrast, we have $U(1)$ gauge instantons~(\ref{Feq})  instead of non-Abelian ones, living on the unique two-cycle of the warped Eguchi--Hanson space. 
In the blow-down limit, these typically give rise to small Abelian instantons sitting at the singularities, a phenomenon which also occurs at the orbifold fixed points of the singular limit of  Bianchi-Sagnotti-Gimon-Polchinski models~\cite{bs,gp}. This indicates that the corrections due to worldsheet instantons wrapping the blown-up $\mathbb{P}^1$ we are looking for are summed up in the $\frac{\pi}{64}$ coefficient of the zero orbit contribution in expression~(\ref{inthresh}). If we were to determine this constant for the theory~(\ref{fullthr}) at arbitrary five-brane charge $\mathcal{Q}_5$ we would see these instanton contribution appear explicitly as $\ee^{-kn}$ corrections.

To conclude the discussion on the zero orbit piece,
let us remark on some technicalities in the determination of the contribution related to the flux, 
{\it i.e.} from the second term on the RHS of eq.~(\ref{thresh5}). From appendix~\ref{Th1}, this part of the zero orbit contribution,
which is separately modular invariant, reads:
\begin{equation}
\Lambda^0_{\text{flux}} =   \frac{T_2}{(24)^2}\int_{\mathscr{F}} \frac{\di^2\tau}{\tau_2^2}\,  \frac{\widehat{F}\big(\widehat{E}_2 E_8 - E_{10}\big)}{\eta^{21}} \,.
\end{equation}
Using Stokes' theorem~(\ref{Jint}) for modular integrals over $\mathscr{F}$, then 
integrating by parts and  remembering that the weight $\nicefrac12$ Maa\ss \ form $\widehat{F}$~(\ref{decomF}) has shadow
$g=-3\sqrt{2} \eta^3$, see~(\ref{shad}), we may rewrite it~\footnote{We are very grateful to J.~Manschot for invaluable help tackling this integral.}, according to~(\ref{outl}):
\begin{equation}\label{int0orb}
\begin{array}{rcl}
{\ds \Lambda^0_{\text{flux}} }
& = &  {\ds
 -\frac{\pi i\, T_2}{3(24)^2} \int_{\mathscr{F}}  \di^2\tau\, \hat F \,\frac{\d}{\d\bar\tau} \Big( \frac{\widehat{E}_2(\widehat{E}_2 E_8 -2 E_{10})}{\eta^{21}} \Big) }\\[12pt]
 & =&{\ds   \frac{\pi\, T_2}{(24)^2} \left( 172 -
  \sqrt{2} \int_{\mathscr{F}}  \frac{\di^2\tau}{\tau_2^2} \, \big(\sqrt{\tau_2} \eta\bar\eta \big)^3  \,\Big( \frac{\widehat{E}_2(\widehat{E}_2 E_8 -2 E_{10})}{\eta^{24}} \Big)
  \right) \,.}
\end{array}
\end{equation}
The first  term on the last line of~(\ref{int0orb}) comes from evaluating the integral~(\ref{I1}) using the standard formula~(\ref{formIr}). The second
integral inside the parenthesis, called $I''_\text{flux}/\pi$~in the appendix eq.~(\ref{I2}), is more involved. Using Poisson resummation, we can reexpress:
\begin{equation}\label{poissresum}
\begin{array}{rcl}
{\ds \big(\sqrt{\tau_2} \eta\bar\eta \big)^3 } & = & {\ds \frac{1}{4\sqrt{2}} \sum_{m,n} (-)^{(m+1)(n+1)} 
\frac{|m +n \tau|^2}{\tau_2} \,  \mathrm{e}^{-\frac{\pi}{2\tau_2} |m+n\tau|^2}  }\\[14pt]
  & = & {\ds -\frac{1}{8\pi} \frac{\partial}{\partial R}\! \left.\left( \frac 1R\, \Gamma_{1,1}(2R) -\Gamma_{1,1}(R) \right) \right|_{R=\frac{1}{\sqrt{2}}}}\,.
\end{array}
\end{equation}
As the rest of the integrand on the last line of expression~(\ref{int0orb}) exhibits a $\frac 1 q$ pole coming from the cusp form $\eta^{24}$ and since the
the radius of the second $\Gamma_{1,1}$ in~(\ref{poissresum}) is fixed at $R=\frac{1}{\sqrt{2}}$, the integration of the $n\neq 0$
terms in the Lagrangian lattice sum~(\ref{poissresum}) are not absolutely convergent, so that one cannot unfold the integral against it.
This is were the novel approach to modular integrals developed in~\cite{Angelantonj:2011br,Angelantonj:2012gw} comes into play: 
by considering the Niebur--Poincar\'e series~\cite{nieb,hej}~\footnote{More recent works on the subject can be found in~\cite{bruinier1,bring,on,bruinier2}.}
\begin{equation}
\mathcal{F}(s,\alpha,r) =  \frac12 \sum_{(c,d)=1} \frac{1}{(c\tau +d)^r} \,\mathcal{M}_{s,r} \!\! \left(\frac{-\alpha \tau_2}{|c\tau + d|^2} \right)
\ee^{\frac a c - \frac{c\tau_1 + d}{c|c\tau + d|^2}}
\end{equation}
which for  the weight values $r \leqslant 0$ of interest are defined by:
\begin{equation}
\mathcal{M}_{s,r} (- y) =  (4\pi y)^{s-\frac r2} \ee^{-2\pi y}  \left._1F_1\right.\!\! \left(s+\tfrac r2; 2s; 4\pi y \right)\,,
\end{equation}
we may rewrite the second part of the integrand in~(\ref{int0orb})  as the linear combination~\cite{Angelantonj:2012gw}:
\begin{equation}\label{Fcomb}
 \frac{\widehat{E}_2(\widehat{E}_2 E_8 -2 E_{10}) }{ \eta^{24} }  = 
 \frac15 \,\mathcal{F}(3,1,0) - 6\,\mathcal{F}(2,1,0)+23\,\mathcal{F}(1,1,0)+432\,.
\end{equation}
By reexpressing the integral  $I''_{\text{flux}}$ in terms of expressions~(\ref{poissresum}) and~(\ref{Fcomb}), and by unfolding the fundamental
domain against the absolutely convergent Niebur--Poincar\'e series in~(\ref{Fcomb}), we obtain, from eq.(4.18) in~\cite{Angelantonj:2012gw} the result
$I''_\text{flux} =40$.
Putting the two pieces of eq.~(\ref{int0orb}) together, we thus arrive at~(\ref{lam0flux}):
\begin{equation}
\Lambda^0_{\text{flux}}= \frac{53\pi}{144}\, T_2 \,.
\end{equation}
Adding up the flux contribution and  the $\nicefrac{1}{6}$ $K3$ contribution appearing in~(\ref{thresh5}), yields the
zero orbit piece~(\ref{fflux}) on the first line of the RHS of expression~(\ref{inthresh}). We refer the reader to the detailed calculation in~(\ref{lam0K3}).

\subsubsection*{Contributions from degenerate orbits}
The first, second and third lines of the RHS of expression~(\ref{inthresh}) capture the contributions from degenerate orbits of $\Gamma$,
determined in appendix~\ref{sec:deg}. These are expressed in terms of generalised Eisenstein series~\footnote{In another context, a close relative of these generalised Eisenstein series with $A\neq 0$ and for $s=\nicefrac 32$ appears in~\cite{RoblesLlana:2006is,Alexandrov:2008gh}, where it captures corrections to the metric of the hypermultiplet  moduli space of type \textsc{iib} compatifications on \textsc{cy} threefolds  due to E(-1) instantons, and to E1 instantons wrapping two-cycles in the geometry. Here however the interpretation is different as we will see.}:
\begin{equation}\label{genEin}
\mathcal{E}_A (U,s) = \frac{1}{2\zeta(2s)} \sum_{(j,p)\neq (0,0)} \frac{U_2^s}{|j+pU|^{2s}} \,\ee^{-2\pi A \frac{|j+p U|}{\sqrt{U_2}}}\,, \qquad \text{Re}\,A \in R_+\,,
\end{equation}
which reduce to the well-known  real analytic Eisenstein series for $A=0$ (see eq.~(\ref{EE})):
\begin{equation}\label{Eisn}
\mathcal{E}_0 (U,s) \equiv E(U,s) \,.
\end{equation}
In particular, upon
patching together 16 local models such as~(\ref{metric2}) into a full-fledged heterotic compactification with flux,
the sums over $(j,p)$ in the (generalised) Eisenstein series~(\ref{genEin}) are expected, like for smooth $T^2\times K3$ models, to  correctly reproduce the double sum over Kaluza--Klein momenta in the open-string channel of the corresponding type  \textsc{i} compactification~\cite{Bachas:1996bp}. 

To be more specific, the second and third term on the first line of~(\ref{inthresh}) are the degenerate orbit contribution 
coming from states localising on the blown-up $\mathbb{P}^1$ which, in accordance with results for compact heterotic
models, are expressed in terms of the real analytic Eisenstein series~(\ref{Eisn}).
In particular, the first contribution proportional to $E(U,1)$ exhibits a logarithmic \textsc{ir}  divergence which can be regularised
as follows:
\begin{equation}\label{E1}
E(U,1) = -\tfrac{3}{\pi} \left(\log|\eta(U)|^4  + \log\big( \mu^2\, T_2\, U_2 \big) \right) + \gamma
\end{equation}
with $\gamma$ a renormalisation scheme depend constant. In particular, if we adopt the regularisation~\cite{Dixon:1990pc}, which introduces
a $\big(1-\ee^{-\frac{N}{\tau_2}}\big)$ regulator in the integrand of~(\ref{EE}) for $r=1$,
and eliminate the \textsc{ir} cutoff by sending $N \rightarrow \infty$, which corresponds to sending $\mu \rightarrow 1$ in expression~(\ref{E1}),
we get:
\begin{equation}
\gamma = 1+ \log \frac{8\pi}{3\sqrt{3}} -\gamma_E\,,
\end{equation}
with $\gamma_E$ being the Euler--Mascheroni constant. Furthermore, the constant multiplying the second term on the first line of~(\ref{inthresh})
is $-\frac{b_{SO(28)}}{4}$ (see discussion about $\beta$-functions for both gauge factors below). Since expression~(\ref{E1})
contains the regulator $-\log \mu^2$ which is an \textsc{ir} effect, we  indeed expect only massless modes to contribute to this constant
factor. In our local model, only localised states (from discrete $\slr_2/U(1)$ representations) give rise to BPS massless modes, and
it can  be checked from~(\ref{deg3}), (\ref{deg4}) and~(\ref{EE}) that these states alone and non non-localised states contribute to the coefficient
in front of $E(U,1)$, in accordance with the fact that $b_{SO(28)}$ counts the number of massless hypermultiplets~(\ref{beta1'}).

The second and third line of~(\ref{inthresh}) are the contributions from non-localised states carrying continuous $\slr_2/U(1)$ representations.
Their coefficients are determined by the Fourier expansion of the following holomorphic (quasi-)modular forms~(\ref{fexp}):
\begin{equation}\label{Fourexp}
\frac{E_2 E_8 -E_{10} }{q^{1/8} \eta^{21}} = \sum_{n=0}^\infty d_1(n)\, q^n\,, \qquad
\frac{E_8}{q^{1/8} \eta^{21}} = \sum_{n=-1}^\infty d_2(n)\, q^n\,.
\end{equation}
To see how the generalised Eisenstein series~(\ref{genEin}) come about, we single out the contribution from bulk states, which is given according to the second line of~(\ref{deg4}) by the following simple integral:
\begin{multline}\label{cdeg}
\Lambda_{\text{deg}}^{\text{c}}[1] = -\frac{T_2}{48}  
\sum_{(j,p)\neq (0,0)} \int_0^\infty \frac{\di \tau_2}{\tau_2^2} e^{-\frac{\pi T_2}{\tau_2 U_2}|j+pU|^2} 
 \sum_{n=0}^\infty (-)^n 
\left( d_1\!\big(\tfrac{n(n+1)}{2}\big) -\frac{3}{\pi\tau_2}  d_2\!\big(\tfrac{n(n+1)}{2}\big) \right) \times\\
\times\, \text{erfc}\big((n+\tfrac12)\sqrt{2\pi\tau_2}\big) \,.
\end{multline}
We now exploit the fact that the series $R(0|\tau)$~(\ref{decomF}) converges absolutely and uniformly for $\tau_2>0$~\cite{zwegers} to
invert the integral and the sum over $n$, and we then use the following integrals:
\begin{multline}
{\ds I_r(a,b)}  \equiv  {\ds \int_0^\infty \frac{\di x}{x^{2+r}} \, \text{erfc}(a\sqrt{x}) \, \ee^{-\frac{b}{x}} \,, \qquad r \geqslant  0 } \\ 
\allowdisplaybreaks
 = {\ds  \left(-\frac{\p}{\p b}\right)^r \int_0^\infty \frac{\di x}{x^2} \, \text{erfc}(a\sqrt{x}) \, \ee^{-\frac{b}{x}}
= \left(-\frac{\p}{\p b}\right)^r \frac{\ee^{-2a\sqrt{b}}}{b} } \hspace{3.9cm}  \\
 =  {\ds 
  \frac{1}{b^{r+1}}
\left( \Gamma(r+1) - \frac{2a\sqrt{b}}{\sqrt{\pi}} \Gamma(r+\tfrac12) \left._1F_2\right.\big(\tfrac12;\tfrac32,\tfrac12-r;a^2 b \big) \,+ \right. } 
\hspace{3cm} \\
 {\ds 
\left. 
+\,(-)^r \frac{a^{2r+2} b^{r+1} \sqrt{\pi}}{ (r+1) \Gamma\big(r+\tfrac32\big)}
 \left._1F_2 \right.\big(r+1;r+\tfrac32,r+2;a^2b \big) 
 \right) \,.
  }
\end{multline}
In particular, expression~(\ref{cdeg}) is given in terms of the following functions:
\begin{equation}
I_0(a,b) = \frac{\ee^{-2a\sqrt{b}}}{b}  \,, \qquad    
I_1(a,b) = \frac{\big( 1+ a\sqrt{b}  \big)}{b^2}\, \ee^{-2a\sqrt{b}} 
\end{equation}
by identifying
\begin{equation}\label{cdegfin}
a= \sqrt{2\pi} \big(n+\tfrac12 \big)\,, \qquad  b = \frac{\pi T_2}{U_2}|j+pU|^2\,,
\end{equation}
so that $\Lambda_{\text{deg}}^{\text{c}}[1]$ precisely gives the second and third line of~(\ref{inthresh}).
For degenerate orbits, we observe that bulk contributions distinguish themselves from the contributions of localised states
by an exponential suppression in~$\mathcal{E}_{(n+\frac12)\sqrt{T_2}}(U,s)$ in the large volume limit of the $T^2$, a novel feature with respect to compact models which appear to be peculiar to local models with non-localised bulk states in their spectrum. 

To try and grasp the physical significance of the degenerate orbit contributions, let us first concentrate on localised states. These give rise
the $E(U,1)$ and $E(U,2)/T_2$ terms in the first line of~(\ref{inthresh}) and are on every count similar
to contributions from degenerate orbits for heterotic $T^2 \times K3$ compactifications~(\ref{refthresh}).
By using the correspondence~\cite{Kaplunovsky:1994fg,Kaplunovsky:1995jw}, they can be shown
to map on the type \textsc{i} side to (perturbative) higher-genus contributions correcting the K\"ahler potential and the $SO(28)$ gauge kinetic function. 
The particulars of this mapping can be understood as follows: under S-duality, the heterotic /  type \textsc{i} string couplings and the $T^2$ K\"ahler modulus
transform (in the $\sigma$-model frame) as:
\begin{equation}\label{sdual}
\lambda_{\text{het}} = \frac{1}{\lambda_\textsc{i}} \,, \qquad 
T_2^\text{het} = \frac{T_2^\textsc{i}}{\lambda_\textsc{i}}\,, \qquad T_1^\text{het} = T_1^\textsc{i}\,,
\end{equation}
with $\lambda_{\text{het}}$ given by the double-scaling parameter~(\ref{DSL}). Then, the expansion in inverse
powers of $T_2$ characteristic of the degenerate orbit contribution is translated to a higher genus expansion
on the type \textsc{i} side~\cite{Bachas:1996bp,bachas1}, with $\lambda_\textsc{i}$ playing the r\^ole of the open-string loop-counting parameter.
Thus, the zero orbit contribution, proportional to $T_2$, corresponds to the leading, $\chi=1$ diagram contribution
in the dual  type \textsc{i} theory,  coming from the disk amplitude, while the $E(U,1)$ term is mapped to $\chi=0$ subleading
corrections corresponding to a combination of the annulus and the M\"obius strip amplitudes, and finally
the $E(U,2)/T_2$ term is translated to a $\chi=-1$ two-loop diagram, such as the disk with two holes.

From this angle, we expect the second and third line of the threshold~(\ref{inthresh}) to encode higher perturbative corrections
on the type~\textsc{i} side due to non-localised bulk states. This should be verified by carrying out in the large $T_2$ limit the appropriate 
expansion of the exponential factor.  In this regard, the observed mixing of the $U$ and $T$ moduli in the $ \ee^{-2\pi(n+\frac12) \sqrt{ \frac{2T_2}{U_2} } |pU_2 +j|}$  factors of the generalised Eisenstein series~(\ref{genEin}) seems at first sight puzzling.
Nevertheless  a similar mixing which occurs in the $\log(\mu^2 T_2 U_2)$ term regularising the \textsc{ir}
divergence in $E(U,1)$~(\ref{E1}) sheds some light on this issue.
Since non-localised state contributions to the threshold corrections act as
regulator of the infinite volume divergence of the underlying \textsc{cft} target-space, we can understand 
this mixing as a distinctive feature of compensating for the holomorphic anomaly of the modified elliptic genus. 
This however remains an analogy since in the first case we regularise an  \textsc{ir} divergence in the effective
field theory with a scale dependent regulator, while in the second case we renormalise the infinite volume divergence of 
the transverse space, where no scale is present to fix a cutoff on the massive  non-localised modes of the theory.

\subsubsection*{Contributions from non-degenerate orbits}
The contribution corresponding to degenerate orbits of $\Gamma$ appear in the last two lines of expression~(\ref{inthresh}), as computed in appendix~\ref{sec:nondeg}. On the type \textsc{i} side, these terms  map to non-perturbative corrections due to E1 instantons wrapping the $T^2$, since they are function
of the induced K\"ahler and complex structure moduli:
\begin{equation}
   \mathcal{T} = kpT \,, \qquad\qquad \mathcal{U} = \frac{j+pU}{k}\,.
\end{equation}
In particular, by using the heterotic / type \textsc{i} map~(\ref{sdual}),  we see that the factor $\ee^{2\pi \mathcal{T}}$ in the fourth line of 
expression~(\ref{inthresh}) is precisely the exponential of the Nambu--Goto action of an E1 string wrapped $N=kp$ times around the $T^2$.
The non-degenerate orbit contribution is also written as an expansion in the inverse volume of the $T^2$ obtained by acting on the elliptic genus with the modular invariant operator
$(-iD)^r (\tau_2\bar\partial_\tau)^r$ which annihilates holomorphic modular forms. Then, this expansion can be elegantly expressed 
in terms of two descendants of the modified elliptic genus, namely:
the weight 0 weak harmonic Maa\ss\, form:
\begin{equation}\label{AK}
\begin{array}{rcl}
{\ds \hat{\mathcal{A}}_K } (\tau)& \doteq  & {\ds (-iD) \tau_2^2 \bar\partial_{\tau} \hat{\mathcal{A}}_{SO(28)} (\tau)} \\[10pt]
& = & {\ds  -\frac{1}{144 \pi\, \eta^{24}} \left( 
\widehat{E}_2E_{10} + 2 E_6^2 +3 E_4^3
- \eta^4\! \left[\widehat{E}_2 E_8 + 4E_{10} + 3 E_8 D_0\right]
\Big(\frac{\widehat F}{\eta} \Big)-
\right.  } \\[14pt]
&& {\ds\hspace{3cm}  \left.  \phantom{\frac{\widehat F}{\eta}} -\frac{\pi}{\sqrt{2}}\! \left[ \widehat{E}_2^2 E_8 +17 \widehat{E}_2 E_{10} -8 E_6^2 -10 E_4^3\right]  (\sqrt{\tau_2} \bar\eta\eta)^3 \right)}
\end{array}
\end{equation}
where the covariant derivative reduces to $D_0(\widehat{F}/\eta)= \pi^{-1}i\partial_{\tau}(\widehat{F}/\eta)$; and the weight $-4$ weak harmonic Maa\ss\, form:
\begin{equation}\label{AH}
\begin{array}{rcl}
{\ds \hat{\mathcal{A}}_H } (\tau) & \doteq &  {\ds  (\tau_2^2 \bar\partial_{\tau})^2 \hat{\mathcal{A}}_{SO(28)} (\tau)}\\[6pt]
  & = &  {\ds \frac{1}{48\sqrt{2} \pi\,\eta^{24}}  \left((\pi\tau_2)^2 \overline{\widehat{E}_2} \big( \widehat{E}_2 E_8 -E_{10}\big) +12 E_8
  \right)  (\sqrt{\tau_2} \bar\eta\eta)^3 \,.} 
\end{array}
\end{equation}
The interpretation of the non-degenerate orbit contribution as E1 multi-instanton corrections on the type \textsc{i} side becomes more manifest if 
we reexpress it  by use of the Hecke operator $H_N$, which acts on a modular form of weight $r$ as follows:
\begin{equation}\label{Hecke}
H_N[\Phi_r](\tau) =  N^{r-1} \sum_{\tiny{\begin{array}{l}  k,p>0 \\ kp=N \end{array}}} \sum_{k>j\geqslant 0} \frac{1}{k^r}\, \Phi_r\!\left(\frac{j+p\tau}{k} \right)
\end{equation}
and thus preserves the space of modular forms of a given weight. 
Adopting a compact notation for the differential operator appearing in~(\ref{inthresh}):
\begin{equation}\label{UD}
\begin{array}{rccc}
\mathfrak{D} :  & \widehat{\mathbb{M}}_r  &  \longrightarrow &  \widehat{\mathbb{M}}_{r-2} \\[8pt]
                           &   \Phi_r(\tau) &  \mapsto &  \tau_2^2 \bar \p_{\tau} \Phi_r(\tau)
  \end{array} \quad ,
\end{equation}
the last two lines of the threshold corrections~(\ref{inthresh}) can be expressed in terms of a sum over Hecke operators~(\ref{Hecke}):
applied to weight 0 weak harmonic Maa\ss\, forms:
\begin{multline}\label{Hecke2}
\Lambda_{SO(28)}^\text{non-deg} = \frac14\sum_{N=1}^\infty \ee^{2\pi i N T} \,H_N\Big[ 
     \hat{\mathcal{A}}_{SO(28)}+\frac{1}{NT_2}\,\hat{\mathcal{A}}^K_{SO(28)} \Big](U)\, + \\[10pt]
       + \frac14\sum_{N=1}^\infty \ee^{2\pi i N T} \, \sum_{r=0}^\infty  \frac{1}{(r+2)!}\frac{1}{(NT_2)^{r+2} } \, 
  H_N\Big[  (-iD)^{r+2} \,\mathfrak{D}^r \!\hat{\mathcal{A}}_{SO(28)}^H\Big] (U)  + \, \text{c.c.}\,.
\end{multline}
This rewriting makes  particularly manifest the non-perturbative nature of these contributions on the type~\textsc{i} side, where they map to multi-instanton corrections, due to E1 instantons wrapping $N$ times the $T^2$. Anti-instanton contributions are also taken into account in
the complex conjugate of this expression, which corresponds to terms with $p<0$ in the sum on the last line of~(\ref{intgr}).
Expressing the instanton sum in terms of a sum over Hecke operators, as in~(\ref{Hecke2}), which by construction preserve the 
weight and modular properties of the forms, has the virtue of making apparent  the invariance of~(\ref{inthresh}) under $SL(2,\zi)_U$,
in keep with the corresponding global symmetry of the background.

\paragraph{The bulk state contributions: finite and large volume expression.} In the following, we will elaborate on some subtleties in the derivation of 
the non-degenerate orbit contribution coming from non-localised bulk states. The details on how to evaluate the integral~(\ref{intgr}) over the double cover of $\mathscr{H}$ can be found
in appendix~\ref{sec:nondeg}. In particular, from the last two lines of~(\ref{Insttot}), we obtain the contribution of bulk states by using~(\ref{BessInst}):
\begin{multline}\label{l2}
\Lambda^{\text{c}}_{\text{non-deg}}[1] =
 - \frac{1}{24} \sum_{k>j\geq 0}\sum_{p > 0} \frac{1}{kp} \sum_{n,m} (-)^n
 \left( e^{2\pi i \mathcal{T}}  \left[    d_1(m) -\frac{3}{\pi\mathcal{U}_2} d_2(m) 
 -\frac{3}{\pi\mathcal{T}_2}\Big( m +\frac{1}{2\pi\mathcal{U}_2}\Big) d_2(m)
\right] \times \right. \\
\times \ee^{2\pi i(m-\frac12n(n+1)) \mathcal{U}} -
2 \ee^{2\pi i \mathcal{T}_1}   \sum_{l=0}^\infty e_{n,l}
\left[  d_1(m) \,(\mathcal{T}_2\,\mathcal{U}_2)^{l+\frac12} \, W_{l,n,m}(\mathcal{T},\mathcal{U}) -  \phantom{\frac3\pi}
\right.\\[10pt]
\left. \left.-  \frac3\pi\, d_2(m) \,(\mathcal{T}_2\,\mathcal{U}_2)^{l-\frac12} \, W_{l-1,n,m}(\mathcal{T},\mathcal{U})  \right] \right)+ \text{c.c.} \,,
\end{multline}
with $d_{i}(n)$, $i=1,2$ the coefficients of the Fourier expansions~(\ref{Fourexp}), while the coefficients:
\begin{equation}
 e_{n,m}= \frac{\pi^m}{m!}\frac{\left(\sqrt{2}(n+\frac12)\right)^{2m+1}}{m+\frac12}
\end{equation}
are obtained by expanding the complementary error function, which is an odd entire series:
\begin{equation}\label{Gexp}
\text{erfc}\big((n+\tfrac12)\sqrt{2\pi\tau_2}\big)=   1-\sum_{m=0}^{\infty}(-)^m e_{n,m}\,\tau_2^{m+\frac12}\,.
\end{equation}
We also use the compact notation:
\begin{equation}\label{Q}
W_{l,n,m}(\mathcal{T},\mathcal{U}) \doteq
\frac{K_l\Big(2\pi \big| \mathcal{T}_2+\big(m-\frac12n(n+1)\big)\mathcal{U}_2\big|\Big)}{\Big| \mathcal{T}_2+\left(m-\frac12n(n+1)\right)\mathcal{U}_2\Big|^{l}}
\end{equation}
given in terms of the modified Bessel functions of the second kind:
\begin{equation}\label{Bess2}
K_{\alpha}(z) =\int_{0}^{\infty} 
\di t\,e^{-z\,\cosh t}\, \cosh \alpha t
 \,, \qquad \text{Re}\,z>0 \,,
\end{equation}
which are clearly even in $\alpha$:
\begin{equation}
K_{-\alpha}(z) = K_{\alpha}(z)\,.
\end{equation}
Now, for $z\in \er$ and $z \gg \big|\alpha^2 - \frac{1}{4}\big|$ the Bessel functions~(\ref{Bess2}) have the series expansion:
\begin{equation}\label{asympt}
K_{\alpha}(z) \cong \sqrt{\frac{\pi}{2z}} \,e^{-z} \sum_{n=0}^{\infty}\frac{\Gamma\left(\alpha+n+\frac12\right)}{n!\Gamma\left(\alpha-n+\frac12\right)}\,\frac{1}{(2z)^n}\,,
\end{equation}
since for $z >0$ it can be shown that by considering an asymptotic expansion where we truncate~(\ref{asympt}) at $n=N$, the remainder $R$ of
this series is bounded by~\cite{gradsh}:
\begin{equation}
|R| < \left| \frac{\Gamma\left(\alpha+2n+\frac12\right)}{(2n)!\Gamma\left(\alpha-2n+\frac12\right)(2z)^{2n}} \right| \,, \qquad
n > \tfrac12 \big( \alpha -\tfrac12\big)
\end{equation}
which is less than the absolute value of the first discarder terms in~(\ref{asympt}). Then, when in particular $z \gg \big|\alpha^2 - \frac{1}{4}\big|$,
we obtain the infinite power series expansion~(\ref{asympt}).

Therefore, by considering a region of the moduli space where $T_2$ is large enough, we can expand~(\ref{l2}) as follows:
\begin{multline}\label{l3}
\Lambda^{\text{c}}_{\text{non-deg}} [1]=
 - \frac{1}{24} \sum_{k>j\geq 0}\sum_{p > 0} \frac{1}{kp} e^{2\pi i \mathcal{T}} 
  \sum_{n,m}
  (-)^n
 \left(  \left[    d_1(m) -\frac{3}{\pi\mathcal{U}_2} d_2(m) 
 -\frac{3}{\pi\mathcal{T}_2}\Big( m +\frac{1}{2\pi\mathcal{U}_2}\Big) d_2(m)
\right] - \right. \\
- \sum_{l=0}^\infty e_{n,l} \,\mathcal{U}_2^{l+\frac12} \sum_{s,r=0}^\infty \frac{(-1)^r}{(4\pi)^s} \frac{1}{s! r!}\frac{1}{ \mathcal{T}_2^{r+s}}
\left[   \frac{\Gamma\big( l+s+r +\tfrac12\big)}{\Gamma\big( l-s+\tfrac12\big)} d_1(m) -
 \frac{\Gamma\big( l+s+r -\tfrac12\big)}{\Gamma\big( l-s-\tfrac12\big)}
\frac{3 d_2(m) }{\pi\mathcal{U}_2} 
\right]  \times \\
 \left.\phantom{\frac{\Gamma}{\Gamma}} \times\, \big(\big[m-\tfrac12n(n+1) \big] \mathcal{U}_2\big)^r \right) \ee^{2\pi i (m-\frac12n(n+1)) \mathcal{U}}
+ \text{c.c.}  \,.
\end{multline}
Then, since for half integer argument the $\Gamma$-function has an expression in terms of double factorials:
\begin{equation}
\Gamma\left(k+\tfrac12\right) = \frac{(2k-1)!! \sqrt{\pi}}{2^k}\,, \qquad
\Gamma\left(\tfrac12-k\right) = (-)^k \frac{2^k\sqrt{\pi}}{(2k-1)!!} \equiv \frac{(-)^k\pi}{\Gamma\left(k+\tfrac12\right)} \,,\quad k\in \en\,,
\end{equation}
we can elegantly express~(\ref{l3}) by means of the modular invariant operator~\cite{bachas1}
\begin{equation}
\Box \equiv  \mathcal{U}_2^2 \p_{\mathcal{U}}  \bar \p_{\mathcal{U}} 
\end{equation}
which annihilates holomorphic modular forms, by observing that:
\begin{equation}\label{l33}
\begin{array}{rcl}
{\ds \ee^{-2\pi i n \mathcal{U}}\mathcal{U}_2^{-(l\pm\frac12)}\,\frac{1}{\pi}\Box\, \mathcal{U}_2^{l \pm \frac12} \ee^{2\pi i n \mathcal{U}}} & = &{\ds 
- \frac{\Gamma\big( l+1 \pm\tfrac12\big)}{\Gamma\big( l\pm\tfrac12\big)} n \mathcal{U}_2 + \frac{1}{4\pi} 
\frac{\Gamma\big( l+1 \pm\tfrac12\big)}{\Gamma\big( l-1\pm\tfrac12\big)}
 } \\[12pt]
{\ds \ee^{-2\pi i n \mathcal{U}}\mathcal{U}_2^{-(l\pm\frac12)}\,\frac{1}{\pi^2}\left(\Box^2 -\frac12\Box\right)\, \mathcal{U}_2^{l\pm\frac12} \ee^{2\pi i n \mathcal{U}}} & = &{\ds 
\frac{\Gamma\big( l+2 \pm\tfrac12\big)}{\Gamma\big( l\pm\tfrac12\big)} \big(n \mathcal{U}_2\big)^2
- \frac{1}{2\pi} \frac{\Gamma\big( l+2 \pm\tfrac12\big)}{\Gamma\big( l-1\pm\tfrac12\big)} n \mathcal{U}_2 
\,+ }\\[12pt] 
&& {\ds  \hspace{3cm}+\, \frac{1}{(4\pi)^2}  \frac{\Gamma\big( l+2 \pm\tfrac12\big)}{\Gamma\big( l-2\pm\tfrac12\big)}
}\\[4pt]
 \vdots&& 
\end{array}
\end{equation}
Then, in the large $T_2$ region, the contribution from bulk states~(\ref{l3}) can be written as follows:
\begin{multline}
\Lambda^{\text{c}}_{\text{non-deg}}[1] = 
 - \frac{1}{24} \sum_{k>j\geq 0}\sum_{p >0} \frac{1}{kp}\,e^{2\pi i\mathcal{T}}  
  \left[ 1 + \frac{1}{\pi \mathcal{T}_2} \Box + \frac{1}{2!} \frac{1}{(\pi \mathcal{T}_2)^2}\left(\Box^2 -\frac12\Box\right) + ... \right] 
 \times\\[6pt]
 \times
 \sum_{n=0}(-)^n \text{erfc}\!\left((n+\tfrac12)\sqrt{2\pi \mathcal{U}_2} \right)
 e^{-\pi i (n+\frac12)^2\mathcal{U}}  \,\frac{\widehat{E}_2(\mathcal{U}) E_8(\mathcal{U})-E_{10}(\mathcal{U})}{\eta(\mathcal{U})^{21}}\,+ \,\text{c.c.} \,.
\end{multline}
The expansion in $\Box$ can alternatively be reorganised as a power expansion in the covariant derivative $\mathfrak{D}$~(\ref{UD}), so that 
by putting together the contribution of localised states~(\ref{Insttot}) containing the Mock modular form $F$~(\ref{mockF}) and the bulk state contribution~(\ref{l33}), we can reconstitute an expression in terms of the Maa\ss\, form $\widehat{F}(\mathcal{U})$~(\ref{decomF}), which is therefore manifestly $SL(2,\zi)_U$ invariant as already discussed:
\begin{equation}\label{l4}
\Lambda^{\text{flux}}_{\text{non-deg}} [1]=
 - \frac{1}{384} \sum_{k>j\geq 0}\sum_{p > 0} \frac{1}{kp}\,e^{2\pi i\mathcal{T}} \sum_{r=0}^{\infty}\frac{1}{r!}\frac{1}{\mathcal{T}_2^r} (-iD)^r(\mathcal{U}_2^2\bar{\partial}_{\mathcal{U}})^r  \left[ \frac{\widehat{F}(\mathcal{U})\,D_8E_8(\mathcal{U})}{\eta(\mathcal{U})^{21}} \right]  \,+ \,\text{c.c.}\,.
\end{equation}
Then, by using formul\ae~(\ref{DE})--(\ref{DE3}), the instanton contributions on the two last lines of~(\ref{inthresh}) can be compactly expressed
in terms of the modified elliptic genus in~(\ref{thresh5}) and its descendants~(\ref{AK}) and~(\ref{AH}). 

In this respect, since the operators $(-iD)^r\mathfrak{D}^r$ annihilate holomorphic modular forms and $\mathcal{U}_2^{-s}$ terms for $s>r$ positive, the contribution  of localised states to the non-degenerate orbit integral stops at the $O(\mathcal{T}_2^{-1})$ term $\mathcal{\hat{A}}^K_{SO(28)}$, so that expansion in powers of $\mathcal{U}_2^{-1}$ is finite and governed by $g_\text{max}$~(\ref{Agross}), whose value is dictated by the number of unbroken supercharges~\cite{bachas1}. In contrast, because they enter into the threshold corrections through the combination $gg^*$ of the shadow function and its transform (see last term in eq.~(\ref{univDD}) with $\widehat{F}=F+g^*$), bulk state contribute
an infinite number of descendants of the modified elliptic genus to the non-degenerate orbit integral, such in particular as the $(-iD)^{r+2}\mathfrak{D}^{r} \mathcal{\hat{A}}^H_{SO(28)}$ terms in eq.~(\ref{inthresh}). The $\mathcal{T}_2^{-1}$ expansion in the large volume limit~(\ref{l3}) indicates that this  is not in contradiction with space-time supersymmetry,
as the highest power of $\mathcal{U}_2^{-1}$ is in this case still $g_\text{max}=1$, as expected from a background preserving $\mathcal{N}_{\textsc{st}}=2$ in four dimensions.

Since the modular invariant descendants of the elliptic genus $(-iD)^r\mathfrak{D}^{r} \mathcal{\hat{A}}_{SO(28)}$ actually  determine corrections to various dimension-eight operators in the effective theory~\cite{bachas1}, this analysis
tends to suggests that  non-localised bulk modes in non-compact heterotic models entail an infinite number of such corrections to
the effective action.

Let us make one final comment about the finite / large volume issue in determining instanton threshold corrections. The large $T^2$ volume limit used
to derive the bulk state contributions on the last two lines  of expression~(\ref{inthresh}) makes manifest, on the type \textsc{i} side, the exponential $\ee^{-S_{\text{E1}}}$ of  $S_{\text{E1}}=2\pi N \!\left(\frac{T_2^\textsc{i}}{\lambda_\textsc{i}} - i T_1^\textsc{i} \right)$ which is the classical action of an E1 instanton wrapping $N$ times the $T^2$. In the finite $T_2 >1$ regime,
we have in contrast to use the more involved expression~(\ref{l2}). There, the exponential of the topological part of the action  $\text{Im}\,S_{\text{E1}} = -\frac{1}{2\pi\alpha'} \int \widehat{B}_\textsc{i}$ is still apparent, while the part of the action depending on the pull-back of the metric on the instanton worldvolume $\text{Re}\,S_{\text{E1}} = \frac{1}{2\pi\alpha'\lambda_\textsc{i}} \int \di\sigma^2\,\sqrt{|\widehat{G}_\textsc{i}|}$  is now apparently entangled with the complex structure modulus, a peculiarity that calls for further investigation, possibly of the the S-dual type~\textsc{i} model.

\subsection{Moduli dependence of the $SU(2)$ threshold corrections}
The $SU(2)$ threshold correction can now be determined very  economically by using the difference~(\ref{diffthresh}), whose RHS can be readily integrated:
\begin{multline}\label{U1mod}
\allowdisplaybreaks
 \Lambda_{SU(2)}[1]   - 2 \Lambda_{SO(28)} [1]   =   12  \int_{\mathscr{F}} \frac{\di^2\tau}{\tau_2}\,\Gamma_{2,2}(T,U) 
  \\  =      4\pi T_2 - 12 \big(  \log|\eta(U)|^4 + \log(\mu^2\,T_2 U_2) + \gamma  \big)  
  + 24 \!\sum_{k>j\geqslant 0} \sum_{p>0} \frac{1}{kp}\,
     \ee^{-2\pi\mathcal{T}_2} \cos\big(2\pi \mathcal{T}_1\big)\,.
 \end{multline}
 Using the $SO(28)$ threshold just computed~(\ref{inthresh}), we get a general formula for the $\beta$-functions:
 \begin{equation}\label{bth}
 b_a = k_a\left( 6-\frac{t_a}{2}\right) \,.
 \end{equation}
As previously seen: $t_{SO(28)}=4$ and $t_{SU(2)}=-44$,  leading to the following $\beta$-functions for the $\mathcal{Q}_5=1$ model :
\begin{equation}\label{hyp2}
b_{SO(28)} = 4\,, \hspace{2cm} b_{SU(2)}  = 56\,.
\end{equation}
This is in perfect agreement with the field theory results~(\ref{hyp}) obtained from the hypermultiplet
counting in table~\ref{tabops}.

\section{The dual type \textsc{i} model}
\label{sec:typeone}
As discussed in the previous section, the gauge thresholds~(\ref{inthresh}) and~(\ref{U1mod}) translate 
 as perturbative and instanton corrections to the K\"ahler and gauge kinetic functions in the S-dual type \textsc{i} model. 
In the case under scrutiny, this theory only has space-time filling D9-branes, which support the gauge group $SO(28) \times SU(2)$.
Half D5-branes at singularities which are usually necessary in orbifold models for anomaly cancellation~\cite{gpsw} are absent here,
since the $A_1$ singularity is resolved and the tadpole equation~(\ref{tad1}) is satisfied by $U(1)$ gauge instantons on the 
blown-up $\mathbb{P}^1$ two-cycle. 

A microscopic description of the type \textsc{i} theory dual to the warped Eguchi--Hanson heterotic background can still 
be hard to come by. Nevertheless, by using the field theory dictionary~\cite{Kaplunovsky:1994fg,Kaplunovsky:1995jw} mapping
heterotic  gauge threshold corrections to type \textsc{i} one-loop corrections to the gauge kinetic functions:
\begin{multline}\label{kapdict}
\left.\frac{4\pi^2}{g^2_a(\mu^2)}\right|_{\text{1-loop}}  =  \left. \text{Re} f_a(M) \right|_{\text{1-loop}}-\frac{b_a}{4} \left[ \log \big(\!-iS+i\bar S \big)  -\log\frac{M_\text{Planck}^2}{\mu^2} \right] + \\
+ \frac14 \left[ c_a \, K(M,\overline{M}) - 2 \sum_{\mathbf{R}} T_a(\mathbf{R}) \,\log \det C_{\mathbf{R}}(M,\overline{M})  \right] \,,
\end{multline}
these corrections can be extracted from the heterotic result, even when the corresponding type~\textsc{i} model is unknown.  In particular, on the RHS of
eq.~(\ref{kapdict}), $K(M,\overline{M})$ is the tree-level K\"ahler potential, $\det C_r(M,\overline{M})$ the determinant of the tree-level K\"ahler
metric for the matter multiplets in the representation $\mathbf{R}$ of the gauge group factor $\mathrm{G}_a$, and the model dependent constants:
\begin{equation}\label{cfunct}
c_a = \sum_{\mathbf{R}} n_{\mathbf{R}}\, T_a({\mathbf{R}})-\,T_a(\text{Adj}_a)\,,
\end{equation}
which for $\mathcal{N}_{\textsc{st}} =2$ theories are equal to $b_a/2$, see~(\ref{beta1}).
 
 For our purpose, we alternatively express the heterotic one-loop gauge thresholds~(\ref{gloop}) in terms of the axio-dilaton multiplet $S$ and the universal function $\Delta_\text{univ}(M,\overline{M})$, as in~(\ref{univ}), and decompose the latter into harmonic functions and a non-harmonic remainder:
\begin{equation}
\Delta_{\text{univ}}(M,\overline{M}) = V(M,\overline{M}) + H(M) + \overline{H(M)}\,.
\end{equation}
Then, by using the tree-level identification of the imaginary part of $S$:
\begin{equation}
\text{Im} \,S =   \frac{M_{\text{Planck}}^2}{M_s^2}\,,
\end{equation}
we recover the corrected K\"ahler potential and gauge kinetic functions in terms of heterotic one-loop quantities~\cite{Kaplunovsky:1995jw,Derendinger,Nilles}:
\begin{subequations}
\begin{align}\label{dict2}
\allowdisplaybreaks
\text{Re}\,f_a|_{\text{1-loop}} = & \; k_a\, \text{Im}\, S  + \frac14 \left(-  c_a  K_{\text{tree}}(M,\overline{M}) +
2 \sum_{\mathbf{R}} T_a(\mathbf{R}) \,\log \det C_{\mathbf{R}}(M,\overline{M}) \,+   \right. \notag\\
&  \left. \hspace{6cm} \phantom{\sum_{\mathbf{R}} }+  \Delta(M,\overline{M})  -  k_a V(M,\overline{M})  \right) \,, \\
K|_{\text{1-loop}} = &\;  K_{\text{tree}}(M,\overline{M})-\log\!\left(-iS+i\bar S -\tfrac12 V(M,\overline{M}) \right) \,. \label{dict3}
\end{align}
\end{subequations}
As seen in the previous section, these corrections have a natural interpretation in the type \textsc{i} S-dual model in terms of perturbative string-loop corrections and non-perturbative corrections due to E1 instantons wrapping the $T^2$.

It has been shown~\cite{Camara:2008zk} that for a general $\mathcal{N}_{\textsc{st}} =2$ heterotic orbifold compactification, the second and third term on the LHS of the first line of~(\ref{dict2})
conspire to cancel the $b_a\log(T_2U_2)$ term in the threshold correction~(\ref{inthresh}) coming from the \textsc{ir} regulator in~(\ref{E1}), due to 
the correlated way the Casimirs $T(\mathbf{R})$ enter into the contribution from the K\"ahler metric of the matter fields and into the functions $c_a$~(\ref{cfunct}) . Thus, the corrections to the gauge kinetic functions
for both $a=\{ SO(28),\, SU(2) \}$ gauge factors
are given by  the harmonic contributions in~(\ref{inthresh}) and~(\ref{U1mod}):
\begin{equation}\label{fa}
f_a = -ik_aS - \frac{k_a(53 -6 t_a)\pi}{144} iT - b_a \log\eta(U)  + \frac{1}{2} \sum_{k>j\geqslant 0} \sum_{p>0} \frac{1}{kp}  \ee^{2\pi i\mathcal{T}} \mathcal{A}_a(\mathcal{U}) \,.
\end{equation}
In particular, they  receive perturbative corrections from the disk and a combination of the annulus and the M\"obius strip diagrams, along with
 E1 instanton contributions, which are given in terms of the holomorphic part of the modified elliptic genera~(\ref{thresh5}) and~(\ref{thresh6}):
\begin{equation}
\mathcal{A}_a(\tau) = -\frac{k_a}{72}\left( E_2 E_{10} -\frac{12+t_a}{24} E_6^2-\frac{12-t_a}{24}E_4^3 - \eta^3 G_1 \big(E_2E_8-E_{10}\big)   \right)\,,
\end{equation}
where we have defined the holomorphic function:
\begin{equation}
G_1(\tau) = F(\tau) -12 \sum_{n=0}^\infty (-) q^{-\frac12 (n+\frac12)^2} \,,
\end{equation}
and the gauge factor dependent coefficients previously determined are summarised in the table:
\begin{equation}
\begin{array}{l  || cc|c}
        & \;k_a \; & \; t_a \;& b_a\\ \hline \hline
 SO(28)\; & 1 & 4 &  4 \\  
 SU(2) & 2 & -44 & \; 56 \;
 \end{array}
\end{equation}

In contrast,  the corrections to the tree level K\"ahler potential of the effective type \textsc{i} theory is given by the non-harmonic real analytic part
of~(\ref{inthresh}), which  for convenience we split into a perturbative and non-perturbative contribution:
\begin{equation}\label{corrK}
K=- \log\big[(T-\overline{T})(\overline{U}-U)\big] - \log(a^2) - \log\!\left(-iS+i\bar S - \frac{V_\text{pert.}(T,U) +V_\text{E1}(T,U)}{2}   \right) \,,
\end{equation}
These corrections originate from higher string-loop and  multi-instanton corrections and yield:
\begin{subequations}
\begin{align}\label{corrV}
\allowdisplaybreaks
{\ds V_\text{pert.}(T,U) }  = &\; {\ds  - \frac{29\pi}{90} \frac{E(U,2)}{T_2}  - 
\frac{1}{2} \sum_{n=0}^\infty   (-)^n \left[  \frac{\pi}{18}d_1\!\big(\tfrac{n(n+1)}{2}\big) \, \mathcal{E}_{(n+\frac12)\sqrt{2T_2}} (U,1) \, - \phantom{\frac{E_{\frac12}}{T_2}}  \right.  }
    \notag \\
   &{\ds \left. - \frac{\sqrt{2} \zeta(3)}{\pi^2} \big(n+\tfrac12 \big)d_2\!\big(\tfrac{n(n+1)}{2}\big)  \, 
   \frac{\mathcal{E}_{(n+\frac12) \sqrt{2T_2}} (U,3/2) }{\sqrt{T_2}} 
     -  \frac{\pi}{90}d_2\!\big(\tfrac{n(n+1)}{2}\big) \, \frac{\mathcal{E}_{(n+\frac12) \sqrt{2T_2}} (U,2) }{T_2}  \right]  \,,}  
\displaybreak[2] \\
{\ds V_\text{E1} (T,U)}  =& {\ds  \sum_{k>j\geqslant 0} \sum_{p>0} \frac{1}{kp}\,
     \ee^{2\pi i\mathcal{T}}
      \left[ \frac{1}{6\pi\eta^{24}(\mathcal{U})} \left( \frac{ E_{10}(\mathcal{U})- \eta^{3}(\mathcal{U})
\widehat{F}(\mathcal{U}) E_8(\mathcal{U}) }{\mathcal{U}_2} - 3\pi \eta^{3}(\mathcal{U})
     \widehat{G}_2 (\mathcal{U})D_8E_8(\mathcal{U}) \right)  
       \right.}
     \notag \\ &
       {\ds \left. \qquad\quad  +\,\frac{\hat{\mathcal{A}}_K(\mathcal{U})}{\mathcal{T}_2} + \sum_{r=0}^\infty  \frac{1}{(r+2)!}\frac{1}{(\mathcal{T}_2)^{r+2} } \, 
   (-iD)^{r+2} (\mathcal{U}_2^2 \bar\partial_{\mathcal{U}})^r 
     \hat{\mathcal{A}}_H(\mathcal{U}) 
       \right]  + \, \text{c.c.}  \,.
     }
\end{align}
\end{subequations}
where we have defined  the non-holomorphic function:
\begin{equation}
\widehat{G}_2(\tau)  =  \sum_{n=0}^\infty(-)^n E\big((n+\tfrac12)\sqrt{2\tau_2}\big) q^{-\frac12 (n+\frac12)^2} \,.
\end{equation}

Note in particular that in these expressions $SL(2,\zi)_T$ modular transformations mix perturbative and instanton corrections, in accordance with
the fact that T-duality is not a symmetry of type \textsc{i} string theory.

\paragraph{Some comments about E1 instanton contributions :}
the analysis in terms of Hecke operators~(\ref{Hecke}) gave us an understanding of the non-perturbative contributions 
in~(\ref{corrK}) and~(\ref{fa}) as coming from E1 instantons wrapping $N$ times the $T^2$, so as multi-instanton corrections.  
Since the $A_1$ singularity is resolved, all potential E1 instantons initially sitting at the fixed point in the orbifold limit 
have been moved away from it. Thus all such instantons present in the blowup regime are either localised on the resolved two 
cycle or at some position in the bulk. Thus they all carry $SO(r)$ Chan--Paton factors. These E1 instantons are characterised  by the following uncharged massless modes:
\begin{itemize}
\item[$\bullet$] in the antisymmetric representation $\mathbf{\frac{r(r-1)}{2}}$: bosonic zero modes $z$ and $\bar z$ and the corresponding fermionic ones  $\lambda^{\alpha, a}$  and $\lambda^{\dot{\alpha}, a}$,
\item[$\bullet$] in the symmetric representation $\mathbf{\frac{r(r+1)}{2}}$: bosonic zero modes $x_\mu$, $\rho$, $\theta$, $\phi$ and $\psi$ and fermionic zero modes  $\chi^{\alpha, a}$ 
and $\chi^{\dot{\alpha}, a}$.
\end{itemize}
The strings extending between an E1 instanton and a stack of $n$ D9-branes produce in addition a bosonic zero mode $\sigma$ in the bifundamental representations $(\mathbf{r},\bar{\mathbf{n}})+(\bar{\mathbf{r}},\mathbf{n})$.

As a final remark, since for instantons to contribute to the gauge kinetic functions they should possess four neutral massless zero modes~\cite{Petersson:2007sc,Akerblom:2007uc,Blumenhagen:2007ip}, we expect most of the zero modes listed above to acquire a mass through the Scherk--Schwarz mechanism.
To determine the surviving zero-modes, one should analyse the subspace of the multi-instanton moduli space for the $N$-instanton contributions~(\ref{Hecke2}), corresponding to
deformations of the instantons along the warped Eguchi--Hanson space, in order to find out when all the components of the multi-instantons
coincide~\cite{GarciaEtxebarria:2007zv,GarciaEtxebarria:2008pi}, for instance on the resolved $\mathbb{P}^1$. This we will not attempt here.

\section{Perspectives}

In order to generalise the results presented in sections~\ref{sec:moddepgen} and~\ref{sec:typeone}, it would be interesting 
to compute explicitely the modular integrals~(\ref{fullthr}) and~(\ref{fullthrU1}) for generic five-brane charge $\mathcal{Q}_5$. This would allow to distinguish explicitely  the contributions from worldsheet instantons wrapped around the $\mathbb{P}^1$, by isolating $\ee^{-kn}$ factors in the tree-level contribution 
to the heterotic gauge thresholds. Then, one could repeat in this case the analysis of section~\ref{sec:typeone} and explicitly
extract perturbative and non-perturbative corrections to the K\"ahler potential and the gauge kinetic functions for
arbitrary five-brane charge $\mathcal{Q}_5$.

In this perspective it would be interesting to be able to cross-check the results we obtain on the type  \textsc{i} side from S-duality
by direct string amplitude computations, along the line of~\cite{Berg:2005ja}, and have an explicit derivation of Chan--Paton factors
attached to the E1-brane instantons, as in~\cite{Camara:2007dy}. This would be particularly attractive
in the present models which allow to go beyond the large volume limit commonly considered for type \textsc{ii} (orientifold) models.
An explicit description of the lifting of fermionic zero modes by instanton effects  in the torsional  local models considered here  
would also be appealing. This analysis would call for a  microscopic understanding of multi-instanton 
effects  originating from E1-branes by adapting the approach~\cite{Blumenhagen:2007bn,GarciaEtxebarria:2007zv,GarciaEtxebarria:2008pi,Blumenhagen:2009qh} to smooth local non-K\"ahler geometries with non-vanishing five-brane charge.

A very important follow-up of this paper would be to consider the situations where the fibration of the two-torus over 
the base is non-trivial, $i.e.$  geometries of the type $T^2 \hookrightarrow \mathcal{M} \to \widetilde{\textsc{eh}}$~\cite{Fu:2008ga}. 
The main novelty in these cases is that the topology of the solution is modified, namely the $\mathbb{P}^1\times T^2$ is replaced 
by $S^3/\mathbb{Z}_N \times S^1$, the first factor being a Lens space. Therefore the E1 instantons can only wrap
a {\it torsional} two-cycle, meaning that the instanton sums should terminate at wrapping number $N-1$. This would be a 
very interesting new effect. Another interesting novelty with non-trivial fibration is that part of the torus moduli are 
restricted to a set of discrete values~\cite{Gukov:2002nw}. The explicite computation of gauge threshold corrections  can be done straightforwardly 
as the worldsheet \textsc{cft} is also known in these cases.

Finally, a most challenging extension of this work is the computation of threshold corrections for 
genuinely {\it compact} torus fibrations  $T^2 \hookrightarrow \mathcal{M} \to \widetilde{K3}$. There, the 
worldsheet \textsc{cft} is not known; one should then rather use the gauged linear sigma model description 
given in~\cite{Adams:2006kb}, or a purely geometrical approach extending ref.~\cite{MohriKawai} to generalised \textsc{cy} 
geometries.  This would open an exciting avenue to tackle phenomenological issues such as
moduli stabilisation due instantonic corrections to the K\"ahler potential, by extending the analysis~\cite{Balasubramanian:2005zx,Conlon:2005ki}
to type~\textsc{i} compactifications on smooth non-K\"ahler spaces supporting line bundles, 
without the restriction of considering a large volume limite thereof. Other applications come
from instanton corrections to gauge kinetic functions, which generically modify gaugino masses and gauge
couplings, and might thus affect the phenomenology of the effective theory. 
Constructing an exact \textsc{cft} description for a full-fledged compactification of the local heterotic torsional models 
examined in the present work would prove particularly relevant to these questions.
We expect to come back to these issues in the next future.


\subsection*{Acknowledgments}

The authors would like to especially thank Carlo Angelantonj, Pablo C\'amara, Jan Manschot, Sameer Murthy, 
Jan Troost and Sanders Zwegers for extremely fruitful discussions and invaluable help. They are also
very much indebted to Emilian Dudas for constant support and help, and acknowledge his 
participation to the initial phase of the project.
They also benefitted from enlightening discussions with Ignatios Antoniadis, Ioannis Florakis, Dumitru Ghilencea, 
Elias Kiritsis and Boris Pioline.
They acknowledge partial  support from  the LABEX P2IO,
the ERC Advanced Investigator 
Grant no. 226371 "Mass Hierarchy and Particle Physics at the TeV Scale"
(MassTeV), the ITN program PITN-GA-2009-237920, the French ANR contracts: TAPDMS 
ANR-09-JCJC-0146 and
05-BLAN-NT09-573739, and the IFCPAR CEFIPRA program 4104-2.

\appendix 

\section{$\boldsymbol{{\mathcal N}=2}$ characters and useful identities}
\label{appchar}
\boldmath
\subsection{$\mathcal{N} =2$ minimal models}\label{sec:minmod}
\unboldmath
The characters of the $\mathcal{N} =2$ minimal models, {\it i.e.}  the supersymmetric $SU(2)_k / U(1)$ gauged \textsc{wzw} model, are conveniently defined through the characters $C^{j\ (s)}_{m}$~\cite{Gepner:1987qi} of the $[SU(2)_{k-2} \times U(1)_2] / U(1)_k$ bosonic 
coset, obtained by splitting the Ramond and Neveu--Schwartz 
sectors according to the fermion number mod 2. Defining $q=\ee^{2\pi i\tau}$ for the complex structure $\tau\in\mathscr{H}$ and $z=\ee^{2\pi i\nu}$ for the elliptic variable $\nu\in\ci$, these characters are determined implicitly through the
identity:
\begin{equation} \label{theta-su2}
\chi_{k-2}^{j} (\nu|\tau)
\Theta_{s,2}(\nu-\nu'|\tau) = \sum_{m \in \zi_{2k}} C^{j\ (s)}_{m} (\nu'|\tau)  \Theta_{m,k} \big(\nu-\tfrac{2\nu'}{k}\big|\tau\big) \, ,
\end{equation}
in terms of the theta functions of $\widehat{\mathfrak{su}(2)}_k$:
\begin{equation}\label{thSU2}
\Theta_{m,k} (\tau,\nu) = \sum_{n}
q^{k\left(n+\tfrac{m}{2k}\right)^2}
z^{k \left(n+\tfrac{m}{2k}\right)}\,,  \qquad m \in \mathbb{Z}_{2k} 
\end{equation}
and $\chi^j_{k-2} $ the characters of the affine algebra $\widehat{\mathfrak{su}(2)}_{k-2}$:
\begin{equation}\label{su2-char}
\chi_{k-2}^j (\nu|\tau) = \frac{\Theta_{2j+1,k} (\nu|\tau)-\Theta_{-(2j+1),k} (\nu|\tau)}{i\vartheta_1(\nu|\tau)}\,.
\end{equation}
We also mention an identity on $\widehat{\mathfrak{su}(2)}_k$ theta functions, which we use in the present work:
\begin{equation}\label{idSU2}
\Theta_{\nicefrac{m}{p},\nicefrac{k}{p}}(\nu|\tau) = \sum_{n\in\zi_p}\Theta_{m+2kn,pk}\big(\tfrac{\nu}{p}\big|\tau\big)\,.
\end{equation}
and another way of writing the $SU(2)_{k-2}$ characters for $\nu=0$:
\begin{equation}\label{chi-lim}
\chi^j_{k-2}(0|\tau) = \frac{{\ds \sum_{n\in\zi} \big(2(k-2)n+2j+1\big)\, q^{(k-2)\left(n+\frac{2j+1}{2(k-2)}\right)^2}}}{{\ds q^{\nicefrac18} \prod_{m=1}^\infty \big(1-q^n \big)^3}}
 = \frac{\Theta'_{2j+1,k-2}(0|\tau)}{\pi i\eta(\tau)^3}\,,
\end{equation}
where $'\equiv \d_\nu$.

Highest-weight representations are labeled by  $(j,m,s)$, corresponding primaries of 
$SU(2)_{k-2}\times U(1)_k \times U(1)_2$. The following identifications apply:
\begin{equation}
(j,m,s) \sim (j,m+2k,s)\sim
 (j,m,s+4)\sim
 \big(\tfrac{k}{2}-j-1,m+k,s+2\big)
\end{equation}
as  the selection rule $2j+m+s =  0  \mod 2$. The spin $j$ is restricted to $0\leqslant j \leqslant \tfrac{k}{2}-1$.  
The conformal weights of the superconformal primary states are:
\begin{equation}
\begin{array}{cclccc}
\Delta &=& {\ds \frac{j(j+1)}{k} - \frac{m^2}{4k} + \frac{s^2}{8}} \ & \text{for} & \ {\ds -2j \leqslant m-s \leqslant 2j} \\[8pt]
\Delta &=& {\ds \frac{j(j+1)}{k} - \frac{m^2}{4k} + \frac{s^2}{8} + \frac{m-s-2j}{2}}
\ & \text{for} & \ {\ds 2j \leqslant m-s \leqslant 2k-2j-4}
\end{array}
\end{equation}
and their $R$-charge reads:
\begin{equation}
Q_R = \frac{s}{2}-\frac{m}{k} \mod 2 \,. 
\end{equation}
\\[4pt]
\underline{ {\it Chiral} primary states:} they are obtained for $m=2(j+1)$ and 
$s=2$ (thus odd fermion number). Their conformal dimension reads:
\begin{equation}
\Delta= \frac{Q_R}{2} = \frac{1}{2} - \frac{j+1}{k}\, .
\end{equation}
\\[4pt]
\underline{{\it Anti-chiral} primary states:} they are obtained for $m=2j$ and $s=0$ 
(thus even fermion number). Their conformal dimension reads:
\begin{equation}
\Delta= -\frac{Q_R}{2} = \frac{j}{k}\, .
\end{equation}
\\[4pt]
Finally we have the following modular S-matrix for the $\mathcal{N}=2$ minimal-model characters:
\begin{equation}
S^{jm s}_{j' m' s'} = \frac{1}{2k} \sin \pi
\frac{(1+2j)(1+2j')}{k} \ e^{i\pi \frac{mm'}{k}}\ e^{-i\pi ss'/2}.
\end{equation}
The usual Ramond and Neveu--Schwarz characters, that we use in the bulk of the paper, are  obtained as:
\begin{equation}
C^{j}_{m} \oao{a}{b} (\nu|\tau)=  e^{\frac{i\pi ab}{2}} \left[ C^{j\, (a)}_{m} (\nu|\tau)
+(-)^b C^{j\, (a+2)}_{m}(\nu|\tau) \right],
\end{equation}
where $a=0$ (resp. $a=1$) denote the \textsc{ns} (resp. \textsc{r}) sector, and characters 
with $b=1$ are twisted by $(-)^F$. They are related to $\widehat{\mathfrak{su}(2)}_k$ characters 
through:
\begin{equation}\label{relation-char-su2}
\chi^j(\nu|\tau)\,  \vartheta \oao{a}{b}  (\nu|\tau)= \sum_{m \in \zi_{2k}} C^j_m\oao{a}{b} (\nu|\tau)\, \Theta_{m,k}(\nu|\tau)\,.
\end{equation}
In terms of those one has the reflexion symmetry:
\begin{equation}
C^j_m \oao{a}{b}(\nu|\tau) = (-)^b C^{\tfrac{k}{2}-j-1}_{m+k} \oao{a}{b}(\nu|\tau)\, . 
\label{reflsym}
\end{equation} 

\boldmath
\subsection{Supersymmetric $\slc$}
\label{sec:supslc}
\unboldmath
The characters of the $\slc$ super-coset
at level $k$ come in different categories corresponding to
irreducible unitary representations of  $SL(2,\mathbb{R})$.

\paragraph{Continuous representations:} they correspond to spin $J = \frac12 + ip$ and  continuous momentum
$p \in \mathbb{R}^+$ states. Their characters are denoted by
 ${\rm ch}_\text{c} (\tfrac{1}{2}+ip,M) \oao{a}{b}$, where the $U(1)_R$ charge of the primary is $Q_R=2M/k$. They read:
\begin{equation}
{\rm ch}_\text{c} (\tfrac{1}{2}+ip,M;\nu|\tau) \oao{a}{b} = \frac{1}{\eta^3 (\tau)} q^{\frac{p^2+M^2}{k}} \vartheta \oao{a}{b} (\nu|\tau)\,
z^{\frac{2M}{k}}\, .
\end{equation}

\paragraph{Discrete representations:} their characters  $\mathrm{ch}_d (J,r) \oao{a}{b}$,
have a real $\slr$ spin in the range $\frac12 < J < \frac{k+1}{2}$. Their  $U(1)_R$  charge reads  $Q_R=2(J+r+\frac a2)/k$,
$r\in \zi$.  Their characters are given by 
\begin{equation}
{\rm ch}_\text{d} (J,r;\nu|\tau) \oao{a}{b} =  \frac{
  q^{\frac{-(J-1/2)^2+(J+r+a/2)^2}{k}}
z^{\frac{2J+2r+a}{k}}}{1+(-)^b \,
z q^{1/2+r+a/2} } \frac{\vartheta \oao{a}{b} (\nu|\tau)}{\eta^3 (\tau)}\,.
\label{idchar}
\end{equation}
\\[4pt]
\underline{{\it Chiral} primaries:} they are obtained for $r=0$, {\it i.e.} $M=J$, in the \textsc{ns} sector  (with even fermion number). Their conformal dimension reads
\begin{equation}
\Delta= \frac{Q_R}{2} = \frac{J}{k}\, . 
\end{equation}
\\[4pt]
\underline{{\it Anti-chiral} primaries:} they are obtained for $r=-1$ (with odd fermion number), and their conformal dimension reads
\begin{equation}
\Delta= -\frac{Q_R}{2} =\frac{1}{2}-\frac{J-1}{k}\, . 
\end{equation}

\paragraph{Identity representation:} this representation corresponds to the the vacuum of the level $k$ super-Liouville theory,  which is both a highest and lowest weight representation  and a chiral and anti-chiral primary with spin and $U(1)_R$ charge $J=0=Q_R$. Its characters are labeled by a discrete charge $r\in\zi$:
\begin{equation}\label{Idrep0}
{\rm ch}_{\text{Id}} (r;\nu|\tau) \oao{a}{b} = \frac{q^{\frac{\left(r+\frac{a+1}{2}\right)\left(r+\frac{a-1}{2}\right)}{k}+r+\frac{a-1}{2}} 
z^{\frac{2\left(r+\frac{a}{2}\right)}{k}+1}}{\left(1+(-)^bzq^{r+\frac{a+1}{2}}\right) \left(1+(-)^bzq^{r+\frac{a-1}{2}}\right)}
\frac{\vartheta \oao{a}{b} (\tau, \nu)}{\eta^3 (\tau)}\,.
\end{equation}
The identity representation in $\slr_k/U(1)$ is non-normalisable.

\subsubsection*{Extended characters} 
Extended characters are defined for $k$ integer by summing
over $k$ units of spectral flow~\cite{Eguchi:2003ik}.\footnote{One can extend their definition to the case of rational $k$, which 
is not useful here.} For instance, the extended discrete characters of charge $r\in \zi_k$ read:
\begin{equation}\label{extdischar}
\begin{array}{rcl}
{\rm Ch}_\text{d} (J,r;\nu|\tau) \oao{a}{b}  & = & {\ds \sum_{w \in \zi} {\rm ch}_\text{d} (J,r+kw;\nu|\tau) \oao{a}{b}  }\\[4pt]
          & = & 
  {\ds \sum_{w\in\zi} \frac{\left(q^{kw+r+\frac{a+1}{2}} z\right)^{\frac{2J-1}{k}}}
  {1+(-)^b \,z q^{kw+r+\frac{a+1}{2}} }\,q^{\frac{\left(kw+r+\frac{a+1}{2}\right)^2}{k}}z^{\frac{2\left(kw+r+\frac{a+1}{2}\right)}{k}}
  \,\frac{\vartheta\oao{a}{b}(\nu|\tau)}{\eta^3 (\tau)} } \\[6pt]
  & =  & 
  {\ds \sum_{w\in\zi} \frac{
  q^{\frac{\left(kw+J+r+\frac{a}{2}\right)^2 - \left(J-\frac12\right)^2}{k}} z^{\frac{2(kw+J+r)+a}{k}}}
  {1+(-)^b \,z q^{kw+r+\frac{a+1}{2}} } 
\,\frac{\vartheta \oao{a}{b} (\nu|\tau)}{\eta^3 (\tau)} } \,.
  \end{array}
\end{equation}
and the extended continuous characters:
\begin{equation}
\begin{array}{rcl}
{\ds {\rm Ch}_\text{c} (\tfrac{1}{2}+ip,M;\nu|\tau) \oao{a}{b}} & = & {\ds   \sum_{w \in \zi} 
{\rm ch}_\text{c} (\tfrac{1}{2}+ip,M+kw;\nu|\tau) \oao{a}{b}} \\[6pt]
 & = &{\ds  \frac{q^{\frac{p^2}{k}}}{\eta^3 (\tau)}\,
\Theta_{2M,k} (\tau,\tfrac{2\nu}{k})\, \vartheta \oao{a}{b} (\nu|\tau)\, ,}
\end{array}
\label{extcontchar}
\end{equation}
where discrete $\mathcal{N}=2$  R-charges are chosen: $2M \in \zi_{2k}$. 

Finally we can also define by the same procedure extended characters for the identity representatation,
with discrete charge $r\in \zi_k$:
\begin{equation}\label{idcont}
{\rm Ch}_{\text{Id}} (r;\nu|\tau) \oao{a}{b} = \sum_{w\in\zi}{\rm ch}_{\text{Id}} (r+kw;\nu|\tau) \oao{a}{b}
\end{equation}
 
Extended characters close under the action of the modular group.
It is worthwhile noting however that although all three kinds of extended characters separately close among themeselves
under a T-transformation, only continuous extended characters~(\ref{extcontchar}) do so under S-transformation:\begin{equation}
{\rm Ch}_{c} (\tfrac{1}{2}+ip,M;-\tfrac{1}{\tau}) \oao{a}{b}
= \frac{1}{2k}\int_0^\infty \!\!\! \di p' \, \cos \frac{4\pi p p'}{k}\!\! \sum_{2M' \in \mathbb{Z}_{2k}} \!\!
e^{-\frac{4i\pi M M'}{k}}
{\rm Ch}_{c} (\tfrac{1}{2}+ip',M';\tau) \oao{b}{-a}\,.
\end{equation}
Extended discrete / identity characters in contrast S-transform in a more involved way into combination of extended discrete / identity characters and extended continuous characters
(see~\cite{Eguchi:2003ik,Israel:2004xj}). 

Finally we mention that the characters of continuous representations in the limit $p\to 0^+$ branch into a linear combination
of characters of discrete boundary representations~\cite{Israel:2004xj}:
\begin{equation}\label{boundrep}
\lim_{p \rightarrow 0^+}{\rm Ch}_{\text{c}} \big(\tfrac12+ip,r+\tfrac12;\nu\big|\tau\big) \oao{a}{b} = {\rm Ch}_{\text{d}} 
\big( \tfrac12,r;\nu\big|\tau\big)  \oao{a}{b}+ {\rm Ch}_{\text{d}} \big( \tfrac{k+1}{2},r;\nu\big|\tau\big) 
 \oao{a}{b}\,, \quad r\in \zi_k \,.
\end{equation}

\section{$\boldsymbol{{\mathcal N}=4}$ characters}
\label{appchar2}
\boldmath
\subsection{Classification of unitary representations}
\label{sec:class}
\unboldmath
Unitary representations exist for discrete values of the central charge $c=6\kappa$ ($\kappa\in\en$),
for which the $\mathcal{N}=4$ super-conformal algebra contains a $\widehat{\mathfrak{su}(2)}_{\kappa}$ affine subalgebra. The highest weight
states are distinguished by their eigenvalue with respect to $L_0$: $h$, and their spin $\widehat{\mathfrak{su}(2)}_{\kappa}$ $I$. Unitarity imposes a bound on $h$: namely $h\geq I$ in the \textsc{ns} sector and $h\geq \nicefrac{\kappa}{4}$ in the \textsc{r} sector. Their characters are discussed in~\cite{Eguchi:1987sm,Eguchi:1987wf,eguchi1}. We  summarised hereafter their distinguishing features.

\paragraph{Massive representations:} these representations have an equal number of bosons and fermions in their grounds states, and thus vanishing Witten index. Their characters are defined in terms of two parameters, $\nu$ and $\mu$, related, respectively, to the spin $I$ and the fermion quantum numbers of a given representation. In the
following, we denote $y=\ee^{2\pi i\mu}$ and $z=\ee^{2\pi i\nu}$.
\\[4pt]
\underline{Ramond sector:}
in the \textsc{r} sector, massive representations exist for $h>\frac{\kappa}{4}$ and in the range $\tfrac12\leq I \leq \frac{\kappa}{2}$:
\begin{equation}\label{masscharr}
{\rm ch}^{\textsc{r}}_{\kappa,h,I}(\nu,\mu|\tau)=q^{h-\frac{I^2}{\kappa+1}+\frac18-\frac{c}{24}}\,
F^{\textsc{r}}(\nu,\mu|\tau)\,\chi_{\kappa-1}^{I-\frac12}(2\nu|\tau)\,.
\end{equation}
with $\chi_{\kappa-1}^{I}$ the bosonic $SU(2)_{\kappa-1}$ characters for $I$ spin representation  defined in eq.~(\ref{su2-char})  and the elliptic function:
\begin{equation}
\begin{array}{rcl}
{\ds F^{\textsc{r}}(\nu,\mu|\tau)}&=&{\ds  z \prod_{n=1}^{\infty} \frac{(1+yzq^n)(1+y^{-1}zq^n)(1+yz^{-1}q^{n-1})(1+y^{-1}z^{-1}q^{n-1})}{(1-q^{n})}} \\[6pt]
 &=& {\ds q^{-\frac18}\,\frac{\vartheta_2(\nu+\mu|\tau)\,\vartheta_2(\nu-\mu|\tau)}{\eta(\tau)^3}\,.}
 \end{array}
\end{equation}
\\[4pt]
\underline{Neveu-Schwarz sector:} in the \textsc{ns} sector, we have the bound $h>I$ and the spin is defined in the range: $0\leq I \leq \tfrac{\kappa-1}{2}$. The characters read:
\begin{equation}\label{masscharns}
{\rm ch}^{\textsc{ns}}_{\kappa,h,I}(\nu,\mu|\tau)=q^{h-\frac{(I+1/2)^2}{\kappa+1}+\frac18-\frac{c}{24}}\,
F^{\textsc{ns}}(\nu,\mu|\tau)\,\chi_{\kappa-1}^{I}(2\nu|\tau)\,.
\end{equation}
with:
\begin{equation}
\begin{array}{rcl}
F^{\textsc{ns}}(\nu,\mu|\tau)& = & {\ds \prod_{n=1}^{\infty} \frac{(1+yzq^{n-\frac12})(1+y^{-1}zq^{n-\frac12})
(1+yz^{-1}q^{n-\frac12})(1+y^{-1}z^{-1}q^{n-\frac12})}{(1-q^n)} }\\[6pt]
    & = & {\ds q^{\frac18}\,\frac{\vartheta_3(\nu+\mu|\tau)\,\vartheta_3(\nu-\mu|\tau)}{\eta(\tau)^3}}
\end{array}
\end{equation}

\paragraph{Massless representations:}
these representations saturate the unitary bounds: $h=\kappa/4$ for the \textsc{r} sector and $h=I$ for the \textsc{ns} sector, and preserve $\mathcal{N}=4$ worldsheet supersymmetry. Their ground states have non-vanishing
Witten index. These representations have been proposed as \textsc{cft} T-dual description of (non)-compact manifolds with $c_1(\mathcal{M})=0$ and produce massless supergravity multiplets.
\\[4pt]
\underline{Ramond sector:}
in the \textsc{r} sector, massless representations saturate the bound $h=\frac{\kappa}{4}$ and exist in the range $0\leq I\leq \frac{\kappa}{2}$:
\begin{equation}\label{masscharr2}
{\rm ch}^{\textsc{r}}_{\kappa,h,I}(\nu,\mu|\tau)=q^{h-\frac{I^2}{\kappa+1}+\frac18-\frac{c}{24}}\,
F^{\textsc{r}}(\nu,\mu|\tau)\,\chi_{\kappa-1}^{(\textsc{r})\,I-\frac12}(2\nu|\tau)\,,
\end{equation}
where $\chi_{\kappa}^{(\textsc{r})\,I}$ are modified $SU(2)_\kappa$ characters for the spin $I$ representation, defined as follows:
\begin{multline}
\chi_{\kappa}^{(\textsc{r})\,I}(2\nu|\tau) =   \frac{z q^{-\frac18}}{i\vartheta_1(2\nu|\tau)}
\sum_{m\in \zi}
q^{(\kappa+2)\left(m+\frac{2l+1}{2(\kappa+2)}\right)^2}\\[4pt]
\times \left(
\frac{z^{2(\kappa+2)\left(m+\frac{I}{(\kappa+2)}\right)}}{(1+yzq^{-m})(1+y^{-1}zq^{-m})}  \right.
-
\left. \frac{z^{-2(\kappa+2)\left(m+\frac{I+1}{(\kappa+2)}\right)}}{(1+yz^{-1}q^{-m})(1+y^{-1}z^{-1}q^{-m})}
\right)
\end{multline}
\\[4pt]
\underline{Neveu-Schwarz sector:} in the \textsc{ns} sector, we have the bound $h=I$ and the spin is defined in the range: $0\leq I \leq \tfrac{\kappa}{2}$. The characters read:
\begin{equation}\label{masscharns2}
{\rm ch}^{\textsc{ns}}_{\kappa,h,I}(\nu,\mu|\tau)=q^{l-\frac{(I+1/2)^2}{\kappa+1}+\frac18-\frac{c}{24}}\,
F^{\textsc{ns}}(\nu,\mu|\tau)\,\chi_{\kappa-1}^{(\textsc{ns})\,I}(\tau,2\nu)\,,
\end{equation}
with the modified $SU(2)_\kappa$ characters:
\begin{multline}
\chi_{\kappa}^{(\textsc{ns})\,I}(2\nu|\tau) =  \frac{z q^{-\frac18}}{i\vartheta_1(2\nu|\tau)}
 \sum_{m\in \zi}
q^{(\kappa+2)\left(m+\frac{2I+1}{2(\kappa+2)}\right)^2} \\[4pt]
\times\left(
\frac{z^{2(\kappa+2)\left(m+\frac{I}{(\kappa+2)}\right)}}{(1+yzq^{\left(m+\frac12\right)})(1+y^{-1}zq^{\left(m+\frac12\right)})}\right.
\left.-
\frac{z^{-2(\kappa+2)\left(m+\frac{I+1}{(\kappa+2)}\right)}}{(1+yz^{-1}q^{(\left(m+\frac12\right)})(1+y^{-1}z^{-1}q^{\left(m+\frac12\right)})}
\right)
\end{multline}

\paragraph{Boundary representations:}
similar to what happens for  $\caln_{\textsc{ws}}=2$ characters~(\ref{boundrep}), we observe a reducibility 
of continuous representations when $h$ reaches its unitary bound:
\begin{equation}
\lim_{h\rightarrow I^+} {\rm ch}_{\kappa,h,I}^{\textsc{ns}}\big(\nu,\mu|\tau \big)
= {\rm ch}_{\kappa,I,I}^{\textsc{ns}}\big(\nu,\mu|\tau \big) +(y+y^{-1})\, {\rm ch}_{\kappa,I+\frac12,I+\frac12}^{\textsc{ns}}\big(\nu,\mu|\tau \big)  + {\rm ch}_{\kappa,I+1,I+1}^{\textsc{ns}}\big(\nu,\mu|\tau \big)\,,
\end{equation}
with $y=\ee^{2\pi i\mu}$. A similar relations holds for characters in the Ramond sector:
\begin{equation}
\lim_{h\rightarrow \frac{\kappa}{4}^+} {\rm ch}_{\kappa,h,I}^{\textsc{r}}\big(\nu,\mu|\tau \big)
= {\rm ch}_{\kappa,\frac{\kappa}{4},I}^{\textsc{r}}\big(\nu,\mu|\tau \big) +(y+y^{-1})\, {\rm ch}_{\kappa,\frac{\kappa}{4},I-\frac12}^{\textsc{r}}\big(\nu,\mu|\tau \big)  + {\rm ch}_{\kappa,\frac{\kappa}{4},I-1}^{\textsc{r}}\big(\nu,\mu|\tau \big)\,.
\end{equation}

\boldmath
\subsection{$\caln=4$ characters at level $\kappa=1$, with $c=6$} \label{kappa1}
\unboldmath
\paragraph{Massive representations:}
in this case, the spin and  takes only two values:  $I=0$ in the \textsc{ns} sector  and $I=\tfrac12$ in the \textsc{r} sector, which label representations of the $\widehat{\mathfrak{su}(2)}_1$  subalgebra characterising the $\mathcal{N}_{\textsc{ws}}=4$ super-conformal algebra at level $\kappa=1$. The corresponding characters are:
\begin{equation}
\begin{array}{ll}
{\ds {\rm ch}^{\textsc{ns}}_{1,h,0}(\nu,\mu|\tau) } = q^{h-\frac14}\,F^{\textsc{ns}}(\nu,\mu|\tau)\,, &\quad h>0  \,.\\[4pt]
{\ds {\rm ch}^{\textsc{r}}_{1,h,\frac12}(\nu,\mu|\tau) } = q^{h-\frac14}\,F^{\textsc{r}}(\nu,\mu|\tau)\,, &\quad h>\tfrac14 \,.
\end{array}
\end{equation}
In particular, when setting $\mu=0$ we have:
\begin{equation}\label{twistcont}
\begin{array}{rclcrcl}
{\ds {\rm ch}^{\textsc{ns}}_{1,h,0}(\nu|\tau) } & = &{\ds  q^{h-\frac18}\,\frac{\vartheta_{3}(\nu|\tau)^2}{\eta(\tau)^3}}\,, 
&\qquad \qquad&
{\ds {\ds {\rm ch}^{\textsc{r}}_{1,h,\frac12}(\nu|\tau) } } & =&{\ds   q^{h-\frac38}\,\frac{\vartheta_{2}(\nu|\tau)^2}{\eta(\tau)^3}\,,}
\\[10pt]
{\ds {\rm ch}^{\widetilde{\textsc{ns}}}_{1,h,0}(\nu|\tau) } & = &{\ds  q^{h-\frac18}\,\frac{\vartheta_{4}(\nu|\tau)^2}{\eta(\tau)^3} \,,} 
&\qquad  \qquad&
{\ds {\ds {\rm ch}^{ \tilde{\textsc{r}} }_{1,h,\frac12}(\nu|\tau) } } & =&{\ds  - q^{h-\frac38}\,\frac{\vartheta_{1}(\nu|\tau)^2}{\eta(\tau)^3}\,.}
\end{array}
\end{equation}

\paragraph{Massless representations:}
The spin has two values $I=0,\tfrac12$ for both the \textsc{r} and \textsc{ns} sector.
For $\mu=0$, the $\mathcal{N}=4$ characters simplify considerably.

\noindent
\underline{Normalisable states:}
\begin{equation}\label{massless1}
\begin{array}{rclcrcl}
{\ds {\rm ch}^{\textsc{ns}}_{1,\frac12,\frac12}(\nu|\tau)} & = & {\ds \sum_{n\in\zi}
\frac{q^{\frac12\left(n-\frac12\right)\left(n+\frac12\right)}z^{n}}{1+ zq^{n-\frac12}}\,\frac{\vartheta_{3}(\nu|\tau)}{\eta(\tau)^3} }\,,
&\qquad &
{\ds {\rm ch}^{\textsc{r} }_{1,\frac14,0}(\nu|\tau)} & = & {\ds \sum_{n\in\zi} 
\frac{q^{\frac12n(n+1)}z^{n+\frac12}}{1+ zq^n}\,\frac{\vartheta_{2}(\nu|\tau)}{\eta(\tau)^3} } \,,
  \\[14pt]
{\ds {\rm ch}^{\widetilde{\textsc{ns}}}_{1,\frac12,\frac12}(\nu|\tau)} & = & {\ds \sum_{n\in\zi}
(-)^n
\frac{q^{\frac12\left(n-\frac12\right)\left(n+\frac12\right)}z^{n}}{1- zq^{n-\frac12}}\,\frac{\vartheta_{4}(\nu|\tau)}{\eta(\tau)^3} }
\,,
 &\qquad &
 {\ds {\rm ch}^{\tilde{\textsc{r}} }_{1,\frac14,0}(\nu|\tau)} & = & {\ds
- \sum_{n\in\zi}  (-)^{n}
\frac{q^{\frac12n(n+1)}z^{n+\frac12}}{1- zq^n}\,\frac{i\vartheta_{1}(\nu|\tau)}{\eta(\tau)^3} } \,.
\end{array}
\end{equation}
 
 \noindent
\underline{Non-normalisable states:}
 \begin{equation}\label{massless12}
\begin{array}{rcl}
{\ds {\rm ch}^{\textsc{ns}}_{1,0,0}(\nu|\tau) } & = & {\ds \sum_{n\in\zi} 
q^{\frac12\left(n-\frac12\right)\left(n+\frac12\right)}z^{n} 
\frac{zq^{n-\frac12}-1}{zq^{n-\frac12}+1}
\,\frac{\vartheta_{3}(\tau,\nu)}{\eta(\tau)^3} } \,,
\\[10pt]
{\ds {\rm ch}^{\widetilde{\textsc{ns}}}_{1,0,0}(\nu|\tau) } & = & {\ds \sum_{n\in\zi} 
(-)^nq^{\frac12\left(n-\frac12\right)\left(n+\frac12\right)}z^{n} 
\frac{zq^{n-\frac12}+1}{zq^{n-\frac12}-1}
\,\frac{\vartheta_{4}(\tau,\nu)}{\eta(\tau)^3} } \,,
\\[14pt]
{\ds {\rm ch}^{\textsc{r}}_{1,\frac14,\frac12}(\nu|\tau)} & = & {\ds  \sum_{n\in\zi} 
q^{\frac12n(n+1)}z^{n+\frac12}
\frac{zq^n-1}{zq^n+1}
\,\frac{\vartheta_{2}(\tau,\nu)}{\eta(\tau)^3} } \,,
\\[14pt]
{\ds {\rm ch}^{\widetilde{\textsc{r}}}_{1,\frac14,\frac12}(\nu|\tau)} & = &- {\ds  \sum_{n\in\zi} 
(-)^n q^{\frac12n(n+1)}z^{n+\frac12}
\frac{zq^n+1}{zq^n-1}
\,\frac{i\vartheta_{1}(\tau,\nu)}{\eta(\tau)^3} } \,.

\end{array}
\end{equation}

\section{Some useful material on modular forms}
\label{AppendixMod}

\boldmath
\subsection*{Jacobi $\vartheta$ functions}
\unboldmath

\begin{equation}
\vartheta\oao{a}{b}(\nu|\tau) = \sum_{n\in \zi} q^{\frac12\left(n-\frac a 2 \right)^2}\, 
\ee^{2\pi i\left( n-\frac a 2\right)\left( \nu-\frac b 2\right)} \,, \qquad a,\,b\in \er
\end{equation}
with $q=\ee^{2\pi i\tau}$. In the Jacobi--Erderlyi notation one has : $\vartheta\oao{1}{1}= \vartheta_1$, 
$\vartheta\oao{1}{0}= \vartheta_2$, $\vartheta\oao{0}{0}= \vartheta_3$ and $\vartheta\oao{0}{1}= \vartheta_4$.

\noindent
Their modular transformations read:
\begin{equation}
\begin{array}{rcl}
{\ds \vartheta\oao{a}{b}(\nu|\tau+1)} & =& {\ds  \ee^{-\frac{i\pi}{4}a(a-2)} \, \vartheta\oao{a}{a+b-1}(\nu|\tau)\,,} \\[15pt]
{\ds \vartheta\oao{a}{b}\!\left(\frac{\nu}{\tau}|-\frac{1}{\tau}\right)} & =& {\ds \sqrt{-i\tau}\, \ee^{\frac{i\pi}{2}ab}\,\ee^{i\pi \frac{\nu^2}{\tau}} \, \vartheta\oao{b}{-a}(\nu|\tau)\,.}
\end{array}
\end{equation}
\paragraph{Some useful identities:}
\begin{eqnarray}   \label{c3}
 & \vartheta_2(0|\tau) \vartheta_3(0|\tau) \vartheta_4(0|\tau)  = 2\eta^3(\tau)\,.
\\[4pt] \label{jacid}
\text{Jacobi identity:} \qquad & \vartheta^4_1(\nu|\tau) - \vartheta^4_2(\nu|\tau) +\vartheta^4_3(\nu|\tau)-\vartheta^4_4(\nu|\tau)  = 0\,.
\\[4pt]
\label{kounnas}
& \vartheta_2(0|\tau)^{12}-\vartheta_3(0|\tau)^{12}+\vartheta_4(0|\tau)^{12}= - 48 \eta(\tau)^{12}\,
\end{eqnarray}

\noindent
We now give explicit expressions for the derivatives of the theta-functions with respect to the variable $\nu$. 
\begin{itemize}
\item[$\bullet $] \underline{First  derivatives in $\nu$:}
\begin{equation}
\p_{\nu} \vartheta_1(\nu|\tau)|_{\nu=0} =  \vartheta'_1(\tau) = 2\pi\eta^3(\tau)
\end{equation}
Also
\begin{equation}\label{der1nu1}
\begin{array}{rcl}
{\ds \frac{1}{\pi} \frac{\p}{\p\nu} \!\left( \frac{\vartheta_1(\nu|\tau)}{\vartheta_4(\nu|\tau)} \right) } &=& {\ds \vartheta_4(0|\tau)^2\, \frac{\vartheta_2(\nu|\tau)\vartheta_3(\nu|\tau)}{\vartheta_4(\nu|\tau)^2} \,,} \\[10pt]
{\ds \frac{1}{\pi} \frac{\p}{\p\nu} \!\left( \frac{\vartheta_2(\nu|\tau)}{\vartheta_4(\nu|\tau)} \right) } &=& {\ds - \vartheta_3(0|\tau)^2\, \frac{\vartheta_1(\nu|\tau)\vartheta_3(\nu|\tau)}{\vartheta_4(\nu|\tau)^2} \,,} \\[10pt]
{\ds \frac{1}{\pi} \frac{\p}{\p\nu} \!\left( \frac{\vartheta_3(\nu|\tau)}{\vartheta_4(\nu|\tau)} \right) } &=& {\ds -\vartheta_2(0|\tau)^2\, \frac{\vartheta_1(\nu|\tau)\vartheta_2(\nu|\tau)}{\vartheta_4(\nu|\tau)^2} \,.} 
\end{array}
\end{equation}
From which we deduce:
\begin{equation}\label{der1nu2}
\begin{array}{rcl}
{\ds \frac{1}{\pi} \frac{\p}{\p\nu} \!\left( \frac{\vartheta_1(\nu|\tau)}{\vartheta_2(\nu|\tau)} \right) } &=& {\ds \vartheta_2(0|\tau)^2\, \frac{\vartheta_3(\nu|\tau)\vartheta_4(\nu|\tau)}{\vartheta_2(\nu|\tau)^2} \,,} \\[10pt]
{\ds \frac{1}{\pi} \frac{\p}{\p\nu} \!\left( \frac{\vartheta_1(\nu|\tau)}{\vartheta_3(\nu|\tau)} \right) } &=& {\ds  \vartheta_3(0|\tau)^2\, \frac{\vartheta_2(\nu|\tau)\vartheta_4(\nu|\tau)}{\vartheta_3(\nu|\tau)^2} \,.}
\end{array}
\end{equation}
\item[$\bullet $] \underline{Second derivatives in $\nu$:}
\begin{equation}
\begin{array}{rcccl}
{\ds  \vartheta_2''(\tau) }&=&  {\ds 4\pi i \p_\tau \vartheta_2(\tau) }  &=& {\ds - \frac{\pi^2}{3} \big( \widehat{E}_2 +\vartheta_3^4 + \vartheta_4^4  \big)\,,}\\[6pt]
{\ds  \vartheta_3''(\tau) }&=&  {\ds 4\pi i \p_\tau \vartheta_3(\tau) }  &=& {\ds - \frac{\pi^2}{3} \big( \widehat{E}_2 +\vartheta_2^4 - \vartheta_4^4  \big)\,,}\\[6pt]
{\ds  \vartheta_4''(\tau) }&=&  {\ds 4\pi i \p_\tau \vartheta_4(\tau) }  &=& {\ds - \frac{\pi^2}{3} \big( \widehat{E}_2 -\vartheta_2^4 - \vartheta_3^4  \big)\,.}
\end{array}
\end{equation}
\end{itemize}

\subsection*{Eisenstein series}
An example of weight $2k >2$ holomorphic modular forms is given by the Eisenstein series:
\begin{equation}\label{eis}
E_{2k}(\tau) =  - \frac{(2k)!}{(2\pi i)^{2k} B_{2k}} \sum_{(m,n)\neq (0,0)} \frac{1}{(m\tau +n)^{2k}}\,,
\end{equation}
where $B_{2k}$ are the Bernoulli numbers.
The holomorphic Eisenstein series $E_2$ diverges and is quasi-modular under $S$-transformation, since it is alternatively given by
the following first derivative:
\begin{equation}
E_2(\tau) = \frac{12}{i\pi}\, \p_\tau\log\eta = 1- 24 \sum_{n=1}^\infty \frac{nq^n}{1-q^n}\,.
\end{equation}
It can nonetheless be regularised by a non-holomorphic deformation:
\begin{equation}
\widehat{E}_2(\tau) =  \frac{3}{\pi^2} \lim_{s\rightarrow 0}\sum_{(m,n)\neq (0,0)} \frac{1}{|m\tau+n|^s(m\tau +n)^{2}}\,.
\end{equation}
In the language of Mock modular forms, this corresponds to a non-holomorphic completion of $E_2$ into the weight 2 Maa\ss \ form
$\widehat{E}_2$, whose shadow function $g(\tau)=12/\pi$ is a constant:
\begin{equation}
\widehat{E}_2 =  E_2 -\frac{3}{\pi \tau_2}\,.
\end{equation}
One can express the Eisenstein series in terms of Jacobi functions:
\begin{equation}
\begin{array}{rcl}
E_4 & = & {\ds \frac12\big( \vartheta_2^8 +  \vartheta_3^8 +  \vartheta_4^8\big) =  1+240 \sum_{n=1}^\infty \frac{n^3 q^n}{1-q^n} \,,} \\[10pt]
E_6 & = & {\ds \frac12 \big( \vartheta_2^4 +  \vartheta_3^4\big) \big(\vartheta_3^4 + \vartheta_4^4\big) \big(\vartheta_4^4 - \vartheta_2^4\big) =  1-504 \sum_{n=1}^\infty \frac{n^5 q^n}{1-q^n} \,,} \\[10pt]
E_8 & \doteq & {\ds E_4^2 = \frac12 \big( \vartheta_2^{16} +  \vartheta_3^{16} +  \vartheta_4^{16}\big)
 =  1+480 \sum_{n=1}^\infty \frac{n^7 q^n}{1-q^n} \,,} \\[10pt]
 E_{10} & \doteq & {\ds E_4E_6 =- \frac12 \big( \vartheta_2^{16} (\vartheta_3^4 +  \vartheta_4^4) +  \vartheta_3^{16} (\vartheta_2^4 -  \vartheta_4^4)-  \vartheta_4^{16}(\vartheta_2^4 +  \vartheta_3^4) \big)
 =  1-264 \sum_{n=1}^\infty \frac{n^9 q^n}{1-q^n} \,.}
\end{array}
\end{equation}
Finally, the unique weight 12 cusp form and the Klein invariant are:
\begin{equation}\label{cuspKlein}
\eta^{24} = \frac{E_4^3-E_6^2}{2^6\cdot 3^3}   =  q - 24 q^2 + 252 q^3 + ... \,, \qquad
j= \frac{E_4^3}{\eta^{24}} = \frac1q + 744 + 196884 q +... \,.
\end{equation}

\subsection*{Modular covariant derivative}
We define the covariant derivative $D_r$ which maps a weight $r$ modular form $\Phi_r$ to a weight $r+2$ modular form as:
\begin{equation}\label{covdevapp}
D_{r}  \Phi_{r}(\tau) \doteq \left( \frac{i}{\pi} \frac{\d}{\d\tau} + \frac{r}{2\pi\tau_2}\right)  \Phi_{r}(\tau)\,,
\end{equation}
and satisfies a Leibniz rule: $D_{r+s}(\Phi_r \Phi_s)= \Phi_sD_r\Phi_r + \Phi_r D_s\Phi_s$, for $\Psi_{r+s} = \Phi_r \Phi_s$ a modular form of weight $r+s$.
In particular we have:
\begin{equation}\label{DE}
 D_2\widehat{E}_2 = \frac16 \big( E_4- \widehat{E}_2^2\big) \,,   \qquad  D_4 E_4 = \frac23 \big( E_6- \widehat{E}_2E_4\big) \,, \qquad
 D_6 E_6 =  E_4^2- \widehat{E}_2E_6 \,. \\
\end{equation}
Combining the above:
\begin{equation}\label{DE2}
{\ds D_8 E_8 = \frac43 \big( E_{10}- \widehat{E}_2E_8\big) } \,,  \qquad   {\ds D_{10}E_{10} = 
\frac13 \big( 2E_6^2 +3E_4^3-5\widehat{E}_2E_{10}\big)  \,, }  \qquad 
{\ds D_{\frac{\alpha}{2}}  \eta^{\alpha} = - \frac{\alpha}{12} \widehat{E}_2\eta^{\alpha}\,.} 
\end{equation}
Then, using the last of the above expresssions:
\begin{equation}
D_{-2} \big(  (\sqrt{\tau_2} \bar\eta)^3 \eta^{-1}\big) = - \frac{(\sqrt{\tau_2} \bar\eta)^3 \widehat{E}_2}{12\eta}\,, \qquad
\text{ since }\;
D_{-\frac32} \big( \sqrt{\tau_2} \bar\eta\big)^3 =0\,.
\end{equation}
Also applying the Cauchy--Riemann operator the weight $\big(-\frac32,0\big)$ modular form $\sqrt{\tau_2} \bar\eta\big)^3$ we get:
\begin{equation}\label{DE3}
\begin{array}{rcl}
{\ds \bar \p \big( \sqrt{\tau_2} \bar\eta\big)^3} & = & {\ds  \bar \p \big( \sqrt{\tau_2}^3\big) \,\bar\eta^3  +  \sqrt{\tau_2}^3\,  \bar \p\bar\eta^3 }\\[6pt]
 & = & {\ds \frac{3i}{4}  \sqrt{\tau_2} \bar\eta^3  +  i\pi\sqrt{\tau_2}^3 \left( \overline{D}_{\frac32} \bar\eta^3 -\frac{3\bar\eta^3}{4\pi\tau_2}  \right) }\\[8pt]
 & = & {\ds  -\frac{i\pi}{4} (\sqrt{\tau_2}\bar \eta)^3 \overline{\widehat{E}}_2  \,,}
\end{array}
\end{equation}
which is a weight $\big(-\frac32, 2 \big)$ modular form.

\section{Elliptic genus of the $\boldsymbol{\slr_k/U(1)}$ \textsc{cft}}
\label{sec:ellgen}
We summarise here the computation of the elliptic genus for the $\slr_k/U(1)$ Kazama--Suzuki model (or 
equivalently $\mathcal{N}=(2,2)$ super-Liouville theory), that was done in the work~\cite{jan}. This elliptic genus is 
defined as usual by the trace:
\begin{equation}\label{ell-Liouv0}
\widehat{\mathcal{Z}}_{k}(\nu|\tau)
=
\text{Tr}_{ \mathcal{H}_{\textsc{r}} \otimes \mathcal{H}_{\bar{\textsc{r}}} }\! \left( 
\ee^{i\pi F}
q^{L_0-\frac{c}{24}} 
\bar{q}^{\bar{L}_0-\frac{\tilde{c}}{24}} z^{J^R}
\right)\,.
\end{equation}
where the trace is over the Ramond sector of the Hilbert space  weighted by the worldsheet fermion number operator $F=J_R + J_{\bar R}$, defined from both left- and right-$U(1)$ R-charge currents of the theory. 

For a non-compact \textsc{cft}, we expect the elliptic genus to receive contribution from both localised and non-localised states, and we split~(\ref{ell-Liouv0}) accordingly
\begin{equation}\label{split}
\widehat{\mathcal{Z}}_{k}(\nu|\tau) = \mathcal{Z}^{\text{d}}_{k}(\nu|\tau) + \mathcal{Z}^{\text{c}}_{k}(\nu|\tau)
\end{equation}
where again c and d refer to continous and discrete $\slr_k/U(1)$ representations.

\subsection*{Discrete representations}
The contribution of discrete representation can be straightforwardly computed either  by a free field calculation or by the algebraic method used in the bulk of this work. By this latter method, we obtain the result by summing all {\it extended} discrete characters (as all spectrally flowed Hilbert spaces must be taken into account for consistency) in the twisted Ramond ground state with $r=-1$ over all possible spin values $1/2\leqslant J \leqslant k/2$:~\footnote{Note that boundary representations must be included: we choose here to include the  $J=\frac12$ representation with weight 1 which is equivalent to summing over both  $J=\big\{\frac12,\,\frac{k+1}{2}\big\}$ with weights $\frac12$.}
\begin{equation}\label{ell-Liouv}
\mathcal{Z}^{\text{d}}_{k}(\nu|\tau)
 =\sum_{2J=1}^{k} \text{Ch}_d(J,-1;\tau,\nu\big) \oao{1}{1}
= -\mathcal{K}_{2k}\left(0,\tfrac{\nu}{k}|\tau \right)\frac{i\vartheta_1(\nu|\tau)}{\eta(\tau)^3}\,.
\end{equation}
Note that because of supersymmetry, $\mathcal{Z}^{\text{d}}_{k}$  only depends on left-moving states. We have also repackaged the result into the higher level Appell function, as is usually done~\cite{Eguchi:2004yi}. This function
is define for $\tau\in \mathscr{H}$ and $u,\, v \in \ci$ with $u+v \in \ci/(\zi\tau + \zi)$:
 \begin{equation}\label{Appell-Lerch}
\mathcal{K}_{K}\left(u,v|\tau\right) = \sum_{n\in\zi} \frac{q^{\frac{K}{2}n^2}b^{Kn}}{1-a\,b\, q^n}
\,, \quad
\text{with } \,a= \ee^{2\pi i u}\,, \; b= \ee^{2\pi i v}\,.
\end{equation}
By resorting to the following identity for geometric series:
\begin{equation}
\frac{1}{1-z^{\frac1k}q^m} = \sum_{n=1}^{k-1} \frac{\left(zq^{km}\right)^{\frac{n}{k}}}{1-zq^{km}}\,.
\end{equation}
we can further cast this result into the well known Appell--Lerch sum of level $2k$ seen previously~(\ref{applerch})
\begin{equation}\label{LiApp}
\mathcal{Z}^{\text{d}}_{k}(\nu|\tau)
 = -z^{-1} A_{2k}\big(\tfrac{\nu}{k},2\nu-k\tau\big|\tau\big)\frac{i\vartheta_1(\tau,\nu)}{\eta(\tau)^3}\,.
\end{equation}
At this stage, it becomes quite obvious, from the mathematical perspective, that any proper derivation of the 
elliptic genus~(\ref{ell-Liouv0}) necessary leads to completing~(\ref{LiApp}) into
a Maa\ss\ form. This computation has been carried out~\cite{jan} and will be briefly presented 
in the following.

\paragraph{Continuous representations:}
As shown~\cite{jan}, we can reformulate the elliptic genus~(\ref{ell-Liouv0}) in terms of a path integral, consisting in a Ray--Singer torsion, a twisted fermion partition functions together with twisted bosonic zero modes.
This path integral reads, in both Lagrangian and (after Poisson resumation) Hamiltonian formulation:
\begin{multline}\label{ray-singer}
\widehat{\mathcal{Z}}_{k}(\nu|\tau)  = 
\sum_{m,w}\int_{0}^1\di s_1\int_{0}^1\di s_2\, \frac{\vartheta_1\left(\tau,s_1\tau+s_2-\frac{k+1}{k}\nu\right)}
{\vartheta_1\left(\tau,s_1\tau+s_2-\frac{1}{k}\nu\right)}
\ee^{-\frac{\pi}{k\tau_2}|(m+ks_2)+(w+ks_1)\tau |^2} z^{\frac{n}{k}} \\
=  \sqrt{k\tau_2}\sum_{m,w}\int_{0}^1\di s_1\int_{0}^1\di s_2\, \frac{\vartheta_1\left(\tau,s_1\tau+s_2-\frac{k+1}{k}\nu\right)}
{\vartheta_1\left(\tau,s_1\tau+s_2-\frac{1}{k}\nu\right)} 
q^{\frac{1}{4k}(kw-n-ks_1)^2} \bar{q}^{\frac{1}{4k}(kw+m+ks_1)^2} z^{\frac{m}{k}}e^{-2\pi i ks_2w}
\end{multline}
In particular, the twist in the fermion partition function not only depends on the $R$-charge but is also due the holonomies $s_i$, $i=1,2$ of the gauge fields on the torus, which shift the left- and right-moving momenta $(n-kw)/\sqrt{2k}$ and  $(n+kw)/\sqrt{2k}$. The bosonic zero modes too are twisted by these holonomies  and thereby couple to the oscillators.

At non-zero $\nu$ the path integral~(\ref{ell-Liouv0}) exhibits poles in the $\vartheta$-function in the denominator which are not cancelled by zeros in the numerator, due to the infinite volume divergence of target-space. This divergence was regularised in holomorphic but non-modular invariant way in the partition function~(\ref{partfunc}), in order to recover an expression which
is interpretable in terms of discrete and continuous extended characters of $\slr_k/U(1)$. Following~\cite{jan}, the opposite choice
will be made here, which is more natural from the elliptic genus perspective, as we expect this index to transform as Jacobi form. 

This regularisation procedure first assumes the range  $|q|<|q^{s_1}z^{-\frac1k}|$ and $|z|\sim 1$, for which
we can disentangle the contribution from discrete representations from that  from states with continuous momenta. Ref.~\cite{jan} then shows how the path integral~(\ref{ray-singer}) splits into two pieces, the first one exactly reproducing the localised contribution~(\ref{ell-Liouv}). By this token, we can identify the remainder as the contribution from continuous representations in the following way. After redefining the right moment $n = m+kw$ and introducing a continous momentum variable $p$ to linearise the dependence in $s_1$, we obtain:
\begin{equation}\label{comp}
\mathcal{Z}^{\text{c}}_{k}(\nu|\tau) \equiv
-\frac{i\vartheta_1(\nu|\tau)}{\pi \eta^3} 
\sum_{n,w} q^{kw^2-nw} z^{-2w+\frac{n}{k}}
\int_{\er-i\varepsilon} \frac{\di p}{2ip+n}
|q|^{\frac{2p^2}{k}+\frac{n^2}{2k}} \,.
\end{equation}
In this expression, the right-moving fermionic and bosonic oscillators have cancelled leaving a measure
over the total right momentum. The integral in $p$ over a continuum of states, weighted by the $U(1)_R$-charge and conformal weights, is evidence showing that the above expressions originates from non-localised states in the spectrum. To support this claim, $\mathcal{Z}^{\text{c}}_{k}$ can indeed be rephrased in terms of a combination of  extended {\it left}-moving continuous $\slr_k/U(1)$ characters:
\begin{equation}\label{comp-nl}
\mathcal{Z}^{\text{c}}_{k}(\nu|\tau) =- \frac{i}{\pi}\sum_n \int_{\er-i\varepsilon} \frac{\di p}{2ip+n} \,\text{Ch}_c\left(\tfrac12+ip,\tfrac{n}{2};\nu\big|\tau\right)\oao{1}{1}\,\bar{q}^{\frac{p^2}{k}+\frac{n^2}{4k}}\,.
\end{equation}
Contrary to what happens in the partition function~(\ref{partfunc}) for example, the sum over the left $U(1)_R$
charges labeling these continuous representations is now weighted differently by {\it right}-moving bosonic zero modes with continuous momentum, this particular asymmetric structure clearly resulting from the non-holomorphic nature of $\mathcal{Z}^{\text{c}}_{k}$.

To make final contact with non-holomorphic Appell--Lech sums~(\ref{Appell-comp}), we compute explicitly the
integral over continuous momenta:
\begin{equation}\label{interf}
\int_{\er-i\varepsilon} \frac{\di s}{2ip+n}
|q|^{\frac{2p^2}{k}+\frac{n^2}{2k}} =  -\frac{\pi}{2}\left(\text{sgn}\left(n+\varepsilon\right)-E\Big(n\sqrt{\tfrac{\tau_2}{k}}\Big)\right)\,,
\end{equation}
the first term on the \textsc{rhs} of the equation being the contribution from continuous representations with
momentum $p\rightarrow 0$.

Plugging  expression~(\ref{interf}) back in~(\ref{comp}), we recover the non-holomorphic completion~(\ref{Appell-comp}) for the level $2k$ Appell--Lerch sum~(\ref{LiApp}). By Putting together the contributions from both
discrete~(\ref{LiApp}) and continous~(\ref{comp}) representations and
by further using the elliptic transformations~(\ref{elltrans}), we find the neat expression for the elliptic genus of the level $k$ super-Liouville theory:
\begin{equation}\label{genn}
\widehat{\mathcal{Z}}_{k}(\nu|\tau) = -\hat{A}_{2k}\left(\tfrac{\nu}{k},2\nu\big|\tau\right)\,
\frac{i\theta_1(\tau,\nu)}{\eta(\tau)^3}\,.
\end{equation}

Furthermore, a glance at the modular and elliptic properties of the non-holomorphic Appell--Lerch sums~(\ref{modA}) and~(\ref{elltrans}) shows that the elliptic genus~(\ref{genn}) transforms as  a Jacobi form
of weight 0 and index $\frac{k^2c}{6}=\frac{k(k+2)}{2}$:
\begin{eqnarray}
\widehat{\mathcal{Z}}_{k}\!\left(\frac{\nu}{d\tau+e}\left|\frac{a\tau +b}{d\tau +e}\right.\right) &=& \ee^{2\pi i\frac{d}{d\tau+e}\frac{c}{6}\nu^2} \hat{\mathcal{Z}}_{k}(\nu|\tau)\,, \qquad \text{for } \left(\begin{matrix} a & b
\\ d & e \end{matrix}\right) \in SL(2,\zi)\,, 
\label{ellM}\\[6pt]
\widehat{\mathcal{Z}}_{k}\left(\nu+\lambda\tau + \mu \right|\tau) & = &
\ee^{\pi i\frac{c}{3}(\lambda+\mu)}
q^{-\frac c6 \lambda^2} z^{-\frac c3 \lambda}
 \widehat{\mathcal{Z}}_{k}(\nu|\tau)\,, \quad \text{for }  \lambda,\,\mu\in k\zi\,. \label{ellSL}
\end{eqnarray}
In particular, the modular transformations of $\widehat{\mathcal{Z}}_{k}$ are those expected from the boundary conditions on the path integral and the factorisation of the $U(1)_R$ current algebra. Also,
the fact that the elliptic transformations~(\ref{ellSL}) hold for $\mu\in k\zi$ is a consequence of 
of $1/k$ quantisation of the $U(1)_R$ charges in the \textsc{ns} sector of the theory, while the
restriction $\lambda  \in k\zi$ is related to the expression of the elliptic genus in terms of extended
$\slr_k/U(1)$ characters, which are constructed by summing over $k$ units of spectral flow (see Appendix~\ref{sec:supslc}). This explains why the index of this non-holomorphic Jacobi form is in fact $\nicefrac{k^2c}{6}$ rather
than $\nicefrac{c}{6}$ as one would naively think from the transformations~(\ref{ellM})--(\ref{ellSL})

\section{Details of $\boldsymbol{SO(28)}$ and $\boldsymbol{U(1)_R}$ threshold calculations for $\boldsymbol{\mathcal{Q}_5=k/2}$}
\label{NLT}
\subsection{Bulk state contributions to the $SO(28)$ threshold corrections}
\label{NLT1}
We consider the contribution to the $SO(28)$ threshold corrections coming from continuous $\slr_k/U(1)$ representations, {\it cf.}
eq.~(\ref{nlocSO28}):
\begin{equation}
\mathcal{R}^{\text{c}}[\mathcal{Q}_5] = \frac{1}{16} \sum_{J=1}^{k/2} \big(\chi_{k-2}^{J-1}+ \chi_{k-2}^{\frac k2-J} \big)
\, \frac{1}{\ell}\sum_{m=0}^{\ell-1}e^{-\pi i\frac{m}{\ell}}  q^{-\frac{1}{k}\left(J-\frac{\ell+1}{2}\right)^2}  
R\!\left(
\big( \tfrac{\ell+1}{2}-J\big)\tfrac{\tau}{\ell} +\tfrac{m}{\ell} \big| \tau 
\right)\,
\frac{D_8 E_8}{\eta^{21}} \,.
\end{equation}
In the following we demonstrate how to derive the second line of formua~(\ref{nlocSO28}).
For $\ell \in 2\en +1$ and $J=1,..., \frac k2$, we have:
\begin{subequations}
\begin{align}
\allowdisplaybreaks
&{\ds \frac{1}{\ell}\sum_{m=0}^{\ell-1}e^{-\pi i\frac{m}{\ell}}  q^{-\frac{1}{k}\left(J-\frac{\ell+1}{2}\right)^2}  
R\!\left(
\big( \tfrac{\ell+1}{2}-J\big)\tfrac{\tau}{\ell} +\tfrac{m}{\ell} \big| \tau 
\right) }  & \notag\\[4pt]
& {\ds \quad  =  \sum_{n\in \zi} (-)^n\left[ \frac{1}{\ell}\sum_{m=0}^{\ell-1}e^{-2\pi i(n+1)\frac{m}{\ell}} \right] 
\left( \text{sgn}\big(n+\tfrac12 \big)- E\big(\big[n+1 +\tfrac{1}{2\ell}(1-2J)\big] \sqrt{2\tau_2}\big) \right) q^{-\frac12 (n+1 +\frac{1}{2\ell}(1-2J))^2}}  & \notag
\\[4pt]
& {\ds  \quad =  \sum_{n\in \zi} (-)^n \sum_{r\in \zi} \delta_{n, -1 \text{ mod }\ell }
\left( \text{sgn}\big(n+\tfrac12 \big)- E\big( \big[n+1 +\tfrac{1}{2\ell}(1-2J)\big] \sqrt{2\tau_2} \big) \right) q^{-\frac12 (n+1 +\frac{1}{2\ell}(1-2J))^2}} &
\notag \displaybreak[2]\\[4pt]
& {\ds \quad =  \sum_{r\in \zi} \Big( \underbrace{\text{sgn}\big(r\ell-\tfrac12 \big) - \text{sgn}\big(r\ell + \tfrac{1}{2\ell} (1-2J) \big)}_{=0}+ }
 & \\[4pt] 
& {\ds \left. \hspace{2cm} + (-)^{\ell r-1} \text{sgn}\big(r\ell + \tfrac{1}{2\ell} (1-2J) \big) \,\text{erfc}\left(\big | r\ell + \tfrac{1}{2\ell} (1-2J) \big| \sqrt{2\pi\tau_2} \right)
 \right) q^{-\frac12 ( \ell r +\frac{1}{2\ell} (1-2J))^2}   } & \notag \\[4pt]
 & {\ds \quad =  \sum_{r\in \zi}   (-)^{r+1} \text{sgn}\big(r-\tfrac12 \big) \,\text{erfc}\big(\tfrac{1}{2\ell} | kr+1-2J| \sqrt{2\pi\tau_2} \big)\,
 q^{-\frac{1}{4k} (kr+1-2J)^2}  
 } &  \notag \\[4pt]
 &  {\ds \quad =  \sum_{n\in \zi}   (-)^{n} \text{sgn}\big(n+\tfrac12 \big) \,\text{erfc}\big(\tfrac{1}{2\ell} | kn+1+k-2J| \sqrt{2\pi\tau_2} \big)\,
 q^{-\frac{1}{4k} (kn+1+k-2J)^2}  \,. } & \notag
\end{align}
\end{subequations}
This is the expression appearing on the second line of eq.~(\ref{nlocSO28}).

\subsection{The $U(1)_R$ threshold corrections}
\label{NLT2}
We give the details of the computation of  the second  derivative
\begin{equation}\d_{\nu}^2 \, \widehat{\mu}
\big(\tfrac{\nu}{\ell}+\big(\tfrac{\ell a+1}{2}-J \big) \tfrac{\tau}{\ell} +\tfrac{b-1}{2}+\tfrac{m}{\ell},\tfrac{\nu}{\ell}+\tfrac{(a-1)\tau 
+(b-1)}{2}\big|\tau \big)\big|_{\nu=0}
\end{equation}
 appearing in expression~(\ref{fullthrU1}). 
First we use relation~(\ref{diffmu}), to compute the directional
derivative for $u,\,v \notin \zi \tau + \zi$:
\begin{equation}\label{dermuapp}
\lim_{\varepsilon \rightarrow 0} \frac{\widehat{\mu}(u+\varepsilon, v+\varepsilon)-\widehat{\mu}(u, v) }{\varepsilon} =  -\frac{\eta^3\, \vartheta_1(u+v)}{\vartheta^2_1(u)\,\vartheta^2_1(v)}\, \vartheta'_1(0) = -\frac{2\pi\,\eta^6\, \vartheta_1(u+v)}{\vartheta^2_1(u)\,\vartheta^2_1(v)}\,.
\end{equation}
For $(a,b)\neq (1,1)$, we use~(\ref{dermuapp}) to compute:
\begin{multline}\label{secdermu}
\left.\frac{\p^2}{\p \nu^2} \widehat{\mu}
\!\left(\tfrac{\nu}{\ell}+\big(\tfrac{\ell a+1}{2}-J \big) \tfrac{\tau}{\ell} +\tfrac{b-1}{2}+\tfrac{m}{\ell},\tfrac{\nu}{\ell}+\tfrac{(a-1)\tau 
+(b-1)}{2} \right)  \right|_{\nu=0}
\\
= -\frac{2\pi\eta^6}{\ell^2}\left. \frac{\p}{\p u} \!\left[ 
\frac{\vartheta_1\!\left( 2u +\big(\tfrac{\ell+1}{2}-J \big) \tfrac{\tau}{\ell} +\tfrac{m}{\ell}+(a-1)\tau + (b-1)\right)}
{\vartheta_1^2
\!\left( u +\big(\tfrac{\ell+1}{2}-J \big) \tfrac{\tau}{\ell} +\tfrac{m}{\ell}+\frac{a-1}{2}\tau + \frac{b-1}{2}\right)
\,\vartheta_1^2\!\left(u+\frac{a-1}{2}\tau + \frac{b-1}{2}\right)}
\right] \right|_{u=0}\\
= - \frac{2\pi\eta^6}{\ell^2}\,(-)^{a+b}\, 
\left.\frac{\p}{\p u} \!\left[ 
\frac{\vartheta_1\!\left( 2u +\big(\tfrac{\ell+1}{2}-J \big) \tfrac{\tau}{\ell}+ \frac{m}{\ell} \right)}
{
\vartheta^2\oao{a}{b}
\!\left( u +\big(\tfrac{\ell+1}{2}-J \big) \tfrac{\tau}{\ell}+ \frac{m}{\ell} \right)
\,\vartheta^2\oao{a}{b}\!\left(u\right)
}
\right] \right|_{u=0}
\hspace{3cm}
 \\
=  -\frac{2\pi\eta^6}{\ell^2}\,(-)^{a+b}\, \frac{1}{\vartheta^2\oao{a}{b}\!\left(0\right)}
\left.\frac{\p}{\p u} \!\left[ 
\frac{\vartheta_1\!\left( 2u +\big(\tfrac{\ell+1}{2}-J \big) \tfrac{\tau}{\ell} + \frac{m}{\ell}\right)}
{
\vartheta^2\oao{a}{b}
\!\left( u +\big(\tfrac{\ell+1}{2}-J \big) \tfrac{\tau}{\ell} + \frac{m}{\ell}\right)
}
\right] \right|_{u=0} \,. \hspace{2.8cm}
\end{multline}
In the last line we used the fact that $\vartheta_i'(0)=0$ for $i=2,3,4$.

Notice that for $\delta \notin \left(\frac{a-1}{2} +\zi\right) \tau + \frac{b-1}{2} + \zi $, we have:
\begin{equation}\label{dermuid}
\left.\frac{\p}{\p u} \!\left[ 
\frac{\vartheta_1(2u +\delta)}{\vartheta\oao{a}{b}^2(u +\delta)} \right] \right|_{u=0} = \frac{2}{\vartheta\oao{a}{b}(\delta)}
\left.\frac{\p}{\p v} \!\left[ 
\frac{\vartheta_1(v)}{\vartheta\oao{a}{b}(v)} \right] \right|_{v=\delta}  = \frac{2\pi \,\vartheta^2\oao{a}{b}(0)\, \vartheta\oao{a+1}{0}(\delta)\,\vartheta\oao{0}{b+1}(\delta)}{\vartheta\oao{a}{b}^3(\delta)} \,.
\end{equation}
To obtain the final identity we used~(\ref{der1nu1}) and~(\ref{der1nu2}).

Plugging~(\ref{dermuid}) into~(\ref{secdermu}), we get for $(a,b)\neq (1,1)$:
\begin{multline}
\left.\frac{\p^2}{\p \nu^2} \widehat{\mu}
\!\left(\tfrac{\nu}{\ell}+\big(\tfrac{\ell a+1}{2}-J \big) \tfrac{\tau}{\ell} +\tfrac{b-1}{2}+\tfrac{m}{\ell},\tfrac{\nu}{\ell}+\tfrac{(a-1)\tau 
+(b-1)}{2} \big| \tau\right)  \right|_{\nu=0} \\
=
-\frac{8\pi\eta^6}{k}\,(-)^{a+b}\,
\frac{\vartheta\oao{a+1}{0}\!\left(\big(\tfrac{\ell+1}{2}-J \big) \tfrac{\tau}{\ell}+ \frac{m}{\ell} |\tau\right)\,\vartheta\oao{0}{b+1}\!\left(\big(\tfrac{\ell+1}{2}-J \big) \tfrac{\tau}{\ell} + \frac{m}{\ell}|\tau\right)}{\vartheta\oao{a}{b}^3\!\left(\big(\tfrac{\ell+1}{2}-J \big) \tfrac{\tau}{\ell}+ \frac{m}{\ell} |\tau\right)}
\,,
\end{multline}
which is the expression appearing on the second line of~(\ref{fullthrU1}).


\section{$\boldsymbol{SO(28)}$ threshold corrections for $\boldsymbol{\mathcal{Q}_5 = 1}$}
\label{Th}
Hereafter, we give the detailed evaluation of the threshold correction~(\ref{thresh5}):
\begin{equation}
\Lambda_{SO(28)}[1] = \frac{1}{8} \int_{\mathscr{F}} \frac{\di^2\tau}{\tau_2}\,\Gamma_{2,2}(T,U)\,\hat{\mathcal{A}}_{SO(28)}
\end{equation}
using the orbit method.

\subsection{Zero orbit} 
\label{Th1}
The zero orbit contribution reads:
\begin{equation}\label{zeor}
\Lambda_{SO(28)} = \frac{T_2}{8} \int_{\mathscr{F}} \frac{\di^2\tau}{\tau_2^2}\,\hat{\mathcal{A}}_{SO(28)}\,.
\end{equation}
To evaluate it, we split the modified elliptic genus~(\ref{thresh5})
\begin{equation}
\hat{\mathcal{A}}_{SO(28)} =  \frac{1}{6}\big( \hat{\mathcal{A}}_\text{K3}[4] +  \hat{\mathcal{A}}_\text{flux} \big)\,,
\end{equation}
in terms of the modified elliptic genus of $K3$~(\ref{refthresh}) and the flux contribution:
\begin{equation}\label{rew}
\hat{\mathcal{A}}_\text{K3}[t] = -\frac{1}{12\eta^{24}} \Big[\widehat{E}_2 E_{10}-\frac{12+t}{24}E_6^2 - \frac{12-t}{24} E_4^3\Big]\,, \qquad
\hat{\mathcal{A}}_\text{flux} = \frac{\widehat{F}\big(\widehat{E}_2E_8 -E_{10}\big)}{12\eta^{21}}\,,
\end{equation}
both of which are separately modular invariant.

\paragraph{Stokes theorem for modular integrals:} to compute integrals of the type:
\begin{equation}
I_r(\Phi)=\int_{\mathscr{F}} \frac{\di^2\tau}{\tau_2^2}\, (\widehat{E}_2)^r \,\Phi(\tau)
\end{equation}
for $\Phi$ a holomorphic modular form of weight $w=-2r$, we use the fact that:
\begin{equation}
\d_{\bar\tau} \widehat{E}_2 = \frac{3i}{2\pi \tau_2^2}
\end{equation}
and that the measure on $\mathscr{H}$ can be rewritten as $\di\tau_1 \wedge \di\tau_2 = \tfrac{1}{2i}\di\tau \wedge \di\bar\tau$, to recast the integral as follows~\cite{borcherds}:
\begin{eqnarray}\label{Jint}
I_r(\Phi) & = & {\ds - \frac{\pi}{3(r+1)} \int_{\mathscr{F}} \di\tau\, \di\bar\tau \, \d_{\bar\tau}\big( (\widehat{E}_2)^{r+1} \Phi \big)=  
 - \frac{\pi}{3(r+1)} \lim_{w \rightarrow \infty}\int_{\mathcal{F}_w} \di\big( (\widehat{E}_2)^{r+1} \Phi\, \di\tau \big) } 
 \notag \\[6pt]
&  = &  {\ds \frac{\pi}{3(r+1)}  \lim_{w \rightarrow \infty} \int_{\tau_1
=- \frac12+iw}^{\tau_1= \frac12+iw} (\widehat{E}_2)^{r+1} \Phi\ =  \frac{\pi}{3(r+1)} 
  \lim_{\tau_2 \rightarrow \infty} (\widehat{E}_2)^{r+1} \Phi |_{q^0}}   \\[6pt]
 &=&  \frac{\pi}{3(r+1)}  \big(  \text{constant term of } (E_2)^{r+1} \Phi \big) \notag
\end{eqnarray}
where we have used Stoke's theorem in the first line, and we have introduced a cutoff
on the fundamental domain: $\mathcal{F}_w = \big\{ |\tau_1| \leqslant \tfrac12 , \, |\tau| \geqslant 1, \, 0  \leqslant \tau_2 \leqslant w\big\}$.

For instance, for a holomorphic modular form with expansion $Q(\tau)=  \sum_{n=-1}^{\infty} c_n q^n$ we have:
\begin{equation}\label{formIr}
I_r(Q) = \frac{\pi}{3(r+1)}  \big( c_0 -24(r+1) c_{-1} \big)\,.
\end{equation}

\boldmath
\subsubsection*{$K3$ contribution}
\unboldmath
To compute the modular integral~(\ref{zeor}) for the K3 contribution~(\ref{rew}), we rewrite:
\begin{equation}
\hat{\mathcal{A}}_\text{K3}[4] = -\frac{1}{12} \left( \widehat{E}_2 Q_1 - Q_2  \right)
\end{equation}
in terms of modular forms with at most a pole of order one:
\begin{equation}
\begin{array}{rcl}
Q_1(\tau) & = & {\ds  \frac{E_{10}}{\eta^{24}} = \frac1q - 240 - 141444 q - 8529280 q^2 + \mathcal{O}(q^3)\,,} \\[10pt]
Q_2(\tau) & = & {\ds  \frac{2 E_6^2 + E_4^3}{3\eta^{24}}  = \frac1q - 408 + 196884 q + 21493760 q^2 + \mathcal{O}(q^3)\,.} 
\end{array}
\end{equation}
The $K3$ contribution to the zero orbit integral can now be computed by formula~(\ref{formIr}):
\begin{equation}\label{lam0K3}
\Lambda_0^{K3} = \frac{T_2}{48} \int_{\mathscr{F}} \frac{\di^2\tau}{\tau_2^2}\,\hat{\mathcal{A}}_{K3}[4] = 
-\frac{T_2}{576} \,\big( I_1(Q_1) - I_0(Q_2) \big)  = -\frac{\pi}{6} T_2 \,.
\end{equation}
\boldmath
\subsubsection*{Flux contribution}
\unboldmath
Next, the flux contribution to the zero orbit integral reads:
\begin{equation}\label{0flux}
\Lambda_0^\text{flux} = \frac{T_2}{48} \int_{\mathscr{F}} \frac{\di^2\tau}{\tau_2^2}\,\hat{\mathcal{A}}_\text{flux} = \frac{T_2}{576} \, I_\text{flux} 
\,,
\end{equation}
in terms of the integral:
\begin{equation}\label{intI}
I_\text{flux}  = \int_{\mathscr{F}} \frac{\di^2\tau}{\tau_2^2}\, \widehat{F} \left(  \frac{\widehat{E}_2 E_8 - E_{10}}{\eta^{21}} \right) \,.
\end{equation}
The integrand is a weight 0 Maa\ss\, form, as $\widehat{F}$ has weight $\nicefrac12$: 
\begin{equation}
\widehat{F}(\tau) = F(\tau) - 6 R(\tau)
\end{equation}
with a Mock modular piece with Fourier expansion:
\begin{equation}
q^{1/8} F(\tau) =  4q^{1/8} \eta(\tau) \sum_{i=2}^4 h_i(\tau) = 1-45q-231q^2 -770q^3 -2277q^4 +\mathcal{O}(q^5)\,
\end{equation}
and a non-holomorphic completion  given by:
\begin{equation}
q^{1/8}R(\tau)= 2 \sum_{n=0}^\infty (-)^n\text{erfc}\big((n+\tfrac12)\sqrt{2\pi\tau_2}\big)   q^{-\frac12n(n+1)}\,.
\end{equation}
The shadow of $F$ is the same as for $12\,\hat{\mu}(\tau,\nu)$,  the non-holomorphic completion of the Appell function.
More precisely, we have:
\begin{equation}\label{shad}
\d_{\bar\tau} \widehat{F}(\tau) = 3\sqrt{2} i \frac{\overline{{\eta}(\tau)}^3}{\sqrt{\tau_2}}\,.
\end{equation}
\paragraph{Computation of $I_f$:}
we use the pocedure outlined above to compute the integral~(\ref{intI}). We rewrite
\begin{equation}\label{outl}
I_\text{flux} =\frac{\pi}{3i} \int_{\mathscr{F}}  \di^2\tau\, \hat F \,\frac{\d}{\d\bar\tau} \left( \frac{(\widehat{E}_2)^2 E_8 -2 \widehat{E}_2 E_{10}}{\eta^{21}} \right)  = I'_\text{flux}  + I''_\text{flux} \,,
\end{equation} 
where the $I_i$ are obtained by integration by parts. The first one is easily computed by using the procedure~(\ref{Jint}):
\begin{equation}\label{I1}
 I'_\text{flux}  = \frac{\pi}{3i} \int_{\mathscr{F}}  \di^2\tau\, \frac{\d}{\d\bar\tau} \left[ \widehat{F} \,\Big( \frac{(\widehat{E}_2)^2 E_8 -2 \widehat{E}_2 E_{10}}{\eta^{21}} \Big)  \right]
  =   \frac{\pi}{6} \left[  F \,\Big( \frac{(E_2)^2 E_8 -2 E_2 E_{10}}{\eta^{21}} \Big) \right]_{q^0}  =  172 \pi
\end{equation}
where we have used 
\begin{equation}
\begin{array}{lcr}
 {\ds \frac{F E_2 E_{10}}{\eta^{21}}  } & = &  {\ds  \frac{1}{q} - 312 - 123180 \,q + 1424120 \,q^2 +\text{O}\big(q^3\big) } \,,\\[10pt]
 {\ds \frac{F(E_2)^2 E_8}{\eta^{21}} } & = &  {\ds \frac{1}{q}  + 408 + 28020 \,q - 3068680 \, q^2 +\text{O}\big(q^3\big) } \,.
 \end{array}
\end{equation}
The second integral $I_2$ requires more care. Using the definition of the shadow~(\ref{shad}), it can be cast into the following form:
\begin{equation}\label{I2}
\begin{array}{rcl}
 I''_\text{flux} &  = & {\ds  -\frac{\pi}{3i} \int_{\mathscr{F}}  \di^2\tau\, \frac{\d  \widehat{F} }{\d\bar\tau} \,\Big( \frac{(\widehat{E}_2)^2 E_8 -2 \widehat{E}_2 E_{10}}{\eta^{21}} \Big) } \\[10pt]
  & = & {\ds    - \sqrt{2} \pi \int_{\mathscr{F}}  \frac{\di^2\tau}{\tau_2^2} \, \big(\sqrt{\tau_2} \eta\bar\eta \big)^3  \,\Big( \frac{(\widehat{E}_2)^2 E_8 -2 \widehat{E}_2 E_{10}}{\eta^{24}} \Big) =  40\pi \,,}
  \end{array}
  \end{equation}
 this integral having been computed in~\cite{Angelantonj:2012gw}, eq.(4.25) therein, by a method developed in~\cite{Angelantonj:2011br}.
 As outlined in section~\ref{sec:moddep}, this yields the result:
 \begin{equation}\label{lam0flux}
  I_\text{flux} = 212\pi \quad \Longrightarrow \quad   \Lambda_0^\text{flux} =  \frac{53\pi}{144}\, T_2\,.
 \end{equation}
 \boldmath
\subsubsection*{Zero orbit result}
\unboldmath
 Putting together (\ref{lam0K3}) and (\ref{lam0flux}) we obtain:
 \begin{equation}\label{fflux}
 \Lambda_{SO(28)}^0 = \Lambda_0^{K3} + \Lambda_0^\text{flux}  = \frac{29\pi}{144}\, T_2 \,.
 \end{equation}
 
\subsection{Degenerate orbits}
\label{sec:deg}
The contribution from degenerate orbits is evaluated over the strip $\mathscr{S} = \big\{ -\frac12 \leqslant \tau_1| < 1/2, \,\tau_2 \geqslant 0\big\}$:
\begin{equation}\label{deg1}
\Lambda_{SO(28)}^\text{deg} =  \frac{T_2}{8}\int_{\mathscr{S}} \frac{\di^2\tau}{\tau_2^2} \sum_{(j,p)\neq (0,0)}
        \ee^{-\frac{\pi T_2}{\tau_2 U_2}|j+pU|^2}\hat{\mathcal{A}}_{SO(28)}\,.
\end{equation}
To compute this integral we now choose to decompose the modified elliptic genus according to discrete and continuous $\slr_k/U(1)$ representations:
\begin{equation}\label{deg2}
\hat{\mathcal{A}}_{SO(28)} =  \hat{\mathcal{A}}_{SO(28)}^\text{d} + \mathcal{R}^\text{c} \,.
\end{equation}
The contribution from the discrete spectrum of states can be expanded as:
\begin{equation}\label{deg3}
\begin{array}{rcl}
\hat{\mathcal{A}}_{SO(28)}^\text{d} & = & {\ds -\frac{1}{72\,\eta^{24}} \Big[\widehat{E}_2 E_{10}-\tfrac{2}{3}E_6^2 - \tfrac13 E_4^3
-\eta^3F(\widehat{E}_2 E_8- E_{10})\Big] \,,} \\[12pt]
 & = & {\ds  8 - \frac{29}{\pi\tau_2} + \text{O}(q)\,.}
\end{array}
\end{equation} 
Notice that there is a subtle cancellation so that  this expression does even have a 'dressed' pole $(\tau_2 q)^{-1}$, unlike $K3$ models.
Defining the Fourier expansion:
\begin{equation}\label{fexp}
\frac{E_2 E_8 -E_{10} }{q^{1/8} \eta^{21}} = \sum_{n=0}^\infty d_1(n)\, q^n\,, \qquad
\frac{E_8}{q^{1/8} \eta^{21}} = \sum_{n=-1}^\infty d_2(n)\, q^n\,.
\end{equation}
the 'zero mode' contribution in $ \mathcal{R}^\text{c}$ reads:
\begin{equation}\label{Rexp}
 \left. \mathcal{R}^\text{c} \right|_{q^0} = -\frac{1}{6} \sum_{n=0}^\infty \text{erfc}\big((n+\tfrac12)\sqrt{2\pi\tau_2}\big)  
\left( d_1\!\big(\tfrac{n(n+1)}{2}\big) -\frac{3}{\pi\tau_2}  d_2\!\big(\tfrac{n(n+1)}{2}\big) \right)
 \,.
\end{equation}
When we integrate~(\ref{deg1}) first over $\tau_1$, only the  $\text{O}(q^0)$ contributions~(\ref{deg3}) and~(\ref{Rexp}) survive. Then using uniform and absolute
convergence of the exponential sum we obtain:
\begin{equation}\label{deg4}
\begin{array}{rcl}
\Lambda_{SO(28)}^\text{deg} 
        &=& {\ds  \frac{T_2}{8} \sum_{(j,p)\neq (0,0)} \int_0^\infty  \frac{\di\tau_2}{\tau_2^2}
       \, \ee^{-\frac{\pi T_2}{\tau_2 U_2}|j+pU|^2}   \left(8 - \frac{29}{\pi\tau_2} \,+  \right. } \\
     && {\ds \left.   -\,\frac{1}{6} \sum_{n=0}^\infty \text{erfc}\big((n+\tfrac12)\sqrt{2\pi\tau_2}\big)  
\left[ d_1\!\big(\tfrac{n(n+1)}{2}\big) -\frac{3}{\pi\tau_2}  d_2\!\big(\tfrac{n(n+1)}{2}\big) \right]
       \right) \,. }
    \end{array}
\end{equation}
We can evaluate the first line of~(\ref{deg4}) by using the integrals:
\begin{equation}\label{EE}
\sum_{(j,p)\neq (0,0)} \int_{0}^\infty   \frac{\di^2\tau}{\tau_2^{1+r}}\,
        \ee^{-\frac{\pi T_2}{\tau_2 U_2}|j+pU|^2}  = \frac{2\zeta(2k) \Gamma(k)}{(\pi T_2)^r} \, E(U,r)\,.
\end{equation}
Combining the ensuing $E(U,1)$ and $E(U,2)$ contributions we get from~(\ref{deg4})
we obtain the second and third terms on the first line of~(\ref{inthresh}), which are the degenerate orbit
contributions of localised states.
Notice that the real analytic Eisenstein series $E(U,1)$ needs to be regularised:
\begin{equation}
E(U,1) = -\tfrac{3}{\pi} \left(\log|\eta(U)|^4  + \log\big( \mu^2\, T_2\, U_2 \big) \right) + \gamma
\end{equation}
with $\mu^2$ an \textsc{ir} regulator and $\gamma$ a renormalisation scheme dependent constant.
This is expression we use in~(\ref{inthresh}).

The bulk state contributions on the second line of~(\ref{EE}) is  treated in details in~(\ref{cdeg})--(\ref{cdegfin}) et seq. .

\subsection{Non-degenerate orbits}
\label{sec:nondeg}

We compute the non-degenerate orbit contribution to the $SO(28)$ threshold:
\begin{equation}\label{thcm}
\Lambda_{SO(28)}^{\text{non-deg}} = 2    \int_{\mathscr{H}} \frac{\di^2\tau}{\tau_2^2}\sum_{k>j\geq 0}\sum_{p\neq 0}
        e^{2\pi kpi T-\frac{\pi T_2}{\tau_2 U_2}|k\tau-j-pU|^2} 
        \hat{\mathcal{A}}_{SO(28)} \,.
\end{equation}
The holomorphic part of the above modified elliptic genus is defined in terms of the following Fourier expansions:
\begin{equation}
\begin{array}{rcl}
{\ds 
-\frac{1}{12\,\eta^{24}} \Big[E_2 E_{10} -\tfrac{2}{3}E_6^2 - \tfrac13 E_4^3
-\eta^3F ( E_2 E_8 - E_{10})\Big]  } &=& {\ds \sum_{n=0}^\infty c_1(n) q^n\,, } \\[10pt]
{\ds  \frac{E_{10}- \eta^3 F E_8 }{\eta^{24}} }&=& {\ds  \sum_{n=0}^\infty c_2(n) q^n\,.}
\end{array}
\end{equation}
Expanding the complementary error function in powers of $\tau_2$:
\begin{equation}
\text{erfc}\big((n+\tfrac12)\sqrt{2\pi\tau_2}\big)=   1-\sum_{m=0}^{\infty}(-)^m e_{n,m}\,\tau_2^{m+\frac12}\,, \qquad \text{with }\;
 e_{n,m}= \frac{\pi^m}{m!}\frac{\left(\sqrt{2}(n+\frac12)\right)^{2m+1}}{m+\frac12}\,,
\end{equation}
and performing the Gaussian integral over $\tau_1$ in~(\ref{thcm}) we get:
\begin{multline}\label{Insttot}
\Lambda_{SO(28)}^{\text{non-deg}} = \frac{\sqrt{T_2 U_2}}{24} \sum_{k>j\geq 0}\frac1k \sum_{p\neq 0} \ee^{2\pi i kp T_1} 
 \int_{0}^\infty \frac{\di \tau_2}{\tau_2^{3/2}}  \\
\left( \sum_n   \ee^{2\pi i n (\frac{j+pU_1}{k})}
 \left[ c_1(n)- \frac{c_2(n)}{4\pi\tau_2}\right] 
\ee^{-\frac{\pi  T_2}{U_2} \left( k+\frac{n}{k}\frac{U_2}{T_2}\right)^2\tau_2 -\frac{\pi p^2T_2U_2}{\tau_2}}  \, -\right. \\
-  \,\sum_{n=0}^\infty  (-)^n \sum_m 
\ee^{2\pi i (m-\frac12n(n+1)) (\frac{j+pU_1}{k})}
 \left[ d_1(m)-\frac{3}{\pi \tau_2}d_2(m) \,-  \right.\\
\left.\left. \,-\sum_{l=0}^\infty e_{n,l} \Big( d_1(m) \tau_2^{l+\frac12} -\frac{3}{\pi} d_2(m) \tau_2^{l-\frac12}\Big) \right] 
\ee^{-\frac{\pi  T_2}{U_2} \left( k+\frac{m-\frac12 n(n+1)}{k}\frac{U_2}{T_2}\right)^2\tau_2 -\frac{\pi p^2T_2U_2}{\tau_2}} 
\right)
\end{multline}
To evaluate the first two lines of the integral in~(\ref{Insttot}), we use the integrals:
\begin{equation}
\begin{array}{rcl}
J_r(a,b) & = &{\ds  \int_0^\infty \frac{\di x}{x^{\frac12+r}}\, \ee^{-ax-\frac{b}{x}} =  \left( -\frac{\p}{\p b}\right)^r J_0(a,b)  } \\[10pt]
 & = & {\ds   \sqrt{\pi}\left( -\frac{\p}{\p b}\right)^r 
\frac{\ee^{-2\sqrt{ab}}}{\sqrt{a}} = 2 \left(\frac{a}{b} \right)^{\frac r2 - \frac14} K_{r-\frac12} (2\sqrt{ab})\,,
\quad r \in \en\,, \quad \text{Re}\,a\,,\; \text{Re}\,b >0
}
\end{array}
\end{equation}
given in terms of the modified Bessel functions of the second kind $K_\alpha(z)$ defined~(\ref{Bess2}).
The integrals of interest here are in particular:
\begin{equation}\label{exbess}
J_1(a,b) = \sqrt{\frac\pi b} \,\ee^{-2\sqrt{ab}}\,, \qquad  J_2(a,b) = \left(\frac{1}{2b} + \sqrt{\frac ab} \right)\sqrt{\frac\pi b} \,\ee^{-2\sqrt{ab}}\,.
\end{equation}

\paragraph{Contribution of discrete representations:} the first line of~(\ref{Insttot}) gives the localised contribution to the non-degenerate orbit integral,
which can be evaluated using~(\ref{exbess}):
\begin{equation}
\begin{array}{rcl}
{\ds \Lambda_{SO(28)}^{\text{non-deg} \,\text{d}} } & = & {\ds  \frac{1}{ 24} 
 \sum_{k>j\geq 0} \sum_{p > 0} \frac{1}{kp} \ee^{2\pi i kp T} \left(   \sum_n \left[ c_1(n) -\frac{k}{4\pi p U_2} c_2(n) \right] \ee^{2\pi n \mathcal{U}} - 
 \right.  }\\[12pt]
 && {\ds \left. \hspace{2cm}- \frac{1}{kp T_2}   \sum_n  \frac{1}{4\pi} \left[ n  + \frac{k}{2\pi p U_2} \right] c_2(n)\,  \ee^{2\pi n \mathcal{U}}
 \right) + \text{c.c.} } \\[12pt]
& = & {\ds  \frac14 \sum_{k>j\geq 0} \sum_{p > 0} \frac{1}{kp} \ee^{2\pi i \mathcal{T}}  \left(\hat{\mathcal{A}}_{SO(28)}^{\text{d}}(\mathcal{U}) +
 \frac{1}{\pi\mathcal{T}_2} \Box \hat{\mathcal{A}}_{SO(28)}^{\text{d}}(\mathcal{U}) \right) + \text{c.c.} }
\end{array}
\end{equation}
with the modular invariant operator:
\begin{equation}
\Box \equiv  \mathcal{U}_2^2 \p_{\mathcal{U}}  \bar \p_{\mathcal{U}} \,.
\end{equation}
\paragraph{Contribution of continuous representations:} the last two lines of~(\ref{Insttot}) are the contributions of bulk states to
the non-degenerate orbit integral. They can be evaluated using~(\ref{exbess}) for the second line while using
\begin{equation}\label{BessInst} 
 \int_0^\infty \frac{\di x}{x^{1-r}}\, \ee^{-ax-\frac{b}{x}} = 2 \left(\frac{b}{a}\right)^{\frac r2} K_r(2\sqrt{ab}) =   -2\left(\frac{\p}{\p a}\right)^r K_0(2\sqrt{ab})\,,
\end{equation}
for the third line of eq.~(\ref{Insttot}). This leads to expression~(\ref{l2}) discussed earlier.

\bibliographystyle{JHEP}
\bibliography{bibbundle2}

\providecommand{\href}[2]{#2}\begingroup\raggedright\begin{thebibliography}{100}

\bibitem{Gross:1985fr}
D.~J. Gross, J.~A. Harvey, E.~J. Martinec, and R.~Rohm, {\it {Heterotic String
  Theory. 1. The Free Heterotic String}},  {\em Nucl. Phys.} {\bf B256} (1985)
  253.

\bibitem{Green:1984sg}
M.~B. Green and J.~H. Schwarz, {\it {Anomaly Cancellation in Supersymmetric
  D=10 Gauge Theory and Superstring Theory}},  {\em Phys. Lett.} {\bf B149}
  (1984) 117--122.

\bibitem{Strominger:1986uh}
A.~Strominger, {\it {Superstrings with Torsion}},  {\em Nucl. Phys.} {\bf B274}
  (1986) 253.

\bibitem{Hull:1986kz}
C.~M. Hull, {\it {Compactifications of the heterotic superstring}},  {\em Phys.
  Lett.} {\bf B178} (1986) 357.

\bibitem{Giddings:2001yu}
S.~B. Giddings, S.~Kachru, and J.~Polchinski, {\it {Hierarchies from fluxes in
  string compactifications}},  {\em Phys. Rev.} {\bf D66} (2002) 106006,
  [\href{http://xxx.lanl.gov/abs/hep-th/0105097}{{\tt hep-th/0105097}}].

\bibitem{Klebanov:2000hb}
I.~R. Klebanov and M.~J. Strassler, {\it {Supergravity and a confining gauge
  theory: Duality cascades and chiSB-resolution of naked singularities}},  {\em
  JHEP} {\bf 08} (2000) 052,
  [\href{http://xxx.lanl.gov/abs/hep-th/0007191}{{\tt hep-th/0007191}}].

\bibitem{Adams:2006kb}
A.~Adams, M.~Ernebjerg, and J.~M. Lapan, {\it {Linear models for flux vacua}},
  {\em Adv. Theor. Math. Phys.} {\bf 12} (2008) 817--851,
  [\href{http://xxx.lanl.gov/abs/hep-th/0611084}{{\tt hep-th/0611084}}].

\bibitem{Adams:2009zg}
A.~Adams and J.~M. Lapan, {\it {Computing the Spectrum of a Heterotic Flux
  Vacuum}},  {\em JHEP} {\bf 1103} (2011) 045,
  [\href{http://xxx.lanl.gov/abs/0908.4294}{{\tt arXiv:0908.4294}}].

\bibitem{McOrist:2010ae}
J.~McOrist, {\it {The Revival of (0,2) Linear Sigma Models}},  {\em
  Int.J.Mod.Phys.} {\bf A26} (2011) 1--41,
  [\href{http://xxx.lanl.gov/abs/1010.4667}{{\tt arXiv:1010.4667}}].

\bibitem{Quigley:2011pv}
C.~Quigley and S.~Sethi, {\it {Linear Sigma Models with Torsion}},  {\em JHEP}
  {\bf 1111} (2011) 034, [\href{http://xxx.lanl.gov/abs/1107.0714}{{\tt
  arXiv:1107.0714}}].

\bibitem{Blaszczyk:2011ib}
M.~Blaszczyk, S.~Nibbelink~Groot, and F.~Ruehle, {\it {Green-Schwarz Mechanism
  in Heterotic (2,0) Gauged Linear Sigma Models: Torsion and NS5 Branes}},
  {\em JHEP} {\bf 1108} (2011) 083,
  [\href{http://xxx.lanl.gov/abs/1107.0320}{{\tt arXiv:1107.0320}}].

\bibitem{Blaszczyk:2011hs}
M.~Blaszczyk, S.~Groot~Nibbelink, and F.~Ruehle, {\it {Gauged Linear Sigma
  Models for toroidal orbifold resolutions}},  {\em JHEP} {\bf 1205} (2012)
  053, [\href{http://xxx.lanl.gov/abs/1111.5852}{{\tt arXiv:1111.5852}}].

\bibitem{Quigley:2012gq}
C.~Quigley, S.~Sethi, and M.~Stern, {\it {Novel Branches of (0,2) Theories}},
  {\em JHEP} {\bf 1209} (2012) 064,
  [\href{http://xxx.lanl.gov/abs/1206.3228}{{\tt arXiv:1206.3228}}].

\bibitem{Adams:2012sh}
A.~Adams, E.~Dyer, and J.~Lee, {\it {GLSMs for non-Kahler Geometries}},
  \href{http://xxx.lanl.gov/abs/1206.5815}{{\tt arXiv:1206.5815}}.

\bibitem{Melnikov:2010pq}
I.~V. Melnikov and R.~Minasian, {\it {Heterotic Sigma Models with N=2
  Space-Time Supersymmetry}},  {\em JHEP} {\bf 1109} (2011) 065,
  [\href{http://xxx.lanl.gov/abs/1010.5365}{{\tt arXiv:1010.5365}}].

\bibitem{Dasgupta:1999ss}
K.~Dasgupta, G.~Rajesh, and S.~Sethi, {\it {M theory, orientifolds and
  G-flux}},  {\em JHEP} {\bf 08} (1999) 023,
  [\href{http://xxx.lanl.gov/abs/hep-th/9908088}{{\tt hep-th/9908088}}].

\bibitem{Fu:2006vj}
J.-X. Fu and S.-T. Yau, {\it {The Theory of superstring with flux on non-Kahler
  manifolds and the complex Monge-Ampere equation}},  {\em J.Diff.Geom.} {\bf
  78} (2009) 369--428, [\href{http://xxx.lanl.gov/abs/hep-th/0604063}{{\tt
  hep-th/0604063}}].

\bibitem{Becker:2006et}
K.~Becker, M.~Becker, J.-X. Fu, L.-S. Tseng, and S.-T. Yau, {\it {Anomaly
  cancellation and smooth non-Kaehler solutions in heterotic string theory}},
  {\em Nucl. Phys.} {\bf B751} (2006) 108--128,
  [\href{http://xxx.lanl.gov/abs/hep-th/0604137}{{\tt hep-th/0604137}}].

\bibitem{Fu:2008ga}
J.-X. Fu, L.-S. Tseng, and S.-T. Yau, {\it {Local Heterotic Torsional Models}},
   {\em Commun. Math. Phys.} {\bf 289} (2009) 1151--1169,
  [\href{http://xxx.lanl.gov/abs/0806.2392}{{\tt arXiv:0806.2392}}].

\bibitem{carlevaro}
L.~Carlevaro, D.~Israel, and P.~M. Petropoulos, {\it {Double-Scaling Limit of
  Heterotic Bundles and Dynamical Deformation in CFT}},  {\em Nucl. Phys.} {\bf
  B827} (2010) 503--544, [\href{http://xxx.lanl.gov/abs/0812.3391}{{\tt
  arXiv:0812.3391}}].

\bibitem{Grana}
M.~Grana, {\it {Flux compactifications in string theory: A comprehensive
  review}},  {\em Phys. Rept.} {\bf 423} (2006) 91--158,
  [\href{http://xxx.lanl.gov/abs/hep-th/0509003}{{\tt hep-th/0509003}}].

\bibitem{Kachru:2003aw}
S.~Kachru, R.~Kallosh, A.~D. Linde, and S.~P. Trivedi, {\it {De Sitter vacua in
  string theory}},  {\em Phys.Rev.} {\bf D68} (2003) 046005,
  [\href{http://xxx.lanl.gov/abs/hep-th/0301240}{{\tt hep-th/0301240}}].

\bibitem{Denef:2005mm}
F.~Denef, M.~R. Douglas, B.~Florea, A.~Grassi, and S.~Kachru, {\it {Fixing all
  moduli in a simple f-theory compactification}},  {\em Adv.Theor.Math.Phys.}
  {\bf 9} (2005) 861--929, [\href{http://xxx.lanl.gov/abs/hep-th/0503124}{{\tt
  hep-th/0503124}}].

\bibitem{Camara:2007dy}
P.~G. Camara, E.~Dudas, T.~Maillard, and G.~Pradisi, {\it {String instantons,
  fluxes and moduli stabilization}},  {\em Nucl.Phys.} {\bf B795} (2008)
  453--489, [\href{http://xxx.lanl.gov/abs/0710.3080}{{\tt arXiv:0710.3080}}].

\bibitem{Argurio:2007qk}
R.~Argurio, M.~Bertolini, S.~Franco, and S.~Kachru, {\it {Meta-stable vacua and
  D-branes at the conifold}},  {\em JHEP} {\bf 0706} (2007) 017,
  [\href{http://xxx.lanl.gov/abs/hep-th/0703236}{{\tt hep-th/0703236}}].

\bibitem{Aharony:2007db}
O.~Aharony, S.~Kachru, and E.~Silverstein, {\it {Simple Stringy Dynamical SUSY
  Breaking}},  {\em Phys.Rev.} {\bf D76} (2007) 126009,
  [\href{http://xxx.lanl.gov/abs/0708.0493}{{\tt arXiv:0708.0493}}].

\bibitem{Aganagic:2007py}
M.~Aganagic, C.~Beem, and S.~Kachru, {\it {Geometric transitions and dynamical
  SUSY breaking}},  {\em Nucl.Phys.} {\bf B796} (2008) 1--24,
  [\href{http://xxx.lanl.gov/abs/0709.4277}{{\tt arXiv:0709.4277}}].

\bibitem{Blumenhagen:2009gk}
R.~Blumenhagen, J.~Conlon, S.~Krippendorf, S.~Moster, and F.~Quevedo, {\it
  {SUSY Breaking in Local String/F-Theory Models}},  {\em JHEP} {\bf 0909}
  (2009) 007, [\href{http://xxx.lanl.gov/abs/0906.3297}{{\tt
  arXiv:0906.3297}}].

\bibitem{Billo:2002hm}
M.~Billo, M.~Frau, I.~Pesando, F.~Fucito, A.~Lerda, {\em et.~al.}, {\it
  {Classical gauge instantons from open strings}},  {\em JHEP} {\bf 0302}
  (2003) 045, [\href{http://xxx.lanl.gov/abs/hep-th/0211250}{{\tt
  hep-th/0211250}}].

\bibitem{Blumenhagen:2006xt}
R.~Blumenhagen, M.~Cvetic, and T.~Weigand, {\it {Spacetime instanton
  corrections in 4D string vacua: The Seesaw mechanism for D-Brane models}},
  {\em Nucl.Phys.} {\bf B771} (2007) 113--142,
  [\href{http://xxx.lanl.gov/abs/hep-th/0609191}{{\tt hep-th/0609191}}].

\bibitem{Ibanez:2006da}
L.~Ibanez and A.~Uranga, {\it {Neutrino Majorana Masses from String Theory
  Instanton Effects}},  {\em JHEP} {\bf 0703} (2007) 052,
  [\href{http://xxx.lanl.gov/abs/hep-th/0609213}{{\tt hep-th/0609213}}].

\bibitem{Florea:2006si}
B.~Florea, S.~Kachru, J.~McGreevy, and N.~Saulina, {\it {Stringy Instantons and
  Quiver Gauge Theories}},  {\em JHEP} {\bf 0705} (2007) 024,
  [\href{http://xxx.lanl.gov/abs/hep-th/0610003}{{\tt hep-th/0610003}}].

\bibitem{Akerblom:2006hx}
N.~Akerblom, R.~Blumenhagen, D.~Lust, E.~Plauschinn, and M.~Schmidt-Sommerfeld,
  {\it {Non-perturbative SQCD Superpotentials from String Instantons}},  {\em
  JHEP} {\bf 0704} (2007) 076,
  [\href{http://xxx.lanl.gov/abs/hep-th/0612132}{{\tt hep-th/0612132}}].

\bibitem{Franco:2007ii}
S.~Franco, A.~Hanany, D.~Krefl, J.~Park, A.~M. Uranga, {\em et.~al.}, {\it
  {Dimers and orientifolds}},  {\em JHEP} {\bf 0709} (2007) 075,
  [\href{http://xxx.lanl.gov/abs/0707.0298}{{\tt arXiv:0707.0298}}].

\bibitem{Blumenhagen:2007zk}
R.~Blumenhagen, M.~Cvetic, D.~Lust, R.~Richter, and T.~Weigand, {\it
  {Non-perturbative Yukawa Couplings from String Instantons}},  {\em
  Phys.Rev.Lett.} {\bf 100} (2008) 061602,
  [\href{http://xxx.lanl.gov/abs/0707.1871}{{\tt arXiv:0707.1871}}].

\bibitem{Billo:2007sw}
M.~Billo, M.~Frau, I.~Pesando, P.~Di~Vecchia, A.~Lerda, {\em et.~al.}, {\it
  {Instantons in N=2 magnetized D-brane worlds}},  {\em JHEP} {\bf 0710} (2007)
  091, [\href{http://xxx.lanl.gov/abs/0708.3806}{{\tt arXiv:0708.3806}}].

\bibitem{Bianchi:2007wy}
M.~Bianchi, F.~Fucito, and J.~F. Morales, {\it {D-brane instantons on the T**6
  / Z(3) orientifold}},  {\em JHEP} {\bf 0707} (2007) 038,
  [\href{http://xxx.lanl.gov/abs/0704.0784}{{\tt arXiv:0704.0784}}].

\bibitem{Marchesano:2009rz}
F.~Marchesano and L.~Martucci, {\it {Non-perturbative effects on seven-brane
  Yukawa couplings}},  {\em Phys.Rev.Lett.} {\bf 104} (2010) 231601,
  [\href{http://xxx.lanl.gov/abs/0910.5496}{{\tt arXiv:0910.5496}}].

\bibitem{Billo':2010bd}
M.~Billo, M.~Frau, F.~Fucito, A.~Lerda, J.~F. Morales, {\em et.~al.}, {\it
  {Stringy instanton corrections to N=2 gauge couplings}},  {\em JHEP} {\bf
  1005} (2010) 107, [\href{http://xxx.lanl.gov/abs/1002.4322}{{\tt
  arXiv:1002.4322}}].

\bibitem{Bianchi:2011qh}
M.~Bianchi, A.~Collinucci, and L.~Martucci, {\it {Magnetized E3-brane
  instantons in F-theory}},  {\em JHEP} {\bf 1112} (2011) 045,
  [\href{http://xxx.lanl.gov/abs/1107.3732}{{\tt arXiv:1107.3732}}].

\bibitem{Aparicio:2011jx}
L.~Aparicio, A.~Font, L.~E. Ibanez, and F.~Marchesano, {\it {Flux and Instanton
  Effects in Local F-theory Models and Hierarchical Fermion Masses}},  {\em
  JHEP} {\bf 1108} (2011) 152, [\href{http://xxx.lanl.gov/abs/1104.2609}{{\tt
  arXiv:1104.2609}}].

\bibitem{Becker:2002nn}
K.~Becker, M.~Becker, M.~Haack, and J.~Louis, {\it {Supersymmetry breaking and
  alpha-prime corrections to flux induced potentials}},  {\em JHEP} {\bf 0206}
  (2002) 060, [\href{http://xxx.lanl.gov/abs/hep-th/0204254}{{\tt
  hep-th/0204254}}].

\bibitem{Balasubramanian:2005zx}
V.~Balasubramanian, P.~Berglund, J.~P. Conlon, and F.~Quevedo, {\it
  {Systematics of moduli stabilisation in Calabi-Yau flux compactifications}},
  {\em JHEP} {\bf 0503} (2005) 007,
  [\href{http://xxx.lanl.gov/abs/hep-th/0502058}{{\tt hep-th/0502058}}].

\bibitem{Conlon:2005ki}
J.~P. Conlon, F.~Quevedo, and K.~Suruliz, {\it {Large-volume flux
  compactifications: Moduli spectrum and D3/D7 soft supersymmetry breaking}},
  {\em JHEP} {\bf 0508} (2005) 007,
  [\href{http://xxx.lanl.gov/abs/hep-th/0505076}{{\tt hep-th/0505076}}].

\bibitem{Harvey:1995fq}
J.~A. Harvey and G.~W. Moore, {\it {Algebras, BPS States, and Strings}},  {\em
  Nucl. Phys.} {\bf B463} (1996) 315--368,
  [\href{http://xxx.lanl.gov/abs/hep-th/9510182}{{\tt hep-th/9510182}}].

\bibitem{jan}
J.~Troost, {\it {The non-compact elliptic genus: mock or modular}},  {\em JHEP}
  {\bf 1006} (2010) 104, [\href{http://xxx.lanl.gov/abs/1004.3649}{{\tt
  arXiv:1004.3649}}].

\bibitem{zwegers}
S.~Zwegers, {\it {Mock Theta Functions}},  {\em Ph.D thesis} (2002).

\bibitem{Zagier2}
D.~Zagier, {\it {Ramanujan's Mock Theta Functions and Their Applications
  [d'apr\`es Zwegers and Bringmann--Ono]}},  {\em S\'eminaire Bourbaki, 60\`eme
  ann\'ee, 2006--2007, no 986}.

\bibitem{Grimm:2010gk}
T.~W. Grimm, A.~Klemm, and D.~Klevers, {\it {Five-Brane Superpotentials,
  Blow-Up Geometries and SU(3) Structure Manifolds}},  {\em JHEP} {\bf 1105}
  (2011) 113, [\href{http://xxx.lanl.gov/abs/1011.6375}{{\tt
  arXiv:1011.6375}}].

\bibitem{murthy}
A.~Dabholkar, S.~Murthy, and D.~Zagier, {\it {Quantum Black Holes, Wall
  Crossing, and Mock Modular Forms}},
  \href{http://xxx.lanl.gov/abs/1208.4074}{{\tt arXiv:1208.4074}}.

\bibitem{Manschot:2011dj}
J.~Manschot, {\it {BPS invariants of N=4 gauge theory on a surface}},
  \href{http://xxx.lanl.gov/abs/1103.0012}{{\tt arXiv:1103.0012}}.

\bibitem{Manschot:2011ym}
J.~Manschot, {\it {BPS invariants of semi-stable sheaves on rational
  surfaces}},  \href{http://xxx.lanl.gov/abs/1109.4861}{{\tt arXiv:1109.4861}}.

\bibitem{Alexandrov:2012au}
S.~Alexandrov, J.~Manschot, and B.~Pioline, {\it {D3-instantons, Mock Theta
  Series and Twistors}},  \href{http://xxx.lanl.gov/abs/1207.1109}{{\tt
  arXiv:1207.1109}}.

\bibitem{Eguchi:2010ej}
T.~Eguchi, H.~Ooguri, and Y.~Tachikawa, {\it {Notes on the K3 Surface and the
  Mathieu group $M_{24}$}},  {\em Exper.Math.} {\bf 20} (2011) 91--96,
  [\href{http://xxx.lanl.gov/abs/1004.0956}{{\tt arXiv:1004.0956}}].

\bibitem{Gaberdiel:2010ca}
M.~R. Gaberdiel, S.~Hohenegger, and R.~Volpato, {\it {Mathieu Moonshine in the
  elliptic genus of K3}},  {\em JHEP} {\bf 1010} (2010) 062,
  [\href{http://xxx.lanl.gov/abs/1008.3778}{{\tt arXiv:1008.3778}}].

\bibitem{Eguchi:2010fg}
T.~Eguchi and K.~Hikami, {\it {Note on Twisted Elliptic Genus of K3 Surface}},
  {\em Phys.Lett.} {\bf B694} (2011) 446--455,
  [\href{http://xxx.lanl.gov/abs/1008.4924}{{\tt arXiv:1008.4924}}].

\bibitem{Cheng:2011ay}
M.~C. Cheng and J.~F. Duncan, {\it {On Rademacher Sums, the Largest Mathieu
  Group, and the Holographic Modularity of Moonshine}},
  \href{http://xxx.lanl.gov/abs/1110.3859}{{\tt arXiv:1110.3859}}.

\bibitem{Cheng:2012tq}
M.~C. Cheng, J.~F. Duncan, and J.~A. Harvey, {\it {Umbral Moonshine}},
  \href{http://xxx.lanl.gov/abs/1204.2779}{{\tt arXiv:1204.2779}}.

\bibitem{Eguchi:2012ye}
T.~Eguchi and K.~Hikami, {\it {N=2 Moonshine}},
  \href{http://xxx.lanl.gov/abs/1209.0610}{{\tt arXiv:1209.0610}}.

\bibitem{Eguchi:2010cb}
T.~Eguchi and Y.~Sugawara, {\it {Non-holomorphic Modular Forms and SL(2,R)/U(1)
  Superconformal Field Theory}},  {\em JHEP} {\bf 1103} (2011) 107,
  [\href{http://xxx.lanl.gov/abs/1012.5721}{{\tt arXiv:1012.5721}}].

\bibitem{Ashok:2011cy}
S.~K. Ashok and J.~Troost, {\it {A Twisted Non-compact Elliptic Genus}},  {\em
  JHEP} {\bf 1103} (2011) 067, [\href{http://xxx.lanl.gov/abs/1101.1059}{{\tt
  arXiv:1101.1059}}].

\bibitem{Sugawara:2011vg}
Y.~Sugawara, {\it {Comments on Non-holomorphic Modular Forms and Non-compact
  Superconformal Field Theories}},  {\em JHEP} {\bf 1201} (2012) 098,
  [\href{http://xxx.lanl.gov/abs/1109.3365}{{\tt arXiv:1109.3365}}].

\bibitem{Ashok:2012qy}
S.~K. Ashok and J.~Troost, {\it {Elliptic Genera of Non-compact Gepner Models
  and Mirror Symmetry}},  {\em JHEP} {\bf 1207} (2012) 005,
  [\href{http://xxx.lanl.gov/abs/1204.3802}{{\tt arXiv:1204.3802}}].

\bibitem{Bachas:1996zt}
C.~Bachas and C.~Fabre, {\it {Threshold effects in open string theory}},  {\em
  Nucl.Phys.} {\bf B476} (1996) 418--436,
  [\href{http://xxx.lanl.gov/abs/hep-th/9605028}{{\tt hep-th/9605028}}].

\bibitem{Antoniadis:1996vw}
I.~Antoniadis, C.~Bachas, C.~Fabre, H.~Partouche, and T.~Taylor, {\it {Aspects
  of type I - type II - heterotic triality in four-dimensions}},  {\em
  Nucl.Phys.} {\bf B489} (1997) 160--178,
  [\href{http://xxx.lanl.gov/abs/hep-th/9608012}{{\tt hep-th/9608012}}].

\bibitem{Kiritsis:1997hf}
E.~Kiritsis and N.~Obers, {\it {Heterotic type I duality in D < 10-dimensions,
  threshold corrections and D instantons}},  {\em JHEP} {\bf 9710} (1997) 004,
  [\href{http://xxx.lanl.gov/abs/hep-th/9709058}{{\tt hep-th/9709058}}].

\bibitem{bachas1}
C.~Bachas, C.~Fabre, E.~Kiritsis, N.~Obers, and P.~Vanhove, {\it {Heterotic /
  type I duality and D-brane instantons}},  {\em Nucl.Phys.} {\bf B509} (1998)
  33--52, [\href{http://xxx.lanl.gov/abs/hep-th/9707126}{{\tt
  hep-th/9707126}}].

\bibitem{Bachas:1997xn}
C.~Bachas, {\it {Heterotic versus Type I}},  {\em Nucl.Phys.Proc.Suppl.} {\bf
  68} (1998) 348--354, [\href{http://xxx.lanl.gov/abs/hep-th/9710102}{{\tt
  hep-th/9710102}}].

\bibitem{Lerche:1998gz}
W.~Lerche, S.~Stieberger, and N.~Warner, {\it {Quartic gauge couplings from K3
  geometry}},  {\em Adv.Theor.Math.Phys.} {\bf 3} (1999) 1575--1611,
  [\href{http://xxx.lanl.gov/abs/hep-th/9811228}{{\tt hep-th/9811228}}].

\bibitem{Lerche:1998nx}
W.~Lerche and S.~Stieberger, {\it {Prepotential, mirror map and F theory on
  K3}},  {\em Adv.Theor.Math.Phys.} {\bf 2} (1998) 1105--1140,
  [\href{http://xxx.lanl.gov/abs/hep-th/9804176}{{\tt hep-th/9804176}}].

\bibitem{Antoniadis:1999ge}
I.~Antoniadis, C.~Bachas, and E.~Dudas, {\it {Gauge couplings in
  four-dimensional type I string orbifolds}},  {\em Nucl.Phys.} {\bf B560}
  (1999) 93--134, [\href{http://xxx.lanl.gov/abs/hep-th/9906039}{{\tt
  hep-th/9906039}}].

\bibitem{Kiritsis:2000zi}
E.~Kiritsis, N.~A. Obers, and B.~Pioline, {\it {Heterotic / type II triality
  and instantons on K(3)}},  {\em JHEP} {\bf 0001} (2000) 029,
  [\href{http://xxx.lanl.gov/abs/hep-th/0001083}{{\tt hep-th/0001083}}].

\bibitem{Conlon:2009kt}
J.~P. Conlon and E.~Palti, {\it {Gauge Threshold Corrections for Local
  Orientifolds}},  {\em JHEP} {\bf 0909} (2009) 019,
  [\href{http://xxx.lanl.gov/abs/0906.1920}{{\tt arXiv:0906.1920}}].

\bibitem{Conlon:2009qa}
J.~P. Conlon and E.~Palti, {\it {On Gauge Threshold Corrections for Local
  IIB/F-theory GUTs}},  {\em Phys.Rev.} {\bf D80} (2009) 106004,
  [\href{http://xxx.lanl.gov/abs/0907.1362}{{\tt arXiv:0907.1362}}].

\bibitem{Berg:2005ja}
M.~Berg, M.~Haack, and B.~Kors, {\it {String loop corrections to Kahler
  potentials in orientifolds}},  {\em JHEP} {\bf 0511} (2005) 030,
  [\href{http://xxx.lanl.gov/abs/hep-th/0508043}{{\tt hep-th/0508043}}].

\bibitem{Berg:2005yu}
M.~Berg, M.~Haack, and B.~Kors, {\it {On volume stabilization by quantum
  corrections}},  {\em Phys.Rev.Lett.} {\bf 96} (2006) 021601,
  [\href{http://xxx.lanl.gov/abs/hep-th/0508171}{{\tt hep-th/0508171}}].

\bibitem{Maldacena:2000kv}
J.~M. Maldacena, H.~Ooguri, and J.~Son, {\it {Strings in AdS(3) and the SL(2,R)
  WZW model. II: Euclidean black hole}},  {\em J. Math. Phys.} {\bf 42} (2001)
  2961--2977, [\href{http://xxx.lanl.gov/abs/hep-th/0005183}{{\tt
  hep-th/0005183}}].

\bibitem{Callan:1991dj}
J.~Callan, Curtis~G., J.~A. Harvey, and A.~Strominger, {\it {World sheet
  approach to heterotic instantons and solitons}},  {\em Nucl. Phys.} {\bf
  B359} (1991) 611--634.

\bibitem{Cecotti:1992qh}
S.~Cecotti, P.~Fendley, K.~A. Intriligator, and C.~Vafa, {\it {A New
  supersymmetric index}},  {\em Nucl. Phys.} {\bf B386} (1992) 405--452,
  [\href{http://xxx.lanl.gov/abs/hep-th/9204102}{{\tt hep-th/9204102}}].

\bibitem{Cecotti:1992vy}
S.~Cecotti and C.~Vafa, {\it {Ising model and N=2 supersymmetric theories}},
  {\em Commun. Math. Phys.} {\bf 157} (1993) 139--178,
  [\href{http://xxx.lanl.gov/abs/hep-th/9209085}{{\tt hep-th/9209085}}].

\bibitem{Abouelsaood:1986gd}
A.~Abouelsaood, J.~Callan, Curtis~G., C.~Nappi, and S.~Yost, {\it {Open Strings
  in Background Gauge Fields}},  {\em Nucl.Phys.} {\bf B280} (1987) 599.

\bibitem{Bachas:1992bh}
C.~Bachas and M.~Porrati, {\it {Pair creation of open strings in an electric
  field}},  {\em Phys.Lett.} {\bf B296} (1992) 77--84,
  [\href{http://xxx.lanl.gov/abs/hep-th/9209032}{{\tt hep-th/9209032}}].

\bibitem{kiritsisbook}
E.~Kiritsis, {\it {String theory in a nutshell}},  {\em Princeton University
  Press} (2007).

\bibitem{Camara:2008zk}
P.~G. Camara and E.~Dudas, {\it {Multi-instanton and string loop corrections in
  toroidal orbifold models}},  {\em JHEP} {\bf 08} (2008) 069,
  [\href{http://xxx.lanl.gov/abs/0806.3102}{{\tt arXiv:0806.3102}}].

\bibitem{Obers:1999um}
N.~Obers and B.~Pioline, {\it {Eisenstein series and string thresholds}},  {\em
  Commun.Math.Phys.} {\bf 209} (2000) 275--324,
  [\href{http://xxx.lanl.gov/abs/hep-th/9903113}{{\tt hep-th/9903113}}].

\bibitem{Obers:1999es}
N.~A. Obers and B.~Pioline, {\it {Eisenstein series in string theory}},  {\em
  Class.Quant.Grav.} {\bf 17} (2000) 1215--1224,
  [\href{http://xxx.lanl.gov/abs/hep-th/9910115}{{\tt hep-th/9910115}}].

\bibitem{Stieberger:1998yi}
S.~Stieberger, {\it {(0,2) heterotic gauge couplings and their M theory
  origin}},  {\em Nucl.Phys.} {\bf B541} (1999) 109--144,
  [\href{http://xxx.lanl.gov/abs/hep-th/9807124}{{\tt hep-th/9807124}}].

\bibitem{Witten:1986bf}
E.~Witten, {\it {Elliptic Genera and Quantum Field Theory}},  {\em Commun.
  Math. Phys.} {\bf 109} (1987) 525.

\bibitem{eguchi1}
T.~Eguchi and A.~Taormina, {\it {On the Unitary Representations of N=2 and N=4
  Superconformal Algebras}},  {\em Phys. Lett.} {\bf B210} (1988) 125.

\bibitem{Eguchi:2004yi}
T.~Eguchi and Y.~Sugawara, {\it {SL(2,R)/U(1) supercoset and elliptic genera of
  non-compact Calabi-Yau manifolds}},  {\em JHEP} {\bf 05} (2004) 014,
  [\href{http://xxx.lanl.gov/abs/hep-th/0403193}{{\tt hep-th/0403193}}].

\bibitem{mordell}
L.~Mordell, {\it {The definite integral $\int\limits_{ - \infty }^\infty
  {\tfrac{{e^{ax^2 + bx} }}{{e^{ax} + d}}da}$ and the analytic theory of
  numbersand the analytic theory of numbers}},  {\em Acta Mathematica} {\bf 61}
  (1933) 323.

\bibitem{Eguchi:1988vra}
T.~Eguchi, H.~Ooguri, A.~Taormina, and S.-K. Yang, {\it {Superconformal
  Algebras and String Compactification on Manifolds with SU(N) Holonomy}},
  {\em Nucl. Phys.} {\bf B315} (1989) 193.

\bibitem{Eguchi:2008gc}
T.~Eguchi and K.~Hikami, {\it {Superconformal Algebras and Mock Theta
  Functions}},  {\em J.Phys.A} {\bf A42} (2009) 304010,
  [\href{http://xxx.lanl.gov/abs/0812.1151}{{\tt arXiv:0812.1151}}].

\bibitem{Eguchi:2009cq}
T.~Eguchi and K.~Hikami, {\it {Superconformal Algebras and Mock Theta Functions
  2. Rademacher Expansion for K3 Surface}},
  \href{http://xxx.lanl.gov/abs/0904.0911}{{\tt arXiv:0904.0911}}.

\bibitem{Hohenegger:2011us}
S.~Hohenegger and S.~Stieberger, {\it {BPS Saturated String Amplitudes: K3
  Elliptic Genus and Igusa Cusp Form}},  {\em Nucl.Phys.} {\bf B856} (2012)
  413--448, [\href{http://xxx.lanl.gov/abs/1108.0323}{{\tt arXiv:1108.0323}}].

\bibitem{Aldazabal:1997wi}
G.~Aldazabal, A.~Font, L.~E. Ibanez, A.~Uranga, and G.~Violero, {\it
  {Nonperturbative heterotic D = 6, D = 4, N=1 orbifold vacua}},  {\em
  Nucl.Phys.} {\bf B519} (1998) 239--281,
  [\href{http://xxx.lanl.gov/abs/hep-th/9706158}{{\tt hep-th/9706158}}].

\bibitem{Grootwp}
L.~Carlevaro and S.~Groot~Nibbelink, {\it {Heterotic models on warped
  Eguchi-Hanson}},  \href{http://xxx.lanl.gov/abs/in preparation}{{\tt in
  preparation}}.

\bibitem{Dixon:1990pc}
L.~J. Dixon, V.~Kaplunovsky, and J.~Louis, {\it {Moduli dependence of string
  loop corrections to gauge coupling constants}},  {\em Nucl.Phys.} {\bf B355}
  (1991) 649--688.

\bibitem{McClain:1986id}
B.~McClain and B.~D.~B. Roth, {\it {Modular invariance for interacting bosonic
  strings at finite temperature}},  {\em Commun.Math.Phys.} {\bf 111} (1987)
  539.

\bibitem{Ditsas:1988pm}
P.~Ditsas and E.~Floratos, {\it {Finite volume temperature closed bosonic
  string in finite volume}},  {\em Phys.Lett.} {\bf B201} (1988) 49--53.

\bibitem{Kutasov:1990sv}
D.~Kutasov and N.~Seiberg, {\it {Number of degrees of freedom, density of
  states and tachyons in string theory and CFT}},  {\em Nucl.Phys.} {\bf B358}
  (1991) 600--618.

\bibitem{Angelantonj:2011br}
C.~Angelantonj, I.~Florakis, and B.~Pioline, {\it {A new look at one-loop
  integrals in string theory}},  \href{http://xxx.lanl.gov/abs/1110.5318}{{\tt
  arXiv:1110.5318}}.

\bibitem{Angelantonj:2012gw}
C.~Angelantonj, I.~Florakis, and B.~Pioline, {\it {One-Loop BPS amplitudes as
  BPS-state sums}},  {\em JHEP} {\bf 1206} (2012) 070,
  [\href{http://xxx.lanl.gov/abs/1203.0566}{{\tt arXiv:1203.0566}}].

\bibitem{Cardella:2008nz}
M.~Cardella, {\it {A Novel method for computing torus amplitudes for Z(N)
  orbifolds without the unfolding technique}},  {\em JHEP} {\bf 0905} (2009)
  010, [\href{http://xxx.lanl.gov/abs/0812.1549}{{\tt arXiv:0812.1549}}].

\bibitem{Cardella:2010bq}
M.~A. Cardella, {\it {Error Estimates in Horocycle Averages Asymptotics:
  Challenges from String Theory}},
  \href{http://xxx.lanl.gov/abs/1012.2754}{{\tt arXiv:1012.2754}}.

\bibitem{Angelantonj:2010ic}
C.~Angelantonj, M.~Cardella, S.~Elitzur, and E.~Rabinovici, {\it {Vacuum
  stability, string density of states and the Riemann zeta function}},  {\em
  JHEP} {\bf 1102} (2011) 024, [\href{http://xxx.lanl.gov/abs/1012.5091}{{\tt
  arXiv:1012.5091}}].

\bibitem{nieb}
D.~Niebur, {\it {A class of non analytic automorphic functions}},  {\em Nagoya
  Math. J.} {\bf 52} (1973) 133--145.

\bibitem{hej}
D.~A. Hejhal, {\it {The Selberg trace formula for PSL(2, R)}},  {\em Lecture
  Notes in Math. Vol 2., Springer, no. 1001} (1983).

\bibitem{bs}
M.~Bianchi and A.~Sagnotti, {\it {Twist symmetry and open string Wilson
  lines}},  {\em Nucl.Phys.} {\bf B361} (1991) 519--538.

\bibitem{gp}
E.~G. Gimon and J.~Polchinski, {\it {Consistency conditions for orientifolds
  and d manifolds}},  {\em Phys.Rev.} {\bf D54} (1996) 1667--1676,
  [\href{http://xxx.lanl.gov/abs/hep-th/9601038}{{\tt hep-th/9601038}}].

\bibitem{bruinier1}
J.~H. Bruinier, {\it {Borcherds products on O(2, l) and Chern classes of
  Heegner divisors}},  {\em Berlin: Springer} (2002).

\bibitem{bring}
K.~Bringmann and K.~Ono, {\it {Arithmetic properties of coefficients of
  half-integral weight Maass-Poincare series}},  {\em Math. Ann. Number 3} {\bf
  337} (2007) 59--612.

\bibitem{on}
K.~Ono, {\it {A mock theta function for the delta-function}},  {\em Berlin:
  Walter de Gruyter} (2009).

\bibitem{bruinier2}
J.~Bruinier and K.~Ono, {\it {Heegner divisors, L-functions and harmonic weak
  Maass forms}},  {\em Ann. Math. Issue 3} {\bf 172} (2010) 2135--2181.

\bibitem{RoblesLlana:2006is}
D.~Robles-Llana, M.~Rocek, F.~Saueressig, U.~Theis, and S.~Vandoren, {\it
  {Nonperturbative corrections to 4D string theory effective actions from
  SL(2,Z) duality and supersymmetry}},  {\em Phys.Rev.Lett.} {\bf 98} (2007)
  211602, [\href{http://xxx.lanl.gov/abs/hep-th/0612027}{{\tt
  hep-th/0612027}}].

\bibitem{Alexandrov:2008gh}
S.~Alexandrov, B.~Pioline, F.~Saueressig, and S.~Vandoren, {\it {D-instantons
  and twistors}},  {\em JHEP} {\bf 0903} (2009) 044,
  [\href{http://xxx.lanl.gov/abs/0812.4219}{{\tt arXiv:0812.4219}}].

\bibitem{Bachas:1996bp}
C.~Bachas and E.~Kiritsis, {\it {F(4) terms in N=4 string vacua}},  {\em
  Nucl.Phys.Proc.Suppl.} {\bf 55B} (1997) 194--199,
  [\href{http://xxx.lanl.gov/abs/hep-th/9611205}{{\tt hep-th/9611205}}].

\bibitem{Kaplunovsky:1994fg}
V.~Kaplunovsky and J.~Louis, {\it {Field dependent gauge couplings in locally
  supersymmetric effective quantum field theories}},  {\em Nucl.Phys.} {\bf
  B422} (1994) 57--124, [\href{http://xxx.lanl.gov/abs/hep-th/9402005}{{\tt
  hep-th/9402005}}].

\bibitem{Kaplunovsky:1995jw}
V.~Kaplunovsky and J.~Louis, {\it {On Gauge couplings in string theory}},  {\em
  Nucl.Phys.} {\bf B444} (1995) 191--244,
  [\href{http://xxx.lanl.gov/abs/hep-th/9502077}{{\tt hep-th/9502077}}].

\bibitem{gradsh}
I.~S. Gradshteyn and I.~M. Ryzhik, {\it {Tables of integrals, series and
  products}},  {\em Academic Press: New York} (1965) 963.

\bibitem{gpsw}
M.~Berkooz, R.~G. Leigh, J.~Polchinski, J.~H. Schwarz, N.~Seiberg, {\em
  et.~al.}, {\it {Anomalies, dualities, and topology of D = 6 N=1 superstring
  vacua}},  {\em Nucl.Phys.} {\bf B475} (1996) 115--148,
  [\href{http://xxx.lanl.gov/abs/hep-th/9605184}{{\tt hep-th/9605184}}].

\bibitem{Derendinger}
J.~Derendinger, S.~Ferrara, C.~Kounnas, and F.~Zwirner, {\it {On loop
  corrections to string effective field theories: Field dependent gauge
  couplings and sigma model anomalies}},  {\em Nucl.Phys.} {\bf B372} (1992)
  145--188.

\bibitem{Nilles}
H.~P. Nilles and S.~Stieberger, {\it {String unification, universal one loop
  corrections and strongly coupled heterotic string theory}},  {\em Nucl.Phys.}
  {\bf B499} (1997) 3--28, [\href{http://xxx.lanl.gov/abs/hep-th/9702110}{{\tt
  hep-th/9702110}}].

\bibitem{Petersson:2007sc}
C.~Petersson, {\it {Superpotentials From Stringy Instantons Without
  Orientifolds}},  {\em JHEP} {\bf 0805} (2008) 078,
  [\href{http://xxx.lanl.gov/abs/0711.1837}{{\tt arXiv:0711.1837}}].

\bibitem{Akerblom:2007uc}
N.~Akerblom, R.~Blumenhagen, D.~Lust, and M.~Schmidt-Sommerfeld, {\it
  {Instantons and Holomorphic Couplings in Intersecting D-brane Models}},  {\em
  JHEP} {\bf 0708} (2007) 044, [\href{http://xxx.lanl.gov/abs/0705.2366}{{\tt
  arXiv:0705.2366}}].

\bibitem{Blumenhagen:2007ip}
R.~Blumenhagen and M.~Schmidt-Sommerfeld, {\it {Gauge Thresholds and Kaehler
  Metrics for Rigid Intersecting D-brane Models}},  {\em JHEP} {\bf 0712}
  (2007) 072, [\href{http://xxx.lanl.gov/abs/0711.0866}{{\tt
  arXiv:0711.0866}}].

\bibitem{GarciaEtxebarria:2007zv}
I.~Garcia-Etxebarria and A.~M. Uranga, {\it {Non-perturbative superpotentials
  across lines of marginal stability}},  {\em JHEP} {\bf 0801} (2008) 033,
  [\href{http://xxx.lanl.gov/abs/0711.1430}{{\tt arXiv:0711.1430}}].

\bibitem{GarciaEtxebarria:2008pi}
I.~Garcia-Etxebarria, F.~Marchesano, and A.~M. Uranga, {\it {Non-perturbative
  F-terms across lines of BPS stability}},  {\em JHEP} {\bf 0807} (2008) 028,
  [\href{http://xxx.lanl.gov/abs/0805.0713}{{\tt arXiv:0805.0713}}].

\bibitem{Blumenhagen:2007bn}
R.~Blumenhagen, M.~Cvetic, R.~Richter, and T.~Weigand, {\it {Lifting
  D-Instanton Zero Modes by Recombination and Background Fluxes}},  {\em JHEP}
  {\bf 0710} (2007) 098, [\href{http://xxx.lanl.gov/abs/0708.0403}{{\tt
  arXiv:0708.0403}}].

\bibitem{Blumenhagen:2009qh}
R.~Blumenhagen, M.~Cvetic, S.~Kachru, and T.~Weigand, {\it {D-Brane Instantons
  in Type II Orientifolds}},  {\em Ann.Rev.Nucl.Part.Sci.} {\bf 59} (2009)
  269--296, [\href{http://xxx.lanl.gov/abs/0902.3251}{{\tt arXiv:0902.3251}}].

\bibitem{Gukov:2002nw}
S.~Gukov and C.~Vafa, {\it {Rational conformal field theories and complex
  multiplication}},  {\em Commun.Math.Phys.} {\bf 246} (2004) 181--210,
  [\href{http://xxx.lanl.gov/abs/hep-th/0203213}{{\tt hep-th/0203213}}].

\bibitem{MohriKawai}
T.~Kawai and K.~Mohri, {\it {Geometry of (0,2) Landau-Ginzburg orbifolds}},
  {\em Nucl.Phys.} {\bf B425} (1994) 191--216,
  [\href{http://xxx.lanl.gov/abs/hep-th/9402148}{{\tt hep-th/9402148}}].

\bibitem{Gepner:1987qi}
D.~Gepner, {\it {Space-Time Supersymmetry in Compactified String Theory and
  Superconformal Models}},  {\em Nucl. Phys.} {\bf B296} (1988) 757.

\bibitem{Eguchi:2003ik}
T.~Eguchi and Y.~Sugawara, {\it {Modular bootstrap for boundary N = 2 Liouville
  theory}},  {\em JHEP} {\bf 01} (2004) 025,
  [\href{http://xxx.lanl.gov/abs/hep-th/0311141}{{\tt hep-th/0311141}}].

\bibitem{Israel:2004xj}
D.~Israel, A.~Pakman, and J.~Troost, {\it {Extended SL(2,R) / U(1) characters,
  or modular properties of a simple nonrational conformal field theory}},  {\em
  JHEP} {\bf 0404} (2004) 043,
  [\href{http://xxx.lanl.gov/abs/hep-th/0402085}{{\tt hep-th/0402085}}].

\bibitem{Eguchi:1987sm}
T.~Eguchi and A.~Taormina, {\it {Unitary representations of N=4 superconformal
  algebra}},  {\em Phys.Lett.} {\bf B196} (1987) 75.

\bibitem{Eguchi:1987wf}
T.~Eguchi and A.~Taormina, {\it {Character formulas for the N=4 superconformal
  algebra}},  {\em Phys.Lett.} {\bf B200} (1988) 315.

\bibitem{borcherds}
R.~Borcherds, {\it {Automorphic forms with singularities on Grassmannians}},
  {\em Invent. Math.} {\bf 132} (1998) 491--562,
  [\href{http://xxx.lanl.gov/abs/alg-geom/9609022}{{\tt alg-geom/9609022}}].

\end{thebibliography}\endgroup

\end{document}